\RequirePackage{fix-cm}

\documentclass[twocolumn,epjc3]{svjour3}  
\smartqed  % flush right qed marks, e.g. at end of proof
\RequirePackage{graphicx}
\RequirePackage{makecell}
\RequirePackage{url}
\RequirePackage{subfigure}
\journalname{}
\begin{document}

\title{XAI-based Feature Selection for Improved Network Intrusion Detection Systems%\thanksref{t1}
}
% \subtitle{Do you have a subtitle?\\ If so, write it here}


\author{Osvaldo Arreche\thanksref{e1,addr1}
        \and
        Tanish Guntur\thanksref{e2,addr2} %etc.
        \and
        Mustafa Abdallah\thanksref{e3,addr3} %etc.
}

\thankstext{e1}{e-mail: oarreche@iu.edu}
\thankstext{e2}{e-mail: tguntur@iu.edu}
\thankstext{e3}{Correspondent e-mail: abdalla0@purdue.edu}


\institute{Purdue School of Engineering and Technology, Purdue University in Indianapolis, IN, USA \label{addr1}
           \and
           Computer Science Department, Purdue University in Indianapolis, IN, USA\label{addr2}
           \and
           Computer and Information Technology Department, Purdue University in Indianapolis, IN, USA\label{addr3}
           % \and
           % \emph{Present Address:} if needed\label{addr4}
}

\date{Received: date / Accepted: date}

\maketitle

\begin{abstract}
Explainability and evaluation of AI models are crucial parts of the security of modern intrusion detection systems (IDS) in the network security field, yet they are lacking. Accordingly, feature selection is essential for such parts in IDS because it identifies the most paramount features, enhancing attack detection and its description. In this work, we tackle the feature selection problem for IDS by suggesting new ways of applying eXplainable AI (XAI) methods for this problem. We identify the crucial attributes originated by distinct AI methods in tandem with the novel five attribute selection methods. We then compare many state-of-the-art feature selection strategies with our XAI-based feature selection methods, showing that most AI models perform better when using the XAI-based approach proposed in this work. By providing novel feature selection techniques and establishing the foundation for several XAI-based strategies, this research aids security analysts in the AI decision-making reasoning of IDS by providing them with a better grasp of critical intrusion traits. Furthermore, we make the source codes available so that the community may develop additional models on top of our foundational XAI-based feature selection framework.
\keywords{Intrusion Detection Systems \and Explainable AI \and  Feature Selection \and Network Security \and CICIDS-2017\and RoEduNet-SIMARGL2021.}
\end{abstract}

\section{Introduction}
\label{sec1}

An intrusion detection system's (IDS) primary goal is to locate and identify any unauthorized activity, including efforts to misuse or abuse computer networks from the inside as well as the outside~\cite{northcutt2002network,vasiliadis2008gnort}. Traditional IDS detects the majority of potential threats by assuming that the behavior of attackers differs considerably from that of normal network traffic. Statistical anomaly~\cite{patcha2007overview,asad2022dynamical} and rule-based misuse models~\cite{ilgun1995state,li2010novel} have been the subject of numerous research studies in the literature to detect intrusions for different applications in real-world systems~\cite{snapp1992dids,jackson1991expert}. Nevertheless, the spread of diverse computer networks and the growing complexity of assaults provide difficulties for intrusion detection, allowing attackers to carry out increasingly complex attacks without being quickly discovered~\cite{muhammad2022intelligent,tabassum2019survey}.

Due to these difficulties, automated intrusion detection systems have been developed with the use of artificial intelligence (AI)~\cite{buczak2015survey,dina2021intrusion}. These artificial intelligence techniques include naive bayes~\cite{amor2004naive,panigrahi2022intrusion}, random forest~\cite{balyan2022hybrid,waskle2020intrusion,negandhi2019intrusion}, artificial neural networks~\cite{kim2017method,tang2020saae}, support vector machines~\cite{tao2018improved,deng2003svm}, decision trees~\cite{ingre2018decision,ferrag2020rdtids,al2021intelligent}, and others. These methods improve the system's capacity to recognize and neutralize harmful activity and allow intrusion detection to be automated.

There are several drawbacks to completely automated AI approaches when it comes to real-world IDS. First of all, these models frequently provide incorrect positive and negative predictions, which might have disastrous effects. False forecasts are especially concerning in large-scale businesses and vital infrastructures where a network breach can cause substantial resource and financial loss. The Twitter hack in July 2020~\cite{attacks2020} and the May 2021~\cite{colonial_pipeline_ransomwareattack} ransomware attack, for instance, demonstrate the possible repercussions of inaccurate IDS forecasts.
Second, the majority of suggested techniques are regarded as black-box prediction models, except decision trees in basic IDS. Put differently, these judgments made by the AI-based IDS are difficult to understand and do not provide an explanation to human security analysts~\cite{arisdakessian2022survey,sabev2020integrated}. These earlier AI-based works have mostly concentrated on classification accuracy, ignoring the requirement for interpretability and comprehension of the AI's behavior.
The aforementioned constraints underscore the necessity of using the nascent domain of explainable AI (XAI) to augment the comprehensibility of AI-generated judgments within IDS~\cite{das2020opportunities}. The IDS can explain the decision-making process by utilizing XAI approaches, which help human analysts comprehend and validate the system's outputs. \sloppy

The use of explainable AI (XAI) in IDS has been the subject of several recent studies~\cite{mahbooba2021explainable,patil2022explainable,islam2019domain,roponena2022towards}. A specific study~\cite{roponena2022towards} presented an abstract idea for an intelligent IDS method that involves the human in the loop. Furthermore, another study that only looked at the NSL-KDD benchmark dataset~\cite{dhanabal2015study} extracted decision rules using a straightforward decision tree algorithm~\cite{mahbooba2021explainable}. Large-scale datasets may not benefit from this strategy since it can produce enormous trees with more difficult-to-understand properties. 
Moreover, these studies did not investigate important aspects of how each feature influences the AI's choice, how important characteristics related to an assault are, how to extract important incursion features, or how to investigate a variety of AI models. We attempt to address this gap in the literature on network intrusion detection with our present research project. Our focus is on the feature selection problem, where we employ XAI techniques to determine the primary intrusion attributes that influence AI model decisions. Additionally, we take into account many critical elements that have been overlooked in previous network intrusion detection research.

The high-level overview of the extensive XAI-based system that this study presents to improve feature selection in IDS is shown in Figure~\ref{fig:Flow_code}. Our system provides specific global feature selection techniques that reflect the joint effect of the primary characteristics retrieved from network traffic on the output of the AI model. Additionally, it presents five unique XAI-based feature extraction techniques that pinpoint important aspects of network penetration. It isolates the key characteristics of each sort of intrusion that make up that specific incursion. Furthermore, by efficiently utilizing XAI approaches, our framework selects the overall most significant features from various AI models for every database, improving feature selection for such AI models in intrusion detection tasks. As such, it can help security analysts comprehend the characteristics of network traffic records.

We perform a thorough analysis of our framework on two different network intrusion datasets. The first dataset (called RoEduNet-SIMARGL2021~\cite{mihailescu2021proposition}) is a recent compilation from the European Union-funded SIMARGL project. Realistic network traffic data, comprising attributes derived from real-time traffic, makes up this dataset. The dataset is invaluable for building deployable network intrusion detection systems because of this feature. The second dataset, CICIDS-2017~\cite{panigrahi2018detailed}, was produced in 2017 by the University of Brunswick's Canadian Institute for Cybersecurity and is a well-known benchmark intrusion detection dataset.
It has a range of assault profiles. We assess the performance of several black-box AI models for every dataset, evaluating these models sequentially based on their performance on several incursion types and the typical overall criteria. The key characteristics for these various models are then extracted using various XAI-based techniques (described in Section~\ref{sec:proposed_feat_selec}). Subsequently, we assess our feature selection techniques by presenting the total feature importance, the contribution of each feature to the AI's judgment, and the most important characteristics related to each sort of intrusion.

We rate the efficacy of our framework by contrasting its results with those of other feature selection methods used in network intrusion detection systems, such as feature correlation~\cite{feat_corr_citation}, Chi-square~\cite{gajawada2019chi}, feature importance~\cite{dutta2018analysing}, information gain~\cite{CICIDS}, K-best~\cite{SENSOR}, and Xplique~\cite{xplique}. Based on this benchmark comparison, we show that most AI models outperform the baseline methods when they use the most important characteristics chosen by our framework.

    \textbf{Overview of Contributions:}
The primary contributions of our study are outlined below. 
\begin{itemize}

    \item We offer a comprehensive architecture to improve feature selection for network intrusion detection systems.
    
    \item We present four different XAI-based feature extraction techniques to discover important network intrusion traits. 
        
    \item We extract significant attributes that are particular to the model and the intrusion for several classes of AI models and various kinds of network intrusions. 
    
    \item We present an empirical comparison of seven black-box AI models on one synthetic and one real-world network intrusion dataset.

    \item We make our source codes available to the community as an XAI-based framework for feature selection in network intrusion detection tasks so anyone may use them and expand upon them with other datasets and models.\footnote{Our source codes may be accessed at: %\mustafa{name of this link should not have ACSAC} 
    \\ \url{https://github.com/ogarreche/XAI_Feature_Selection}}
\end{itemize}

\textbf{Paper Structure:}  This is how the rest of the paper is structured. Section~\ref{sec:background} describes the problem definition and the initial steps in detecting network intrusions. The principal elements of our framework are outlined in Section~\ref{sec:framwork}. The evaluation of the suggested feature selection methodology using network intrusion datasets is presented in Section~\ref{sec:evaluation}. The previous related work is displayed in Section~\ref{sec:related_work}. Limitations and related Discussions are presented in Section~\ref{sec: Discussion}. Conclusion and prospects for the study are presented in Section~\ref{sec: conclusion}.

\section{The Problem Statement}\label{sec:background}

This section explains the basic ideas behind network intrusion detection, the problems with black-box AI, and why feature selection must be done using XAI-based techniques.

\subsection{Network Intrusion Types}

Many forms of network intrusions are often encountered. In this work~\cite{strom2018mitre}, we examine the primary network attacks inside the standard MITRE ATT\&CK architecture. As a result, the following categories are used to group the network traffic:

\textbf{Benign traffic:} Benign traffic is the regular data flow that is gathered from the network.

\textbf{Network Service Discovery / PortScan (PS) [T1046, MITRE ATT\&CK ID]:} The goal of the intrusion is for the attacker to identify the victim's computer. Finding weak spots and potential entry sites is frequently the initial stage of an attack. Without actually completing the connection, the functionality is to send the victim a connection solicitation. But in this attack, requests are sent to a variety of ports, and the ports that respond with a message are mapped as potential sites of entry~\cite{lee2003detection}.

\textbf{Network Denial of Service [MITRE ATT\&CK ID: T1498] / Denial of Service (DoS):} The goal of this kind of assault is to make the target inaccessible to the network. A common illustration of this kind of attack is when a hacker persistently sends requests to establish a connection with a server. Nevertheless, the server never hears back from the origin after accepting the solicitation and sending an acknowledgment in the hopes of receiving a response. Consequently, until the server goes down, the memory for these open interactions is kept open and fully utilized. For further information on certain kinds of DoS assaults, see~\cite{CICIDS,SENSOR}.

\textbf{Brute Force [T1110, MITRE ATT\&CK ID]:} This type of attack involves the attacker trying every possible password to gain access to the victim's network. Frequently, this technique is combined with the ability to guess the most popular passwords. When a person uses a password that is weak or often guessed, it can become effective~\cite{CICIDS}.

\textbf{Initial Access/Web Attack [MITRE ATT\&CK ID: TA0001, T16 59, T1189]:} This is a kind of online assault where the attacker takes advantage of security holes in websites. An attacker may, for instance, use a software defect, misconfiguration, or malfunction to exploit a public-facing application and obtain access to the program's underlying instance or container. Drive-by Compromise assaults are also included in the category of web attacks~\cite{Web_attack_Ref2}; however, web attacks (such as SQL injection and cross-site scripting) usually do not grant initial access to a remote server~\cite{strom2018mitre}.

\textbf{First Access/Infiltration [MITRE ATT\&CK ID: TA0001]:} This attack happens when someone tries to obtain initial access to a system or application. Attacks of this kind can employ a wide range of strategies, including targeted spearfishing and taking advantage of flaws on web servers that are visible to the public. This assault can build its base by anything from a straightforward password reset to ongoing access via legitimate accounts and third-party remote services.

\textbf{Infrastructure Compromise/Botnet [MITRE ATT\&CK ID: T1 584.005, T1059, T236, T1070]:} This type of automated assault is carried out via remotely operated machines that have been taken over by the attacker, where programs, or bots, replicate and imitate human behavior~\cite{stone2009your}. Its scripting approach makes it possible to scale and deploy it easily, which makes it the perfect tool for hitting several assault locations at once. As a result, it is a popular form of network assault on many computer networks.
%\vspace{-1mm}

\subsection{Intrusion Detection Systems}

Critical infrastructure across a range of industries is seriously threatened by the previously described sophisticated network attacks~\cite{khan2021m2mon,hussain2021noncompliance}. These dangers undermine trust in the security of stored data in addition to jeopardizing service availability~\cite{hussain2021noncompliance}. IDS is used in computer network systems to detect and reduce harmful activity carried out by external and internal intruders~\cite{mirzaei2021scrutinizer}. Traditional IDS are often designed with the understanding that the behaviors of an intruder differ significantly from those of a legal user and that many unauthorized acts are detectable~\cite{lukacs2015strongly}. Researchers have investigated the application of AI models to automate the detection of network intrusions~\cite{buczak2015survey,dina2021intrusion}, using recent advances in AI.

\subsection{The drawbacks of Black-box AI Models}

An important contribution to automating intrusion detection has been made by AI models. Nonetheless, because of the intricate interactions between the features in the models that go into producing their results, they are inherently black-box because of their innate complexity, understanding how and why the AI model produces its outputs is quite difficult. The black-box problem affects many AI models, such as deep neural networks and random forests. Though they have been successfully applied in many areas and have a high prediction accuracy, it is still difficult to understand the reasoning behind the models' actions, especially when there are faults or blunders. Particularly in safety-critical applications such as network security via intrusion detection systems (IDS), these mistakes might have serious repercussions.
To overcome these difficulties, it is important to offer justifications for the choices made by AI models in IDS~\cite{botacin2021challenges}. Extraction and comprehension of the primary features influencing these AI models' actions in IDS constitute a primary step toward meeting this demand.

\vspace{-1mm}

\subsection{Explainable AI}

The area of explainable AI (XAI) has just come into existence as a means of addressing the black-box character of most AI models. With a particular focus on comprehending the elements that go into a model's decision-making process in tasks including classification or regression, XAI seeks to shed light on the interpretability of AI models and it includes several subfields, such as methods that provide explanations for the behavior of AI models at the local level (unique to individual instances) or the global level (across several data instances). These methods can be model-agnostic, which means they can explain any kind of AI model, or model-specific (i.e., they are made to explain particular kinds of AI models). 
Furthermore, XAI uses explanation techniques such as feature value evaluation to provide a feature priority list for an AI model, or building surrogate models to explain complicated models using simpler ones.

\subsection{Benefits of XAI for Network IDS}

The existing AI-based intrusion detection system produces output that is difficult for human analysts to comprehend. Analysts must thus manually review a large number of records in order to spot unusual activity. Furthermore, they encounter difficulties in locating and resolving problems with the AI models themselves. Therefore, in the field of network intrusion detection, obtaining high-accuracy results and also elucidating the behavior and decision-making process of the AI system becomes paramount. 
Trust in the AI algorithms will be established if security analysts can understand the judgments that the algorithms make. Given this, our XAI-based approach, which selects intrusion features using black-box XAI techniques, represents a significant advancement in the area.

After outlining the essential prerequisites for network intrusion detection and the necessity of improving feature selection via the use of XAI-based methodologies, we go on to describe the key elements of our XAI-based framework for choosing primary intrusion features in network intrusion detection assignments.

\section{Framework}\label{sec:framwork}

\begin{figure}[t]
   \centering
\includegraphics[width=0.95\linewidth]{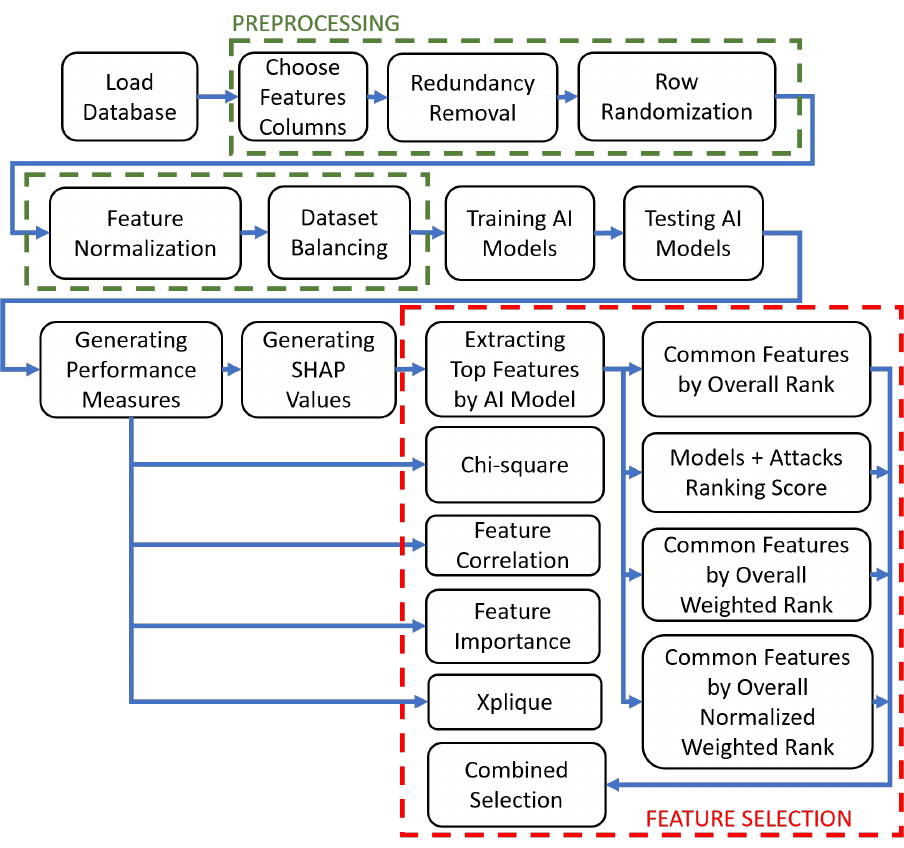}

\vspace{-3mm}
   \caption{A summary of our feature selection approach for network intrusion detection based on XAI.}
   \label{fig:Flow_code}
   \vspace{-3mm}
\end{figure}

\subsection{XAI-based Workflow Sections for Choosing Features}

The low-level elements of our XAI-based feature selection process are now explained. The various elements (depicted in Figure~\ref{fig:Flow_code}) are elucidated thereafter.

\textbf{Intrusion Dataset Loading:} Our framework's first step is to load the data from the database. We start with two well-known network intrusion detection datasets: CICIDS-2017~\cite{panigrahi2018detailed} and RoEduNet-SIMARGL2021~\cite{mihailescu2021proposition}.

\textbf{Extraction of Features:} The next phase in our framework is to choose a basic set of features from the database's log files after the database has been loaded.
The AI model's results are significantly influenced by the feature extraction method since some characteristics may represent the nature of network traffic more accurately than others. To extract a basic set of characteristics for each network intrusion dataset, we employed the same methods as in prior works~\cite{mihailescu2021proposition,panigrahi2018detailed}.

\textbf{Reduction of Redundancy:} After loading the database and finishing the basic feature extraction, the next step is to remove duplicate items from the traffic. Redundancy is frequently present in network traffic~\cite{kwon2018understand}, and eliminating it is essential to avert the intrusion detection AI model's possible underperformance. We just need to remove every row that is precisely the same to do this. Moreover, we randomize the rows as a preprocessing step before training. This stage is required since attack labels may have been used to arrange the data sequentially. By arranging the rows in a random order, you can make sure that there is no bias introduced into the training data due to entry order.

\textbf{Data Distribution:} We also handle the problem of data imbalance in our architecture. Both of our datasets show an imbalance where the usual traffic greatly surpasses the attack traffic, as seen in Table~\ref{tbl:samples_distributions_datasets}. Remarkably, while developing strong AI models for network security, the issue of imbalanced network traffic types frequently arises~\cite{barradas2021flowlens}. We use the random oversampling technique~\cite{moreo2016distributional} to balance the dataset to address this problem. 
By oversampling the cases of the minority class, this strategy makes sure that all classes of network traffic are balanced. Our datasets are huge, and the random oversampling approach is successful in balancing them despite its simplicity~\cite{moreo2016distributional}. The AI model may learn from a representative collection of data by creating a more balanced distribution of cases across different classes through data balancing.

\textbf{Normalization of Features:} We use a conventional feature normalization step (min-max feature scaling) to avoid scale differences between distinct traffic features. By bringing all characteristics to a uniform numerical scale, this procedure makes sure that there are no magnitude differences throughout the dataset. We provide a consistent representation that makes training and analyzing the AI model easier by normalizing the characteristics.

\textbf{Training Black-box AI Models:} After the database is ready, we begin training the AI model. To do this, we separated the data into a training set, which comprised 70\% of the entire set, and a testing set, which included the remaining 30\%. Notably, we have included seven popular AI classification models at this stage: k-nearest neighbors (KNN), LightGBM, Adaptive Boosting (AdaBoost), support vector machine (SVM), random forest (RF), deep neural network (DNN), and multi-layer perceptron (MLP). Every model is trained using a unique set of parameters for optimal outcomes. We have carefully adjusted each model's parameters to maximize its effectiveness (see~\ref{app:ai_models_hyperparams} for a comprehensive setup).

\textbf{Evaluating Black-box AI Models:}
In our architecture, testing each of the seven AI models with untested data comes after they have been trained. Next, we proceed to use the test data that has not been seen to examine how well these models work. To assess each AI model's efficacy for this performance study, we first construct a confusion matrix for it. From there, we extract several common indicators. Accuracy, precision, recall, F1-score, Matthews correlation coefficient, balanced accuracy, and the area under the ROC curve are some examples of these measurements. To evaluate the computing efficiency of each AI model, we also provide the training and testing run times.

\textbf{XAI Global Feature Selection:} 
It is essential to remember that the AI models we have developed and assessed for our framework fall under the category of ``black-box" models. Therefore, it becomes imperative to explain these models together with their related characteristics (traffic log data) and labels (attack kinds). Consequently, using XAI approaches for feature selection is the next step in our architecture. For every characteristic, we produce global significance values. This is accomplished by creating a variety of graphs that let us infer how each attribute affects the AI model's decision-making process.
These realizations help to improve comprehension and expectations regarding the behavior of the AI model in IDS. We use SHAP global graphs, which are based on the theory of Shapley values~\cite{lundberg2017unified}, for this purpose. The Shapley value for a given feature is estimated using coalitions, which are graphs that show the average marginal contribution of each feature value over all conceivable feature combinations.

After outlining the primary elements of our system, we go into further depth about the suggested feature selection techniques and the baseline techniques we employed in our current study.

\vspace{-2mm}

\subsection{IDS Feature Selection Methods}\label{sec:proposed_feat_selec}

Starting with an XAI methodology termed SHAP~\cite{lundberg2017unified}, we first describe our suggested feature selection techniques.

\textbf{Model Specific Features:} The global rankings of features for the examined datasets were acquired using SHAP. For every AI model, the most important features are taken from this rank. Since we employ seven distinct AI models, we extract the top-ranked attributes for each model that influence the algorithm's choice of which incursion class to detect.

\textbf{Common Features by Overall Rank:} By ranking each feature independently of other models, this technique produces a single feature rank that applies to all models. To do this, the average rank of every single characteristic across all AI models is determined.

\textbf{Common Features by Overall Weighted Rank:} This approach expands on the earlier one. The distinction is that, as opposed to sequential numbering, it considers the correctness of each AI model and the SHAP values for every feature. The product of the SHAP value and the accuracy of the AI model for that feature determines the feature's relevance. Next, a weighted sum that takes into account the accuracy of the AI model as well as the SHAP relevance of each feature is averaged to rank the features.

\textbf{Common Features by Overall Weighted Normalized Rank:} With one exception, this approach is identical to the previous one. Every SHAP value is normalized. It was discovered throughout the trials that some models, such as LightGBM, provide SHAP values that are significantly larger than those of other models. To prevent this kind of bias, the normalizing step was included.

\textbf{Models + Attacks Ranking Score:} This method extracts significant intrusion features via selecting the top-$k$ ranked features across all different AI models and all different intrusion types.  
 Suppose the set of AI intrusion detection models is denoted by $\mathcal{M}$ in which each entry $m \in \mathcal{M}$ represents one black-box AI model and that the set of intrusion types  be given by 
 $\mathcal{A}$ in which $a \in \mathcal{A}$ represents one intrusion class.
We calculate the overall ranking score of each feature (given by $r_i$) as follows. 
$ 
r_i = \frac{1}{2}\left(\frac{\sum_{m \in \mathcal{M}} r_{im}}{|\mathcal{M}|} + \frac{\sum_{a \in \mathcal{A}} r_{ia}}{|\mathcal{A}|}\right), 
$  where $r_{im}$ and $r_{ia}$ are the ranks of feature $i$ for model $m \in \mathcal{M}$ and intrusion $a \in \mathcal{A}$, respectively. The overall ranking score of a feature $i$ ($r_i$) is given by the weighted sum of both the feature rank across all AI models and across all intrusion types. We then chose the $k$ features with the lowest rank value. Note that the lower the $r_i$ value, the higher the feature rank.

\textbf{Combined Selection:} Using this approach, we assign a weight to each feature based on how frequently it appears in the top-$k$ features across all feature selection techniques that have been suggested. Put differently, the choice of a characteristic in this case is determined by its overall significance when compared to all other suggested approaches.

\subsection{Traditional Feature Selection Methods}
Next, we provide popular feature selection techniques that serve as benchmarks for our suggested techniques in this study. Most of these techniques do not require prior model training.

\textbf{Chi-square~\cite{gajawada2019chi}}: This is a well-known statistical technique that considers two occurrences and calculates their independence. An attribute is more important when its chi-square value is higher.

\textbf{Feature Correlation~\cite{feat_corr_citation}:} The correlation between each pair of attributes is measured using this approach. When two traits are closely connected, only one of them may take their position. 

\textbf{Feature Importance~\cite{dutta2018analysing}:} To assess the significance of features on an artificial classification problem, this approach uses a forest of trees to determine the features' value for a dataset.

\textbf{Information Gain~\cite{CICIDS}:}
This approach evaluates a dataset's properties according to information theory's notion of information gain. The entropy of the target variable is measured both before and after the dataset is divided based on a certain characteristic to determine the information gain.

\textbf{K-best~\cite{brownlee2019choose}:} 
The top-K most informative features from a dataset are to be chosen using this feature selection technique. It is a simple method that chooses the $K$ characteristics with the highest scores after ranking the features according to a given criterion.

\subsection{Feature Selection Methods from Xplique}

Next, we provide the primary feature selection techniques that we employed in our study using Xplique~\cite{xplique}, a recent XAI tool. It can be described as a Python feature selection toolkit package that focuses on explaining complex TensorFlow-based neural network models. Users can choose from a variety of XAI methods in this library to test various approaches or assess how well different methodologies are explained.

\vspace{-2mm}

    \begin{itemize}
    
        \item \textbf{Saliency~\cite{saliency}:}  
        Saliency maps draw attention to the key areas in an input image or set of data that affect how the model predicts a certain class. The back-propagation approach is used internally by these maps, and it focuses on determining the class derivatives while taking the input data into account. They aid in visualizing the input components that have an impact on the AI model's conclusion. A saliency map representing the pixels associated with the item or class in the picture or input data, respectively, is the end product.

        \item \textbf{Integrated Gradients~\cite{IG}:}
        One way to link the model's prediction to its input attributes is to use integrated gradients. The process computes the gradients between the input and the baseline. Afterward, the attributions are obtained by integrating such gradients along this path. It is regarded as one of the deep neural network techniques with the highest degree of reliability~\cite{IG}.

        \item \textbf{Occlusion~\cite{DeconvNet-Occlusion-}:} 
        The Occlusion approach is concealing or deleting portions of the input in a methodical manner and then monitoring the effect on the model's prediction. The procedure checks to see if the model bases its prediction on the context—that is, the data around the item. It accomplishes this by methodically applying a gray square or pre-set value to the relevant area of the input data and tracking the correctness. Therefore, it is helpful to know which areas are essential to the model's conclusion.

        \item \textbf{SmoothGrad~\cite{smoothgrad}:} 
        Deep neural networks can use the SmoothGrad approach to lower noise in their sensitivity maps. To create a smoother representation, it entails creating many sounds, adding them to the relevant pictures, calculating the saliency map for each, and then averaging them. This procedure enhances the coherence and sharpness of sensitivity maps~\cite{smoothgrad} while lowering visual noise.

        \item \textbf{VarGrad~\cite{vargrad}:} 
        Similar to SmoothGrad, VarGrad adds a variation term that is used to produce explanations. The explanation variance that results from introducing noise to the input is called VarGrad. VarGrad captures higher-order partial derivatives, which sets it apart from SmoothGrad~\cite{vargrad}.

        \item \textbf{SquareGrad~\cite{squaregrad}:}
        An expansion of SmoothGrad, SquareGrad incorporates squared Gaussian gradients into the noise-adding averaging procedure. The primary distinction between the original SmoothGrad approach and SquareGrad is that the former squares each estimate before averaging it. The feature magnitude may matter more than the direction~\cite{squaregrad}.

        \item \textbf{DeconvNet~\cite{DeconvNet-Occlusion-}:}  
        A method for displaying characteristics that convolutional neural networks (CNNs) learn is called DeconvNet, also known as Deconvolutional Network. It maps the activations back to the input space using a technique akin to deconvolution. Rebuilding feature activations using processes including unpooling, rectification, and filtering allows DeconvNet to recreate the activation of the ConvNet layer. The DeconvNet-obtained visualization sheds light on the roles played by intermediate feature layers and the AI classifier~\cite{DeconvNet-Occlusion-}.

        \item \textbf{GradientInput  ~\cite{GradientInput}:} 
        By multiplying the input to a neuron by the output gradient regarding that neuron, this approach determines the gradient input. In this way, it takes into account both the sign and intensity of the input and gives information about how relevant the input characteristics are to a neural network. It is an effective tool for examining network activity because of its level of detail, which highlights important characteristics and their effects on the network.

         \item \textbf{LIME~\cite{lime}:}  
         Through the process of introducing perturbations around the sample of interest and abstracting its behavior, an approach known as Local Interpretable Model-Agnostic Explanations (LIME) with a local scope builds a surrogate model around the region of interest. Because of this, such a model is an easier-to-understand version of the original model while maintaining the same behavior in the relevant domain. This has the benefit of being more explainable than the original model.

    \end{itemize}

Drawing on Xplique's several feature selection techniques, we provide a straightforward voting procedure that applies to each of these approaches, as elaborated below.
\begin{itemize}
    \item \textbf{Voting:} 
    Our suggested voting approach is based on a frequency analysis of all the earlier feature selection techniques that Xplique was used to assess. The method uses a point system where the least significant feature is assigned a score of one, and each subsequent characteristic's value is increased by one until it reaches the highest point. Thus, out of 10 attributes, the least significant is assigned a score of one, while the most significant is assigned a value of ten. Because there are nine Xplique approaches, we add up all the points that each feature has received. Next, we arrange the features in descending order to determine which characteristics are the most important using this straightforward voting method.

\end{itemize}

\vspace{-1mm}

\subsection{Feature Explanation} 
Our framework's last part entails taking specific measurements out of the overall feature descriptions. In particular, we concentrate on extracting characteristics that are distinct to the incursion and traits that are specific to the model. The most crucial elements of every AI model are referred to as model-specific characteristics. The relevance of these traits in impacting the model's decision-making process is what led to their identification. We can learn more about the precise traits that influence the AI model's predictions by obtaining model-specific attributes. Conversely, intrusion-specific traits are those that are most closely linked to a given kind or class of network intrusion.
These characteristics are essential to comprehending what makes the various forms of intrusions unique. We are able to determine the essential characteristics that aid in distinguishing and identifying particular kinds of network intrusions by extracting information specific to intrusions.

\textbf{Importance of Features for Different Attacks:}
The creation of a list of the best characteristics for every kind of assault is a significant result of our system. Security analysts may use this information to narrow down certain log data aspects that are indicative of various attack types, which is useful when they are looking into suspicious network traffic.
This feature importance analysis may also help modify current intrusion detection systems (IDS) by offering user-friendly dashboards that are customized to the unique attack patterns seen in an organization's historical data. Remember that the seven major attack types in the two datasets we analyzed in our work—normal traffic, DoS, PortScanning, Brute Force, Infiltration, Web assault, and Bot—are the subject of our present analysis. Using two datasets, CICIDS-2017 and RoEduNet-SIMARGL2021, we want to have a better knowledge of the feature importance for each sort of assault. 
It is important to note that the CICIDS-2017 dataset includes information on all attack types, but the RoEduNet-SIMARGL2021 dataset only includes information on DoS and Port Scan attacks combined with regular traffic. The distribution of samples across various attack labels and the number of samples available in each dataset are shown in Table~\ref{tbl:samples_distributions_datasets}.

\textbf{Comprehensive List of Top Network Intrusion Features:}
We provide the full list of important network intrusion features for the two datasets this study examines, together with the relevant explanations, to help readers better comprehend these aspects. The rest of the article will make frequent references to these feature lists. In particular, a thorough explanation of every feature in the RoEduNet-SIMARGL2021 network intrusion dataset can be found in Table~\ref{tbl:feature_list_sensor}. In a similar vein, each feature in the CICIDS-2017 network intrusion dataset is described in Table~\ref{tbl:feature_list_CICIDs}.

\begin{table}[hbt!] 
\centering
\caption{An explanation of the RoEduNet-SIMARGL2021 dataset's primary features~\cite{flow1234}.} 
\vspace{1mm}
\resizebox{\columnwidth}{!}{
\begin{tabular}{ll}
\hline
RoEduNet-SIMARGL2021 Features               & Explanation      \\                               \hline                                     FLOW\_DURATION\_MILLISECONDS & Flow duration in milliseconds                         \\
PROTOCOL\_MAP                                                           & IP protocol name (tcp, ipv6, udp, icmp)                                                                   \\
TCP\_FLAGS                                                              & Cumulation of all flow TCP flags \\

TCP\_WIN\_MAX\_IN            & Max TCP Window (src-\textgreater{}dst)       \\

TCP\_WIN\_MAX\_OUT           & Max TCP Window (dst-\textgreater{}src) \\

TCP\_WIN\_MIN\_IN            & Min TCP Window (src-\textgreater{}dst) \\

TCP\_WIN\_MIN\_OUT   & Min TCP Window (dst-\textgreater{}src)       \\

TCP\_WIN\_SCALE\_IN  & TCP Window Scale (src-\textgreater{}dst)   \\

TCP\_WIN\_MSS\_IN  & TCP Max Segment Size (src-\textgreater{}dst) \\

TCP\_WIN\_SCALE\_OUT & TCP Window Scale (dst-\textgreater{}src)   \\
SRC\_TOS                                                                & TOS/DSCP (src-\textgreater{}dst)             \\
DST\_TOS                                                                & TOS/DSCP (dst-\textgreater{}src)  \\

FIRST\_SWITCHED& SysUptime of First Flow Packet\\

LAST\_SWITCHED & SysUptime of Last Flow Packet\\
TOTAL\_FLOWS\_EXP & Total number of exported flows \\
L4\_SRC\_PORT & TCP/UDP source port number \\ %(e.g., FTP, Telnet)  \\
FLOW\_ID & ID used to distinguish data records  \\

\end{tabular}
}
\label{tbl:feature_list_sensor}
\vspace{-3mm}
\end{table}

% \vfill

\begin{table}[hbt!]
%\vspace{-3mm}
\centering

\caption{The primary characteristics of the benchmark CICIDS-2017 dataset are described.~\cite{ahlashkari_2021}.}
\vspace{1mm}
\resizebox{\columnwidth}{!}{
\begin{tabular}{ll}
\hline
\begin{tabular}[c]{@{}c@{}}CICIDS-2017  Features\end{tabular} & Explanation                                                                                                   \\ \hline
Packet Length Std                                         & Standard deviation  length of a packet                                      \\
Total Length of Bwd Packets                               & Total size of packet in backward direction                          \\
Subflow Bwd Bytes                                         & Average number of bytes in backward sub flow        \\
Destination Port                                          & Destination Port Address                                  \\
Packet Length Variance                                    & Variance length of a packet                               \\
Bwd Packet Length Mean                                    & Mean size of packet in backward direction                           \\
Avg Bwd Segment Size                                      & Average size observed in the backward direction                     \\
Bwd Packet Length Max                                     & Maximum size of packet in backward direction                     \\
Init\_Win\_Bytes\_Backward                                & Total number of bytes in initial backward window \\
Total Length of Fwd Packets                               & Total packets in the forward direction                                                                                \\
Subflow Fwd Bytes                                         & Average number of bytes in a forward sub flow  \\
Init\_Win\_Bytes\_Forward                                 & Total number of bytes in initial forward window \\
Average Packet Size                                       & Average size of packet                                                                                                \\
Packet Length Mean                                        & Mean length of a packet                                                                                               \\
Max Packet Length                                         & Maximum length of a packet  \\
FIN Flag Count             & Number of packets with Finish (FIN) flag   \\
Down/Up Ratio & Download to upload ratio \\
Flow IAT Std &  Std. deviation of inter-arrival times within a flow \\
Bwd Packets/s & Number of backward packets per second \\
\hline
\end{tabular}
}
\label{tbl:feature_list_CICIDs}
\vspace{2mm}
\end{table}

%\vspace{-2mm}

\section{Evaluation}\label{sec:evaluation}

\begin{table}[hbt!]
\centering
\caption{Our two datasets' summary and statistics (size, number of attacks, and distribution of attack kinds) are provided.}
% \vspace{-3mm}
\resizebox{0.95\columnwidth}{!}{
\begin{tabular}{|l|c|c|}
\hline
                            & \textbf{CICIDS-2017} & \textbf{RoEduNet-SIMARGL2021} \\ \hline
\textbf{Number of Labels}   & 7               & 3               \\ \hline
\textbf{Number of Features} & 78              & 29              \\ \hline
\textbf{Samples Quantity}   & 2,775,364       & 31,433,875      \\ \hline
\hline
\textbf{Normal}       & 84.442\%        & 62.20\%                              \\ \hline
\textbf{DoS}          & 9.104\%         & 24.53\%                              \\ \hline
\textbf{PortScan}     & 5.726\%         & 13.27\%                              \\ \hline
\textbf{Brute Force}  & 0.498\%         & -                                    \\ \hline
\textbf{Web Attack}   & 0.157\%         & -                                    \\ \hline
\textbf{Bot}          & 0.071\%         & -                                    \\ \hline
\textbf{Infiltration} & 0.001\%         & -                                    \\ \hline
\end{tabular}
}
\label{tbl:samples_distributions_datasets}
\vspace{1mm}
\end{table}

We provide a thorough examination of our evaluation results in this part to answer the following important queries:

\begin{itemize}
\vspace{-1mm}

\item How well do black-box AI models perform on the two network intrusion datasets that are being examined?

\item How can XAI help identify the key characteristics of various AI models for intrusion detection? 

\item Which characteristics are essential for identifying and categorizing every kind of intrusion?

\item Do many kinds of AI models share any common, significant intrusive features?

\item What effect does feature selection have on how well AI models identify network intrusions?

\end{itemize}

\subsection{Description of Datasets}

\textbf{RoEduNet-SIMARGL2021~\cite{mihailescu2021proposition}:} 
This dataset comes from the SIMARGL project, which was a joint venture between the Romanian Education Network (RoEduNet) and the European Union, funded under the Horizon program. It consists of real-time traffic analysis features together with legitimate network traffic data.  The information is organized according to a data schema that is similar to Netflow~\cite{claise2004cisco}, a network protocol that CISCO created to record and observe network flows.

\textbf{CICIDS-2017~\cite{panigrahi2018detailed}:} 
Created in 2017 by the University of Brunswick's Canadian Institute for Cybersecurity, this dataset acts as a standard for intrusion detection. It includes six different types of attack profiles, including brute force, heartbleed, botnet, denial of service, portscan, web assault, and infiltration attack. The dataset includes background traffic created by a B-Profile system~\cite{sharafaldin2018towards}, which extracts different user behaviors depending on network protocols, to provide a realistic environment.

\textbf{Dataset Summary and statistics:} Table~\ref{tbl:samples_distributions_datasets} displays the number of attack kinds (labels), the distribution (\%) for each attack type, and the size of each dataset (samples amount).

\vspace{-1mm}

\subsection{Design of Experiments}

\textbf{Coding Resources:} For the construction and analysis of our various black-box AI techniques, we employed several open-source toolboxes (including Keras, Scikit-Learn, Pandas, and Matplotlib). 

\textbf{XAI Methods:} The following XAI tools were employed in our assessment:

\textbf{(a) SHAP~\cite{lundberg2017unified}:} We employed the widely utilized Shapley Additive exPlanations (SHAP) XAI toolkit~\cite{lundberg2017unified}. For our AI models, we generated global features and associated techniques using that SHAP tool.

\textbf{(b) Xplique~\cite{xplique}:}
For feature selection, we additionally utilized the Python toolkit Xplique library~\cite{xplique}, emphasizing explainability and comprehension of complex neural network models.  This collection of techniques includes descriptions, examples, and links to official publications for a wide range of approaches (e.g., Saliency, Occlusion, Integrated-Gradients, SmoothGrad, VarGrad, SquareGrad, GuidedBackprop, DeconvNet, among others).  It enables the user to assess model explanations based just on TensorFlow or try other approaches.

\textbf{AI Methods:} 
We assess black-box AI methods and various components of our proposed XAI-based feature selection framework for a better understanding of network intrusion detection tasks by pairing seven popular AI classification algorithms: deep neural network (DNN)~\cite{tang2020saae}, random forest (RF)~\cite{waskle2020intrusion}, AdaBoost (ADA)~\cite{yulianto2019improving}, k-nearest neighbor (KNN)~\cite{li2014new}, support vector machine (SVM)~\cite{tao2018improved}, multi-layer perceptron (MLP)~\cite{MEBAWONDU2020e00497}, and light gradient-boosting machine (LightGBM)~\cite{jin2020swiftids}.

\textbf{Hyper-parameters:} We list the primary hyper-parameter selections in~\ref{app:ai_models_hyperparams} for each of the many AI models that were employed in this study.

\textbf{Black-box AI Metrics:} We use a commonly used set of criteria for assessing intrusion detection and classification issues to evaluate the performance of the black-box AI models. The measures that fall under this category include balanced accuracy (Bacc), Matthews correlation coefficient (MCC), accuracy (Acc), precision (Prec), recall (Rec), and F1-score (F1).
The effectiveness of the AI model is also determined by calculating the AucRoc (area under the ROC curve) score.

\textbf{Selecting Features for XAI Techniques Evaluation:} We produce a variety of feature selection techniques for evaluating XAI approaches to offer performance insights. Global summary charts, which provide a summary of feature rank throughout the dataset, are one of these techniques. Furthermore, we ascertain the significance of each feature for every form of assault, pinpointing the primary attributes that serve as indicators of the existence of particular attack kinds. Additionally, we examine the overall feature rank for each dataset, emphasizing the top characteristics that are shared by several AI models. These techniques offer a thorough assessment of the XAI-based feature selection techniques put forward in this study.

\vspace{-2mm}

\subsection{Findings of the Evaluation Results}
%%%%%%%%%%%%%%%%%%%%%%%%%%%%%%%%%%%%%%%%%%%%%%%%%%%%%%%%%%%%%%%%

%%%%%%%%%%%%%%%%%%%%%%%%%%%%%%%%%%%%%%%%%%%%%%%%%%%%%%%%%%%%%%%%
\textbf{Superiority of our Feature Selection Methods:} 
The comparison results between all feature selection methods (i.e., baseline methods and our proposed feature selection methods) are displayed in Table~\ref{tbl:feat_select_scoring}. All performance metrics (Acc, Prec, F1, MCC, Bacc, and AucRoc) for the CICIDS-2017 and RoEduNet-SIMARGL2021 datasets are analyzed. It displays which feature methods—here, 5 and 10—performed better overall for the various values of the top characteristics. The way we calculate scores is as follows. The best technique receives three points, the second best receives two, and the third best receives one. Which feature selection techniques work best overall are determined by adding together these scores, or weights. 
The most accurate techniques are the best ones; in the event of a tie in accuracy, we use the next metric (such as precision, recall, etc.). Our feature selection methods outperform baseline methods, as shown in Table~\ref{tbl:feat_select_scoring}, which has the best overall score in 11 out of 12 based on the previously mentioned ranking, which is based on the best three methods for each of the four setups that are taken into consideration ($k = 5$ for CICIDS-2017, $k = 10$ for CICIDS-2017, $k = 5$ for RoEduNet-SIMARGL2021, and $k = 10$ for RoEduNet-SIMARGL2021).

%Score feature importance
\begin{table*}[hbt!]
\caption{A comparison of our suggested feature selection methods' results with baseline approaches using our weighted scoring system for our two datasets. For every configuration, the top three feature selection techniques are indicated (in bold).}

% \vspace{-3mm}
\resizebox{\textwidth}{!}{%
\begin{tabular}{ c c }
%\hline
\begin{tabular}{|l|c c c c|}
\hline
Weights                                             & 3     & 2      & 1     &                \\ \hline
\textbf{CICIDS-2017 (k = 5)}                                             & First & Second & Third & Score \\ \hline
\textbf{Common Features by Overall Rank}       & 1     & 1      & 1     & \textbf{6}     \\ \hline
\textbf{Model Specific Features}                             & 0     & 2      & 1     & 5     \\ \hline
\textbf{Chi-square}                                          & 1     & 1      & 0     & 5     \\ \hline
\textbf{Feature Correlation}                                 & 1     & 0      & 1     & 4     \\ \hline
\textbf{Feature Importance}                                  & 0     & 1     & 1     & 3     \\ \hline
\textbf{Models + Attacks Ranking Score}                     & 0     & 0      & 1     & 1     \\ \hline
\textbf{Common Features by Overall Weighted Rank}           & 2     & 1      & 0     & \textbf{8}     \\ \hline
\textbf{Common Features by Overall Normalized Weighted Rank}  & 2     & 0      & 3     & \textbf{9}    \\ \hline
\textbf{Combined Selection} &  0  & 1      & 0     & 2    \\ \hline
\hline
\end{tabular}
&
\begin{tabular}{|l| c c c c|}
\hline
Weights                                             & 3     & 2      & 1     &                \\ \hline
\textbf{RoEduNet-SIMARGL2021   (k = 5)}                                           & First & Second & Third & \textbf{Score} \\ \hline
\textbf{Common Features by Overall Rank}                     & 1     & 1      & 0     & 5     \\ \hline
\textbf{Model Specific Features}                             & 0     & 1      & 2     & 4     \\ \hline
\textbf{Chi-square}                                          & 0     & 0      & 2     & 2     \\ \hline
\textbf{Feature Correlation}                                 & 0     & 1      & 1     & 3     \\ \hline
\textbf{Feature Importance}                                  & 1     & 3      & 0     & 9    \\ \hline
\textbf{Models + Attacks Ranking Score}                      & 1     & 2      & 4     & \textbf{11}    \\ \hline
\textbf{Common Features by Overall Weighted Rank}            & 1     & 2      & 4     & \textbf{11}    \\ \hline
\textbf{Common Features by Overall Normalized Weighted Rank} & 1     & 2      & 4     & \textbf{11}    \\ \hline
\textbf{Combined Selection} & 4     & 1      & 0     & \textbf{14}    \\ 
\hline

\end{tabular}
\\
\begin{tabular}{|l|c c c c|}
\hline
Weights                                             & 3     & 2      & 1     &       \\ \hline
\textbf{CICIDS-2017 (k = 10)}                                    & First & Second & Third & Score \\ \hline
\textbf{Common Features by Overall Rank}                     & 1     & 5      & 1     & \textbf{14}    \\ \hline
\textbf{Model Specific Features}                             & 1     & 2      & 2     & 9     \\ \hline
\textbf{Chi-square}                                          & 2     & 2      & 1     & \textbf{11}    \\ \hline
\textbf{Feature Correlation}                                 & 1     & 2      & 0     & 7     \\ \hline
\textbf{Feature Importance}                                  & 1     & 1      & 2     & 7     \\ \hline
\textbf{Models + Attacks Ranking Score}                      & 2     & 2      & 1     & \textbf{11}     \\ \hline
\textbf{Common Features by Overall Weighted Rank}            & 1     & 0      & 3     & 6     \\ \hline
\textbf{Common Features by Overall Normalized Weighted Rank} & 1     & 2      & 1     & 8     \\\hline
\textbf{Combined Selection} & 1     & 0      & 1     & 4    \\ 
\hline
\end{tabular}
&
%\hline
\begin{tabular}{|l|c c c c|}
\hline
Weights                                             & 3     & 2      & 1     &       \\ \hline
   \textbf{RoEduNet-SIMARGL2021 (k = 10)}  & First & Second & Third & Score \\ \hline
\textbf{Common Features by Overall Rank}                     & 0     & 1      & 0     & 2     \\ \hline
\textbf{Model Specific Features}                             & 0     & 1      & 0     & 2     \\ \hline
\textbf{Chi-square}                                          & 1     & 1      & 1     & 6     \\ \hline
\textbf{Feature Correlation}                                 & 1     & 2      & 1     & 8    \\ \hline
\textbf{Feature Importance}                                  & 1     & 2      & 0     & 7    \\ \hline
\textbf{Models + Attacks Ranking Score}                     & 1     & 1      & 1     & 6    \\ \hline
\textbf{Common Features by Overall Weighted Rank}            & 2     & 2      & 1     & \textbf{11}    \\ \hline
\textbf{Common Features by Overall Normalized Weighted Rank} & 2     & 1      & 1     & \textbf{9}    \\ \hline
\textbf{Combined Selection} & 4     & 1      & 2     & \textbf{16}    \\ 
\hline
\end{tabular}
%&
\\
%\hline
\end{tabular}
}
\label{tbl:feat_select_scoring}
%\vspace{-2mm}
\end{table*}

\textbf{Overall Black-box AI Model Performance:} The overall effectiveness of our various black-box AI models for IDS is then displayed utilizing various feature selections (all features, top-15, top-10, and top-5). For the RoEduNet-SIMARGL2021 and CICIDS-2017 datasets, respectively, these performances are displayed in Tables~\ref{tbl:overall_performance_sensors}-\ref{tbl:overall_performance_15_CICIDS}. They display the many metrics that we gathered for various AI models. We provide the top results for several AI models in bold language. 
The RF, KNN, and SVM models consistently exhibit the greatest performance over most settings in both datasets. Conversely, MLP performs the poorest overall across various trial outcomes for both datasets. Lastly, for both datasets, LightGBM and ADA perform moderately.

\begin{table}[hbt!]
\centering
%\vspace{-2mm}
\caption{Overall results for the RoEduNet-SIMARGL2021 dataset for AI models with various feature configurations.}
\vspace{1mm}
\resizebox{\columnwidth}{!}{
\begin{tabular}{cccccccc}
\hline
\textbf{AI Model (k = all)} & \textbf{Acc}  & \textbf{Prec} & \textbf{Rec}  & \textbf{F1}   &  \textbf{Bacc} &  \textbf{Mcc}  & \textbf{AucRoc} \\ \hline
RF       & \textbf{0.999} & \textbf{0.999} & \textbf{0.999} & \textbf{0.999} & \textbf{0.999} & \textbf{0.999} & \textbf{0.999}   \\
ADA      &  0.765 & 0.647 & 0.647 & 0.647 & 0.735 &  0.471 &  0.516   \\
DNN              & 0.777 &  0.666 &  0.666 &  0.666 & 0.749 & 0.499 & 0.505 \\ 
% DNN      &  0.559 & 0.339 & 0.339 & 0.339 & 0.504 &  0.008 & nan   \\
SVM              & 0.915 & 0.897 & 0.897 & 0.897 & 0.923 & 0.846 & 0.971 \\
% SVM      & 0.985 & 0.978 & 0.978 & 0.978 & 0.984 &  0.968 & nan      \\
KNN      & 0.998    &  0.998    & 0.998    &  0.998    &0.998    & 0.997    & \textbf{0.999}     \\
% KNN              & 0.9988 & 0.9981 & 0.9981 & 0.9981 & 0.9985 & 0.9971 & 0.9997 \\ 
MLP      &  0.555    &  0.333   &  0.333  &  0.333 & 0.500  & 0.000 & 0.993 \\
LightGBM      & \textbf{0.999 }   &\textbf{ 0.999  }  & \textbf{0.999}    & \textbf{0.999}    & \textbf{0.999}    & \textbf{0.999}    & \textbf{0.999} \\
\hline
 \textbf{AI Model (k = 15)}  & \textbf{Acc}  & \textbf{Prec} & \textbf{Rec}  & \textbf{F1}   & \textbf{Bacc} & \textbf{Mcc}  & \textbf{AucRoc} \\ \hline
RF       & 0.998 & 0.997 & 0.997 & 0.997 & 0.998 & 0.995 & \textbf{0.999}   \\
ADA      & 0.745 &  0.617 &  0.617 &  0.617 &  0.713 &  0.426 &  0.566   \\
DNN      &  0.564 & 0.346 & 0.346 & 0.346 & 0.509 & 0.019 & 0.392  \\
SVM      &  0.940 &  0.910 & 0.910 & 0.910 &  0.932 &  0.865 & 0.972  \\
KNN      &  \textbf{0.999} & \textbf{0.998} & \textbf{0.998} & \textbf{0.998} &  \textbf{0.999} &  \textbf{0.998} & 0.998   \\
MLP      & 0.555 &  0.333 &  0.333 &  0.333 & 0.500 & 0.000 & \textbf{0.999} \\
LightGBM &  \textbf{0.999} & \textbf{0.998} &\textbf{ 0.998} & \textbf{0.998} & \textbf{0.999} & \textbf{0.998} & \textbf{0.999} 
%\\ XGBoost & 0.62 & 0.43 & 0.43 & 0.43 & 0.57 & 0.15 & 0.58
\\
\hline
\textbf{AI Model (k = 10)} & \textbf{Acc}  & \textbf{Prec} & \textbf{Rec}  & \textbf{F1}   &  \textbf{Bacc} &  \textbf{Mcc}  & \textbf{AucRoc} \\ \hline
RF       &  0.997 & 0.996  & 0.996 & 0.996 & 0.997 & 0.995 &  \textbf{0.999}  \\
ADA      & \textbf{0.999} & \textbf{0.999} & \textbf{0.999} &\textbf{0.999} &\textbf{0.999} & \textbf{0.999} &\textbf{0.999}       \\
DNN      & 0.560 & 0.340 & 0.340 & 0.340 & 0.505 & 0.011 & 0.988   \\
SVM     & 0.956 &  0.934 & 0.934 & 0.934 & 0.951 & 0.902 & 0.921      \\
KNN      &  \textbf{0.999} & \textbf{ 0.999} &  \textbf{0.999} &  \textbf{0.999}&  \textbf{0.999} &  \textbf{0.999}& 0.996       \\
MLP     &  0.555 & 0.333 & 0.333 & 0.333 & 0.500 &  0.000 & 0.995  \\
LightGBM     & \textbf{0.999} & \textbf{0.999} & \textbf{0.999}  & \textbf{0.999}  & \textbf{0.999}  & \textbf{0.999 } & \textbf{0.999} 
\\
\hline
\textbf{AI Model (k = 5)} & \textbf{Acc}  & \textbf{Prec} & \textbf{Rec}  & \textbf{F1}   &  \textbf{Bacc} &  \textbf{Mcc}  & \textbf{AucRoc} \\ \hline
RF       & 0.921 & 0.881 & 0.881 & 0.881 & 0.911 & 0.822 & 0.985   \\
ADA      & \textbf{0.999} & \textbf{0.999} & \textbf{0.999} & \textbf{0.999} & \textbf{0.999} & \textbf{0.999} & 0.487      \\
DNN      & 0.470 & 0.206 & 0.206 & 0.206 &  0.404 & -0.190 & 0.506   \\
SVM     & 0.933 & 0.899 & 0.899 & 0.899 & 0.925 &  0.849  & 0.901  \\
KNN      & 0.997 & 0.995 & 0.995 & 0.995 &  0.996 & 0.993 & 0.992  \\
MLP     &  0.555 & 0.333 & 0.333 & 0.333 &  0.500 & 0.000 & 0.992  \\
LightGBM     & \textbf{0.999} & 0.998 & 0.998 & 0.998 & 0.998 & 0.997 & \textbf{0.999}
\end{tabular}
}
\label{tbl:overall_performance_sensors}
\vspace{-3mm}
\end{table}

\begin{table}[hbt!]
\centering
% \vspace{-2mm}
\caption{Overall results for the CICIDS-2017 dataset for AI models with various feature configurations.}
\vspace{1mm}
\resizebox{\columnwidth}{!}{
\begin{tabular}{cccccccc}
\hline
\textbf{AI Model (k = all)} & \textbf{Acc} & \textbf{Prec} & \textbf{Rec} & \textbf{F1} & \textbf{Bacc} & \textbf{Mcc} & \textbf{AucRoc} \\ \hline
RF              & 0.988       & 0.961         & 0.961        & 0.961       &0.977        & 0.954       &  0.420          \\
% ADA             & 0.762  &  0.168   & 0.168 & 0.168 & 0.514&  0.029 & nan          \\
ADA              & 0.750 & 0.127 & 0.127 & 0.127 & 0.491 & 0.017 & 0.964 \\
% DNN             & 0.755  &  0.142 &  0.142 &  0.142 & 0.500 & 0.001  & nan    \\
DNN              & 0.762 & 0.168 & 0.168 & 0.168 &  0.515 & 0.030 &  0.975 \\
% SVM             &  0.975 &  0.914 & 0.914 &  0.914 & 0.949 &  0.899 & nan    \\
SVM              & 0.961 & 0.866 & 0.866 & 0.866 & 0.922 & 0.844 & 0.867 \\
% KNN             & \textbf{0.999}  & \textbf{0.996}  &\textbf{ 0.996}& \textbf{0.996} &  \textbf{0.997} & \textbf{0.995} & nan          \\
KNN              & \textbf{0.999} & \textbf{0.992} & \textbf{0.992} & \textbf{0.992} &  \textbf{0.995} & \textbf{0.990} & 0.997 \\ 
% MLP             &  0.760  &  0.161 & 0.161 &  0.161   & 0.510 & 0.021 & nan \\
MLP              &  0.714 & 0.001 & 0.001 & 0.001 & 0.416 & 0.166 & 0.996 \\
LightGBM   & 0.981 &  0.936  &   0.936 & 0.936 & 0.963 &0.926 & \textbf{0.999 } \\
\hline
\textbf{AI Model (k = 15)}   & \textbf{Acc}  & \textbf{Prec} & \textbf{Rec}  & \textbf{F1}   & \textbf{Bacc} & \textbf{Mcc}  & \textbf{AucRoc} \\ \hline
RF       & 0.979 & 0.928 & 0.928 & 0.928 &  0.958 & 0.916 & 0.678   \\
ADA      &  0.762 & 0.169 & 0.169 & 0.169 &  0.515 & 0.030 & 0.972   \\
DNN      & 0.740 & 0.091 & 0.091 & 0.091 & 0.469 &  -0.060 & 0.937   \\
SVM      & 0.933 & 0.765 & 0.765 & 0.765 & 0.863 & 0.726 & 0.795  \\
KNN      & \textbf{0.999} & \textbf{0.997} & \textbf{0.997} & \textbf{0.997} & \textbf{0.998} &  \textbf{0.996} & \textbf{0.998}   \\
MLP      & 0.759 &  0.158 &  0.158 &  0.158 & 0.508 & 0.017 & 0.991   \\
LightGBM & 0.872 & 0.552 & 0.552 &0.552 & 0.738 & 0.477 &  0.981    
%\\ XGBoost  &   0.93   & 0.28     &0.28      & 0.28     & 0.62     &0.24      &0.53     
\\
\hline
\textbf{AI Model (k = 10)} & \textbf{Acc}  & \textbf{Prec} & \textbf{Rec}  & \textbf{F1}   &  \textbf{Bacc} &  \textbf{Mcc}  & \textbf{AucRoc} \\ \hline
RF        &  0.992 & 0.971 &  0.971 &  0.971 & 0.983 & 0.967 & 0.679\\
ADA      &  0.753 & 0.132 & 0.132 & 0.132 &  0.494 &  -0.012 & 0.960      \\
DNN      & 0.767 & 0.184 & 0.184 & 0.184 &  0.524 & 0.048 & 0.929    \\
SVM      & 0.929 & 0.750 & 0.750 & 0.750 & 0.854 & 0.708 & 0.626   \\
KNN      &  \textbf{0.999} & \textbf{0.997} & \textbf{0.997} &\textbf{0.997} &  \textbf{0.998} & \textbf{0.997} & \textbf{0.998}       \\
MLP     & 0.759 & 0.158 & 0.158 & 0.158 & 0.509 &   0.018 & 0.987  \\
LightGBM     &  0.872 & 0.552 & 0.552 & 0.552 & 0.739 & 0.478 &  0.986
\\
\hline
\textbf{AI Model (k = 5)} & \textbf{Acc}  & \textbf{Prec} & \textbf{Rec}  & \textbf{F1}   &  \textbf{Bacc} &  \textbf{Mcc}  & \textbf{AucRoc} \\ \hline
RF       & 0.989 & 0.964 & 0.964 & 0.964 & 0.979 & 0.958 & 0.524   \\
ADA      & 0.762 &  0.168 &  0.168 &  0.168 &  0.514 & 0.029 &  0.957       \\
DNN      & 0.774 & 0.210 & 0.210 & 0.210 &  0.539 & 0.078 &  0.883       \\
SVM      & 0.910 & 0.686 & 0.686 & 0.686 &0.817& 0.634 & 0.558   \\
KNN      & \textbf{0.999} & \textbf{0.996} &  \textbf{0.996} &  \textbf{0.996} & \textbf{0.998} &\textbf{0.996} & 0.996       \\
MLP     & 0.714 & 0.153 & 0.153 & 0.153 &  0.506 &  0.012 & 0.977  \\
LightGBM     &0.971 & 0.899 & 0.899 & 0.899 & 0.941 & 0.883 & \textbf{ 0.997} 
\end{tabular}
}
\label{tbl:overall_performance_15_CICIDS}
% \vspace{-2mm}
\end{table}

\textbf{Impact of Feature Selection on Black-box AI Models:} We next go over the thorough performance analysis we conducted on our various black-box AI models for IDS using the four feature selection setups discussed earlier. In particular, the RoEduNet-SIMARGL2021 dataset performance metrics are presented in Table~\ref{tbl:overall_performance_sensors}. We find that after being trained on all characteristics, five of the AI models—ADA, SVM, KNN, RF, and DNN—perform worse. Furthermore, we note that the MLP and LightGBM models perform identically when given varying feature sets; that is, the feature selection does not affect these models' results.

Table~\ref{tbl:overall_performance_15_CICIDS} for the CICIDS-2017 dataset demonstrates that three AI models—ADA, SVM, and KNN—perform better when feature selection is used during training. However, the DNN and MLP do marginally worse when it comes to feature selection. This experiment demonstrates how crucial it is to carefully choose characteristics to improve AI model performance. It also demonstrates how the AI model and the dataset's properties affect this kind of feature selection (for example, DNN and MLP models require all attributes to perform better in our configuration).

It is crucial to emphasize that these broad performance measures do not offer in-depth information on how well an AI model functions in various attack kinds, which we cover next.

\textbf{Variation of Best AI Model across Attack Types:} 
The accuracy of each AI model for distinct attack types in both datasets is shown in Tables~\ref{tbl:acc_attack_sensor}-\ref{tbl:acc_attack_CICIDS}. Tables show that most AI models do not always perform better than other models in all attack classes in the datasets.
This underlines how crucial it is to comprehend each AI model's decision-making process for various attack scenarios, which is the main focus of our ongoing research.

%attack all features
\begin{table}[hbt!]
\centering
\caption{Accuracy for the RoEduNet-SIMARGL2021 dataset broken down by attack type (normal, DoS, and Port Scan).}
\vspace{1mm}
\resizebox{\columnwidth}{!}{
\begin{tabular}{cccc|ccc}
\hline
\textbf{AI Model} & \textbf{Normal} & \textbf{DoS}  & \textbf{PS} &  \textbf{Normal} & \textbf{DoS}  & \textbf{PS} \\ \hline
& \multicolumn{3}{c|}{\textbf{k $=$ all}} &  \multicolumn{3}{c}{\textbf{k $=$ 15}}\\ \hline
RF                                                    & \textbf{0.999} & \textbf{0.999} & \textbf{0.999} &   0.997 &  0.997 & \textbf{0.999} \\
ADA                                                   &  0.719    & 0.860   & 0.714  & 0.758    &  0.832    &  0.644   \\
DNN                                                   & 0.340 & 0.339 & 0.333 & 0.354 & 0.346 & 0.992 \\
SVM                                                   & 0.991 &  0.978 & 0.987 & 0.977 & 0.910 & 0.933    \\
KNN                                                   & \textbf{0.999} & 0.998 & \textbf{0.999}  & \textbf{0.999} & \textbf{0.999} & \textbf{0.999}       \\
MLP   &  0.666 & 0.333 &  0.666                                                &  0.666 & 0.333 & 0.666     \\
LightGBM                                               & \textbf{0.999}& \textbf{0.999}& \textbf{0.999} &   0.998 & \textbf{0.999} &  0.998 \\
\hline
%\textbf{AI Model k=15} & \textbf{Normal} & \textbf{DoS}  & \textbf{PS} \\ \hline
%\textbf{AI Model} & \textbf{Normal} & \textbf{DoS}  & \textbf{PS} \\ \hline
& \multicolumn{3}{c|}{\textbf{k $=$ 10}} &  \multicolumn{3}{c}{\textbf{k $=$ 5}}\\ \hline
RF                                                    & 0.998 & 0.996 & 0.998  & 0.993 & 0.881 & 0.888 \\
ADA                                                   & \textbf{0.999} & \textbf{0.999} & \textbf{0.999}   & \textbf{0.999} & \textbf{0.999} & \textbf{0.999}\\
DNN                                                   & 0.356 & 0.340 & 0.984   & 0.261 &  0.837 & 0.312 \\
SVM                                                   & 0.986 &  0.934 & 0.948  & 0.911 & 0.899 & 0.989\\
KNN                                    &  \textbf{0.999} &  \textbf{0.999} &  \textbf{0.999} &  0.998 & 0.995 & 0.997 \\
MLP                                     & 0.666 &  0.333 &  0.333                &  0.666 &  0.333 &  0.333 \\
LightGBM                           & \textbf{0.999}  & \textbf{0.999}  & \textbf{0.999}                       & 0.998 & 0.998 & \textbf{0.999} \\
\hline
\end{tabular}
}
\label{tbl:acc_attack_sensor}
%\vspace{-1mm}
\end{table}

% acc per attack cicids all feat
\begin{table}[hbt!]
\centering

% \vspace{-2mm}
\caption{Accuracy for the CICIDS-2017 dataset with all features broken down by attack type (normal, DoS, Brute Force, Web attack, Infiltration, Bot, and Port Scan).}
% \vspace{-3mm}
\resizebox{\columnwidth}{!}{
\begin{tabular}{cccccccc}
\hline
\textbf{AI Model k=all} & \textbf{Norm} & \textbf{DoS}  & \textbf{B. Force} & \textbf{Web} & \textbf{Infilt.} & \textbf{Bot}  & \textbf{PS}   \\ \hline
RF                                                        & 0.961 & 0.996 & \textbf{0.999}   & 0.967    & \textbf{0.999}   & \textbf{0.999} & \textbf{0.999}\\
ADA                                                       & 0.852  &  0.885 & 0.857 & 0.857 & 0.857& 0.698  & 0.328 \\
DNN                                                       & 0.772 & 0.227 & 0.183   & 0.857    & 0.857   & 0.857 & 0.857\\
SVM                                                       & 0.964    & 0.968    & 0.969      & 0.963       & 0.985      &0.992    & 0.987    \\
KNN                                                       & \textbf{0.999} & 0.997 & \textbf{0.999}    & \textbf{0.999} & 0.996  & \textbf{0.999} & \textbf{0.999} \\
MLP                                                       & 0.859 & 0.661 & 0.716    & 0.665    & \textbf{0.999}   & 0.646 & 0.776 \\ 
LightGBM                                                  & 0.936 & \textbf{0.999} & \textbf{0.999}    & \textbf{0.999}   & 0.998   & 0.952 & 0.986\\
\hline
\textbf{AI Model k=15} & \textbf{Norm} & \textbf{DoS}  & \textbf{B. Force} & \textbf{Web} & \textbf{Infilt.} & \textbf{Bot}  & \textbf{PS}   \\ \hline
RF                                                        & 0.933 & 0.995 & 0.996 &  0.972 & \textbf{0.999} &  0.961 &  0.998 \\
ADA                                                       & 0.729  & 0.847 &  0.857 & 0.857 & 0.857 & 0.705 &  0.484 \\
DNN                                                       & 0.857 & 0.142 &  0.856  &  0.857 &   0.856 &  0.857 &  0.856\\
SVM                                                       &  0.891 & 0.935 & 0.941 & 0.904 & 0.955 &  0.991 & 0.912 \\
KNN                                                       & \textbf{0.999}& \textbf{0.997} & \textbf{0.999} & \textbf{0.975} & \textbf{0.999} &\textbf{0.997}& \textbf{0.999}\\
MLP                                                       & 0.829 &  0.833 &  0.775 & 0.665 & 0.830  & 0.716 & 0.664 \\ 
LightGBM                                                  &  0.809 &0.985 & 0.996 & 0.740   & 0.857 &  0.716 & 0.998
%XGBoost          
\\

\hline
\textbf{AI Model k=10} & \textbf{Norm} & \textbf{DoS}  & \textbf{B. Force} & \textbf{Web} & \textbf{Infilt.} & \textbf{Bot}  & \textbf{PS}   \\ \hline
RF                                                        &  0.992 & \textbf{0.998} &  0.995 &   0.992 & 0.983 & 0.992 &  0.990\\
ADA                                                       &  0.709 & 0.714 & 0.857 & 0.857 &  0.857 & 0.762 & 0.508\\
DNN                                                       & 0.173 &  0.858 & 0.857 & 0.857 & 0.857 &  0.857 & 0.857\\
SVM                                                       & 0.901 & 0.918 &  0.943 &  0.910 & 0.921 & 0.913 & 0.992\\
KNN                                                       & \textbf{0.999} & \textbf{0.998} & 0.997 & \textbf{0.999} &  \textbf{0.999} & \textbf{0.999} &  \textbf{0.999}\\
MLP                                                       &  0.829 &  0.833 &  0.775 & 0.666 & 0.831 & 0.716 &  0.664\\
LightGBM                                                  & 0.820 & 0.716 & \textbf{0.999} & 0.730 & 0.857 & 0.996 &  0.985       
\\

\hline
\textbf{AI Model k=5} & \textbf{Norm} & \textbf{DoS}  & \textbf{B. Force} & \textbf{Web} & \textbf{Infilt.} & \textbf{Bot}  & \textbf{PS}   \\ \hline
RF                                                        & 0.983 &  0.997 &  0.998 & 0.979 &  0.980 & 0.998 & 0.990\\
ADA                                                       & 0.731 & 0.813 & 0.857 & 0.857 & 0.857 & 0.709 & 0.510\\
DNN                                                       & 0.857 & 0.797 &  0.080 &  0.849 & 0.854 & 0.826 & 0.840\\
SVM                                                       & 0.868 & 0.885 & 0.882 & 0.874 &  0.939 & 0.990 & 0.932 \\
KNN                                                       & \textbf{0.999} & \textbf{0.999} & \textbf{0.999} &  \textbf{0.997} &  \textbf{0.997} & \textbf{0.999} &  \textbf{0.999}\\
MLP                                                       & 0.879 &  0.833 &  0.774 & 0.775 & 0.666 & 0.714 & 0.664\\
LightGBM                                                  & 0.899 & 0.995 &   \textbf{0.999} &  \textbf{0.997} &   0.922 & \textbf{0.999} &  0.985       
\\
\hline
%XGBoost                                                 & 0.97 & 0.74 & 0.75    & 0.74    & 0.99   & 0.77 & 0.74
\end{tabular}
}
\label{tbl:acc_attack_CICIDS}
\vspace{-1mm}
\end{table}

\textbf{Enhancement of Attack Detection under Feature Selection:} 
The quantification of AI model enhancements in detecting assaults (provided by many AI models with optimal performance) under feature selection is presented in Table~\ref{tbl:enhancement_under_feature_selection}. For instance, we have two models that performed better with $k = 5$ and five models that performed better with $k = 15$ for CICIDS-2017 in the `Brute Force' assault. We have two models that perform better when $k = 5$, three models that perform better when $k = 10$, one model that performs better when $k = 15$, and one model that performs better when $k = all$ for RoEduNet-SIMARGL2021 in the `Port Scan' attack.

\begin{table}[hbt!]
\centering
\caption{Quantification of AI model improvements in attack detection under feature selection (based on the number of AI models performing at the highest level).}
% \vspace{-3mm}
\resizebox{\columnwidth}{!}{
\begin{tabular}{lcccc}
\hline
\textbf{CICIDS-2017}  & \textbf{k=5} & \textbf{k=10} & \textbf{k=15} & \textbf{All} \\ \hline
Normal           & 1            & 1             & 5            & 0            \\
DoS              & 1           & 1             & 3             & 2            \\
Brute Force      & 2            & 0             & 5             & 0            \\
Web Attack       & 0            & 3             & 3             & 1            \\
Infiltration     & 0            & 1             & 4             & 2            \\
Bot              & 2            & 0             & 3             & 2            \\
Port Scan        & 1            & 0             & 6             & 0            \\ \hline
\textbf{RoEduNet-SIMARGL2021} & \textbf{k=5} & \textbf{k=10} & \textbf{k=15} & \textbf{All} \\ \hline
Normal           & 1            & 2             & 2            & 2     \\
DoS              & 1            & 2             & 3             & 1     \\
Port Scan        & 2            & 3             & 1             & 1       \\  
\hline
\end{tabular}
}
\vspace{-1mm}
\label{tbl:enhancement_under_feature_selection}
\end{table}

\textbf{False Positive Rates:} 
The analysis of the false positives for the CICIDS-2017 and RoEduNet-SIMARGL2021 datasets, Table~\ref{tbl:False_Positive_Rate}, clearly shows differences in the performances of the different methods. The DNN approach has a notable false positive rate of 34.00\% in the RoEduNet-SIMARGL2021 dataset, indicating a large room for improvement. It is noteworthy to observe that, with a much reduced false-positive rate of 2.46\%, our technique performs noticeably better in the CICIDS-2017 dataset. In the CICIDS-2017 dataset, the LightGbm technique shows a false positive rate of 2.50\%; in comparison, the RoEduNet-SIMARGL2021 dataset records an incredibly low false-positive rate of just 0.0033\%. 
Moreover, there is a discernible performance gap between the two datasets for the ADA algorithm, suggesting that different datasets require different AI techniques for intrusion detection.

\begin{table}[hbt!]
\centering
\caption{False positive rates (\%) for various AI models used in intrusion detection for both datasets.} 
\vspace{1mm}
\resizebox{1\columnwidth}{!}{% Resize table to fit within text width
\begin{tabular}{ccc}%{lS[table-format=2.4]S[table-format=2.4]}
\hline
{\textbf{Algorithm}} & {\textbf{False Positives CICIDS-2017 (\%)}} & {\textbf{False Positives SIMARGL (\%)}} \\
\hline

DNN     & 2.46  & 34.0000 \\
KNN     & 0.11  & 0.0390  \\
RF      & 0.07  & 0.0190  \\
SVM     & 2.19  & 0.3200  \\
MLP     & 0.18  & 0.1800  \\
ADA     & 4.94  & 12.5600 \\

LightGBM & 2.50  & 0.0033  \\
\hline

\end{tabular}
}
\label{tbl:False_Positive_Rate}
% \vspace{-3mm}
\end{table}

\textbf{The importance of Features via Explainable AI:}  
We now demonstrate the significance of the top features that influence each AI model's choice for the two network incursion datasets.

\begin{figure*}[hbt!]
\centering

\begin{minipage}{.3\textwidth}
  \centering
  \includegraphics[width=1\linewidth]{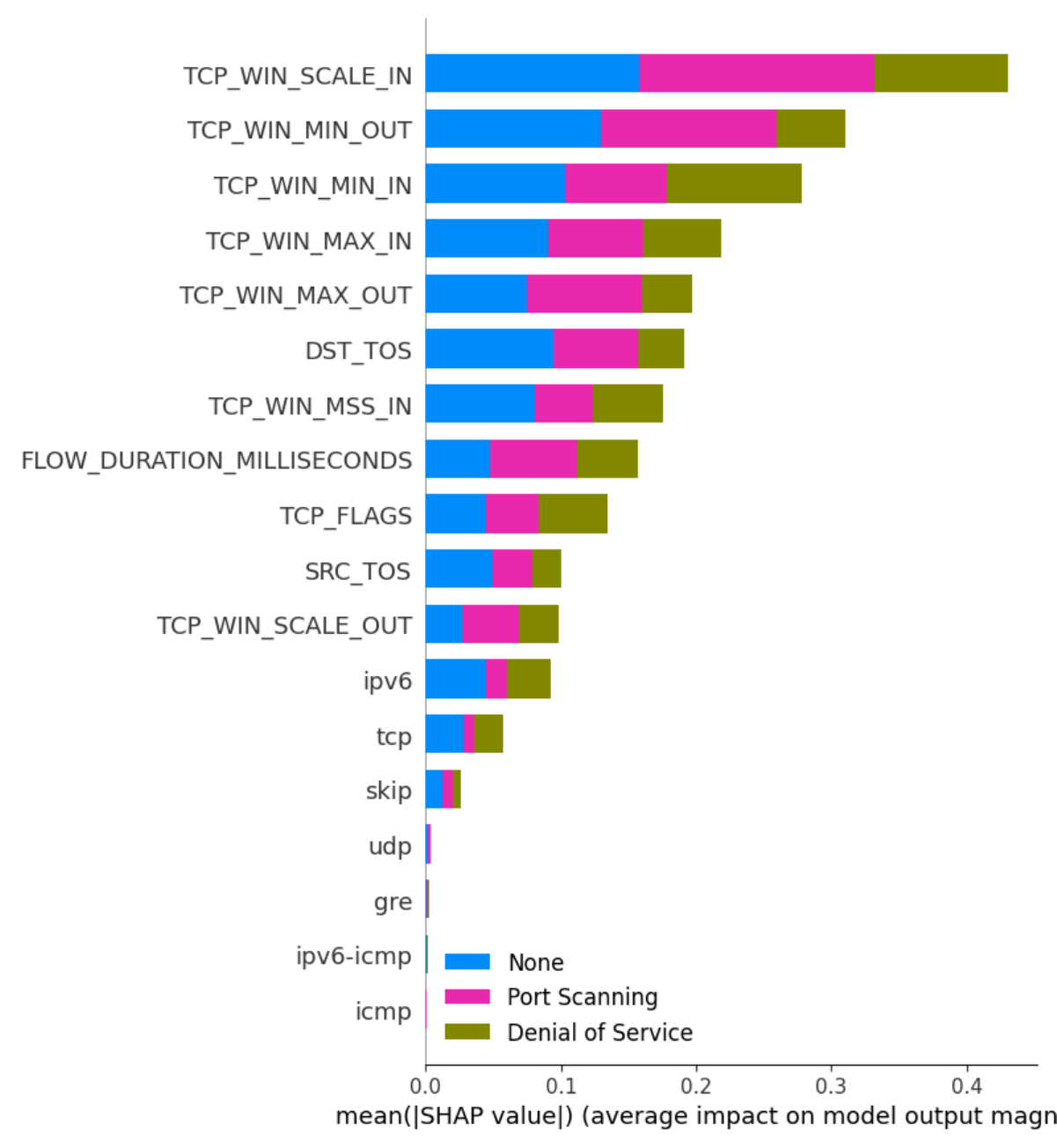}
  {(a) DNN}
  \label{fig:DNN_shap_15_summary}
\end{minipage}%
\begin{minipage}{.3\textwidth}
  \centering
  \includegraphics[width=1\linewidth]{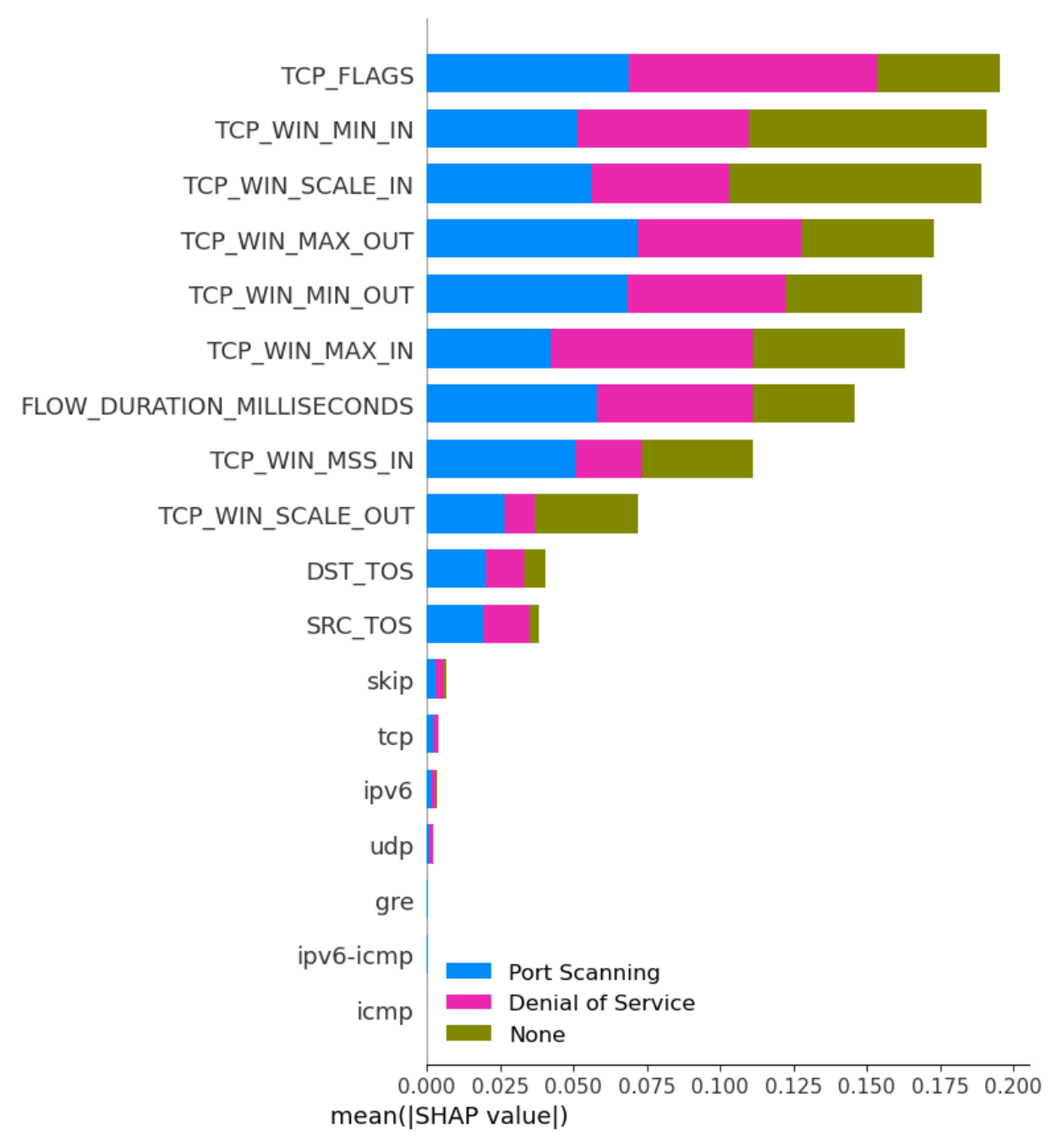}
  {(b) RF}
  \label{fig:RF_shap_15_summary}
\end{minipage}

\begin{minipage}{.3\textwidth}
  \centering
  \includegraphics[width=1\linewidth]{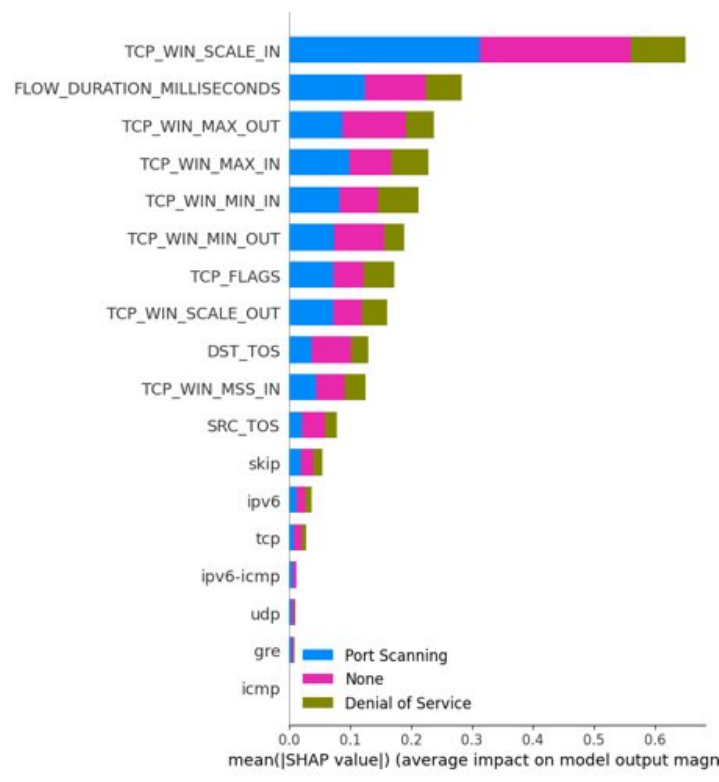}
  {(c) LightGBM}
  \label{fig:Lightbm_shap_15_summary}
\end{minipage}%
\begin{minipage}{.3\textwidth}
  \centering
  \includegraphics[width=1\linewidth]{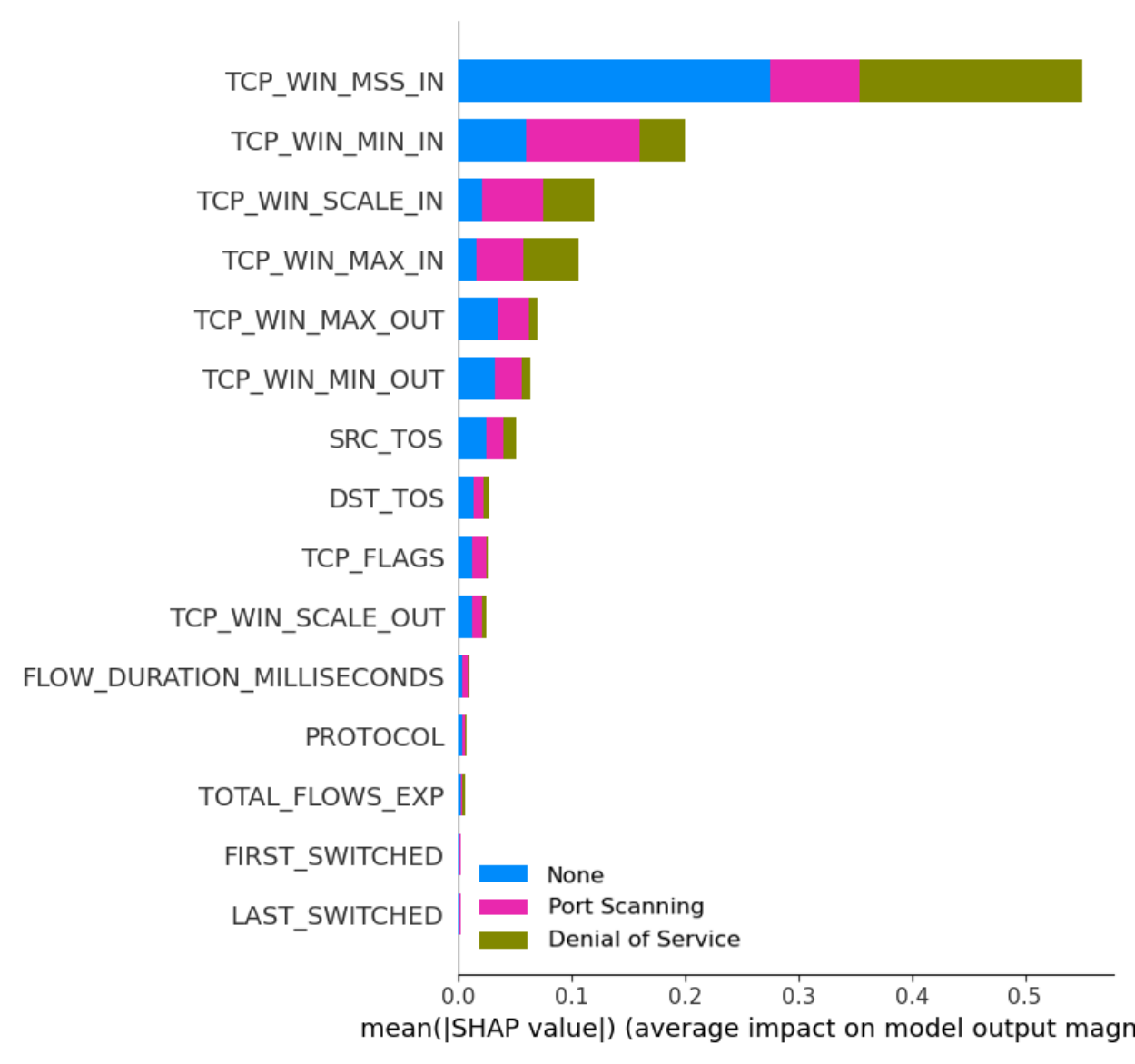}
  {(d) SVM}
  \label{fig:SVM_shap_15_summary}
\end{minipage}

\begin{minipage}{.3\textwidth}
  \centering
  \includegraphics[width=1\linewidth]{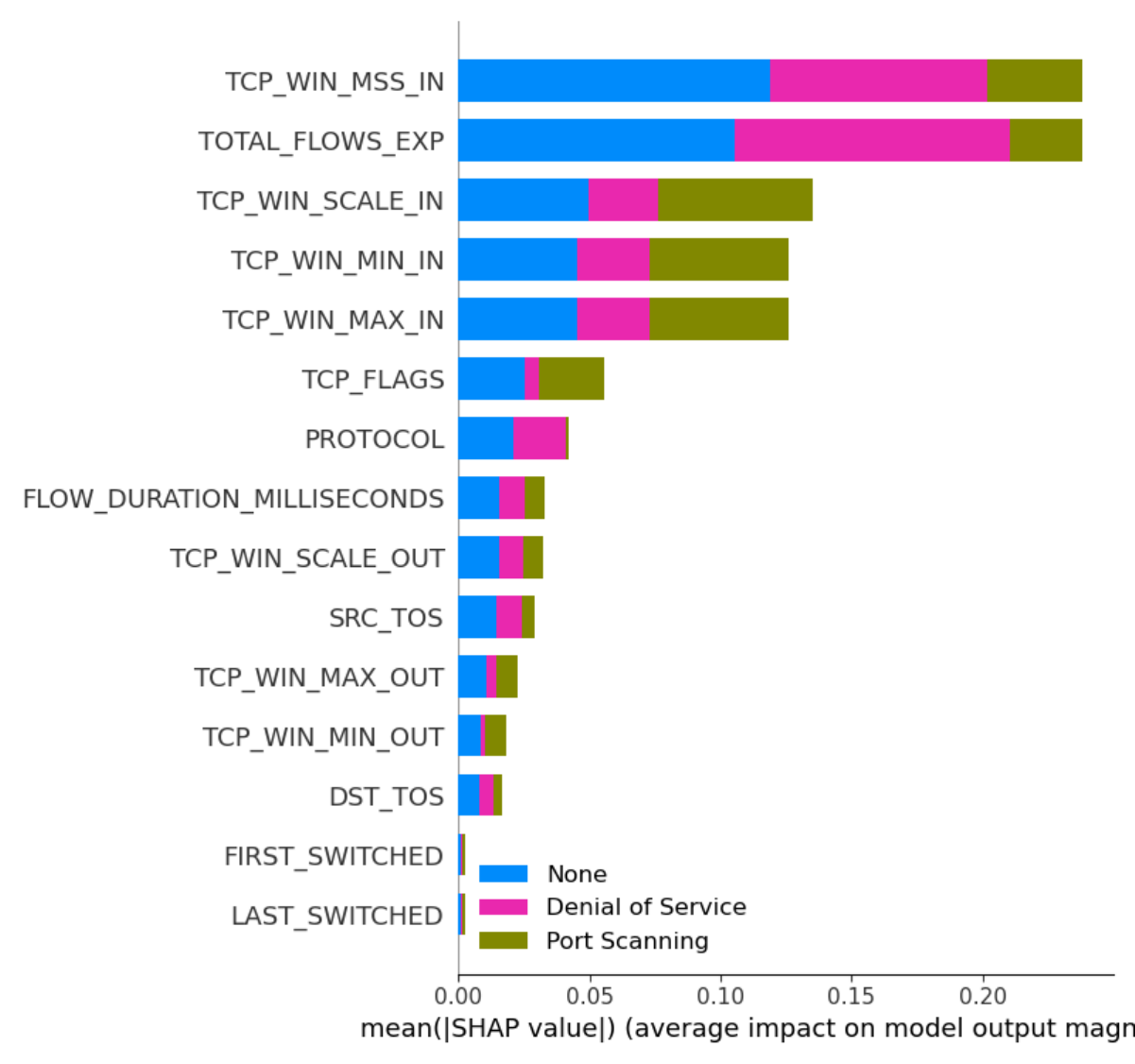}
  {(e) KNN}
  \label{fig:KNN_shap_15_summary}
\end{minipage}%
\begin{minipage}{.3\textwidth}
  \centering
  \includegraphics[width=1\linewidth]{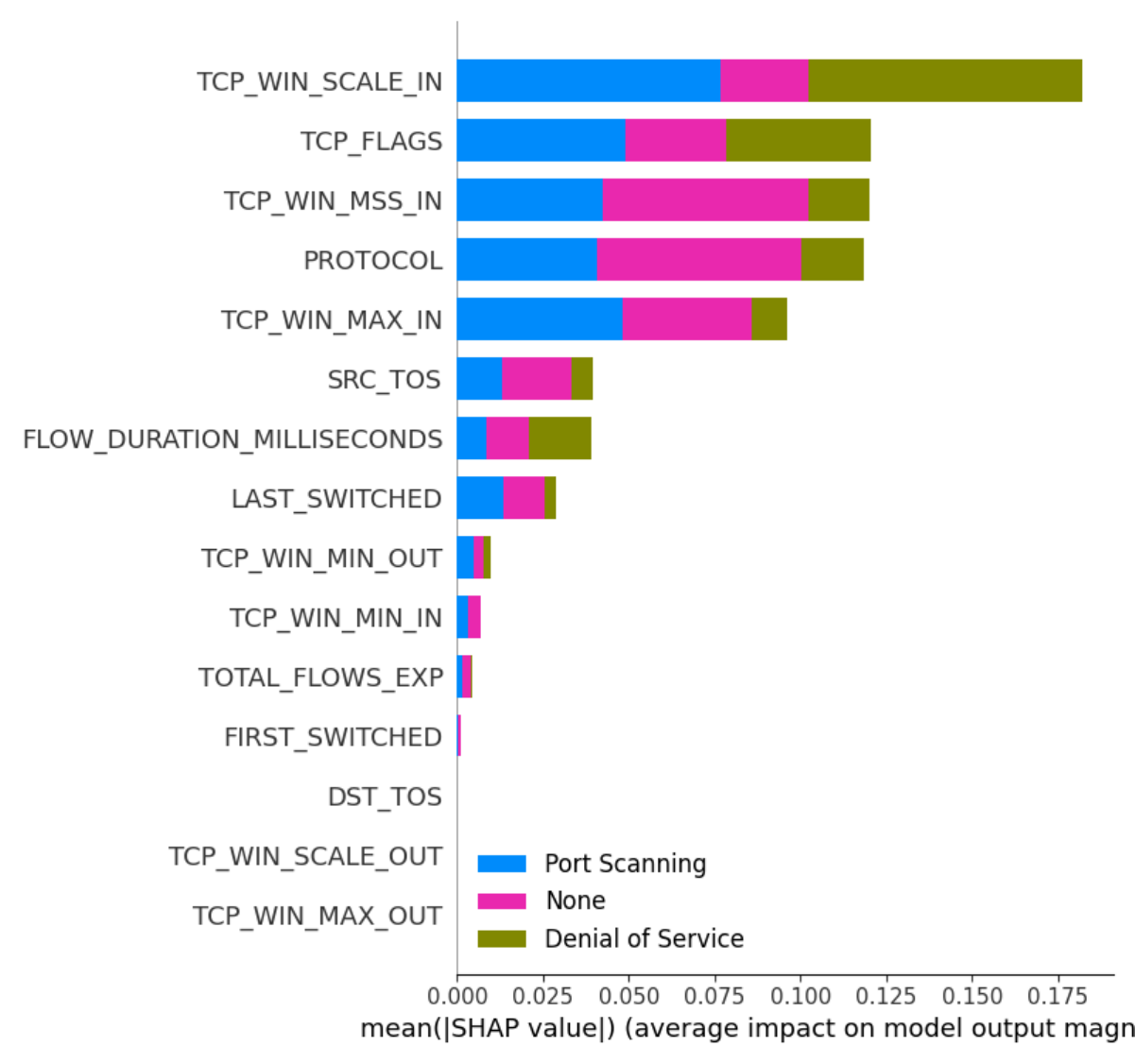}
  {(f) ADA}
  \label{fig:summary_ada_sensor}
\end{minipage}%
\begin{minipage}{.3\textwidth}
  \centering
  \includegraphics[width=1\linewidth]{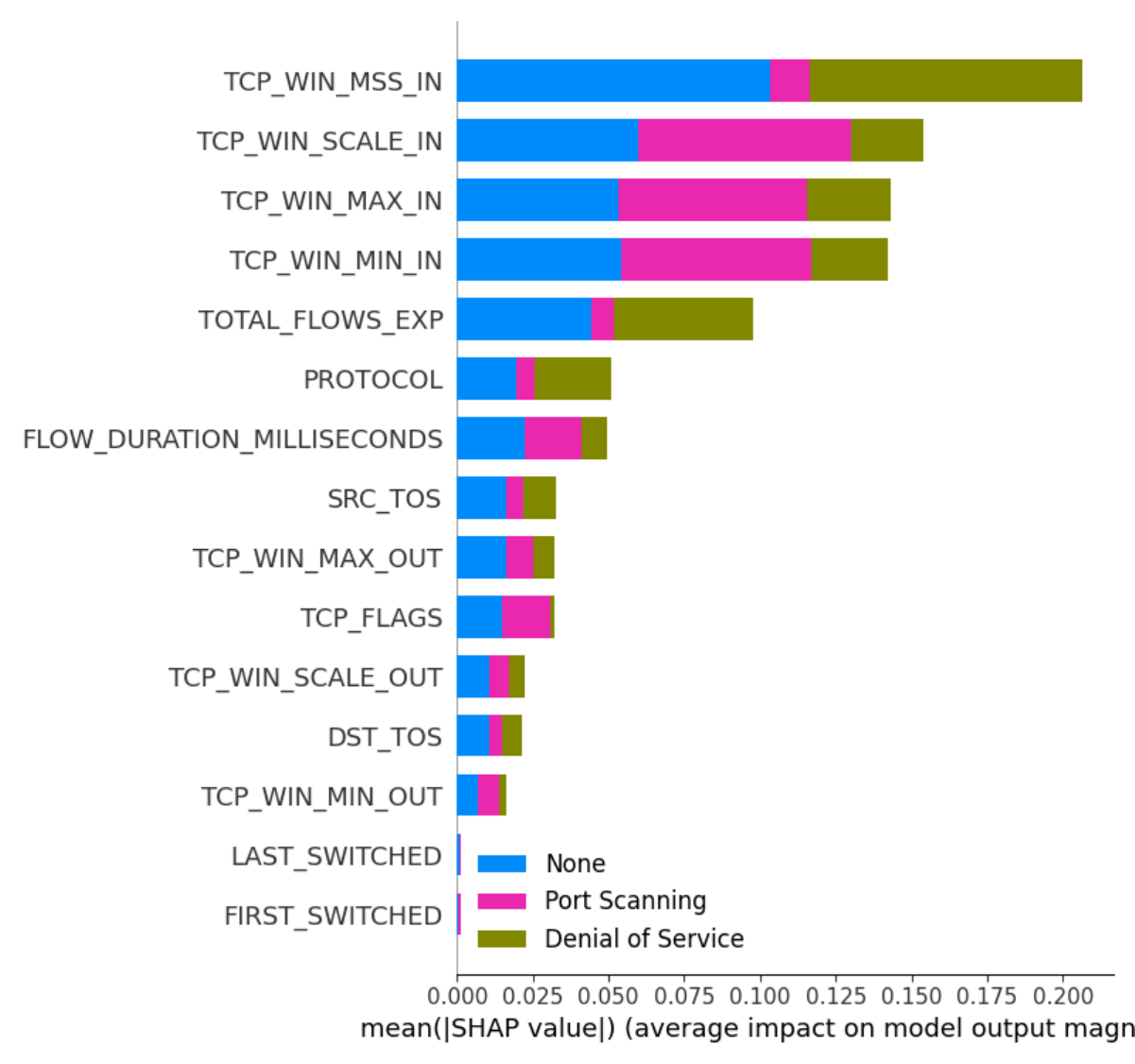}
  {(g) MLP}
  \label{fig:summary_mlp_sensor}

\end{minipage}%

 \caption{For every intrusion detection AI model for the RoEduNet-SIMARGL2021 dataset, global summary graphs (by SHAP) are provided. It illustrates the relative weights of several attributes, with varying hues signifying the relevance of every kind of attack.}

  % \caption{Global summary plots (by SHAP) for each intrusion detection AI model for the RoEduNet-SIMARGL2021 dataset. It visualizes the importance levels of different features, with different colors indicating the significance of each attack type.}
\label{fig:summary_plots_sensors_dataset}
\end{figure*}

\begin{figure*}[hbt!]
\centering

\begin{minipage}{.3\textwidth}
  \centering
  \includegraphics[width=1\linewidth]{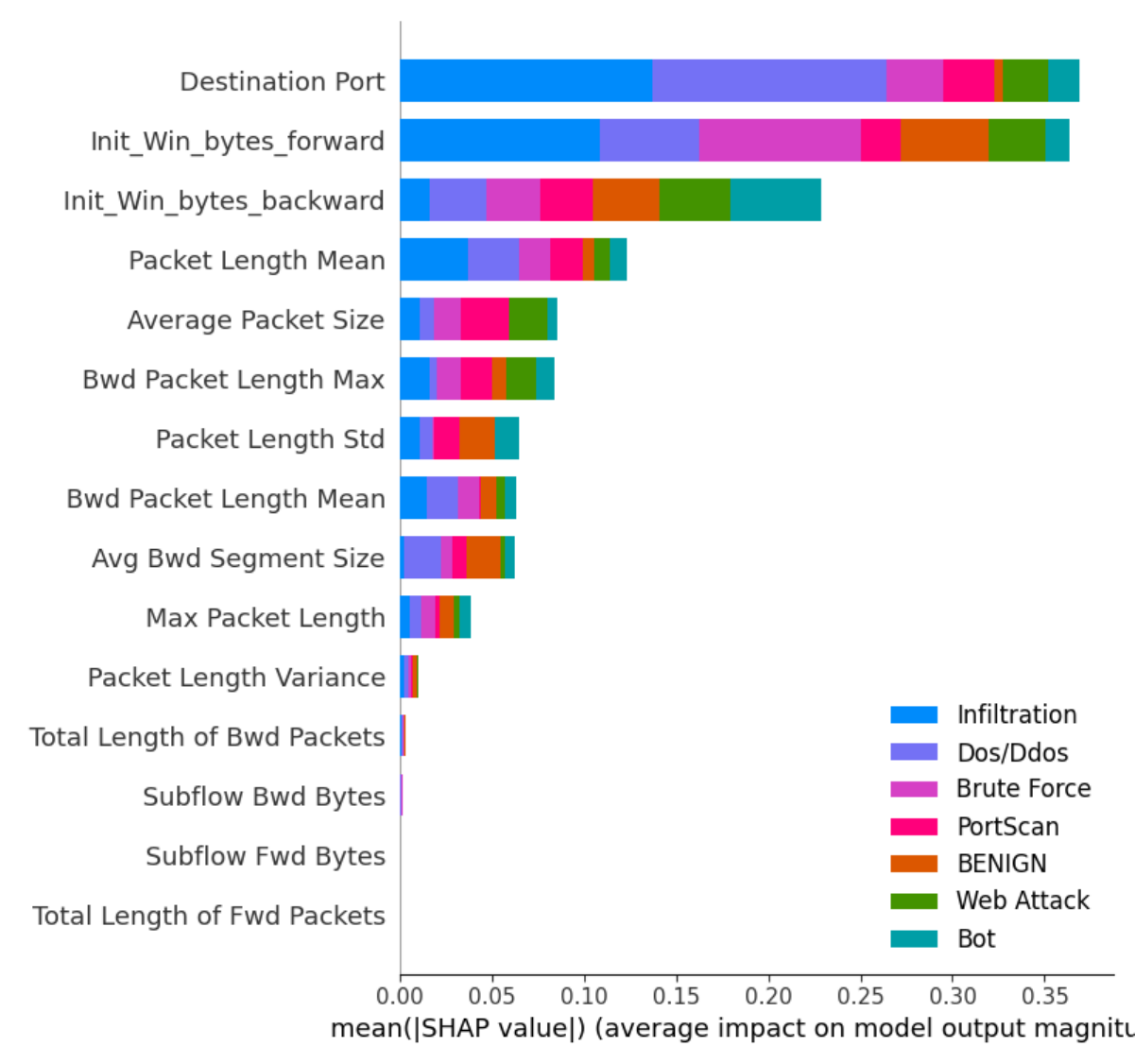}
  {(a) DNN}
  \label{fig:DNN_shap_15_summary_cicids}
\end{minipage}%
\begin{minipage}{.3\textwidth}
  \centering
  \includegraphics[width=1\linewidth]{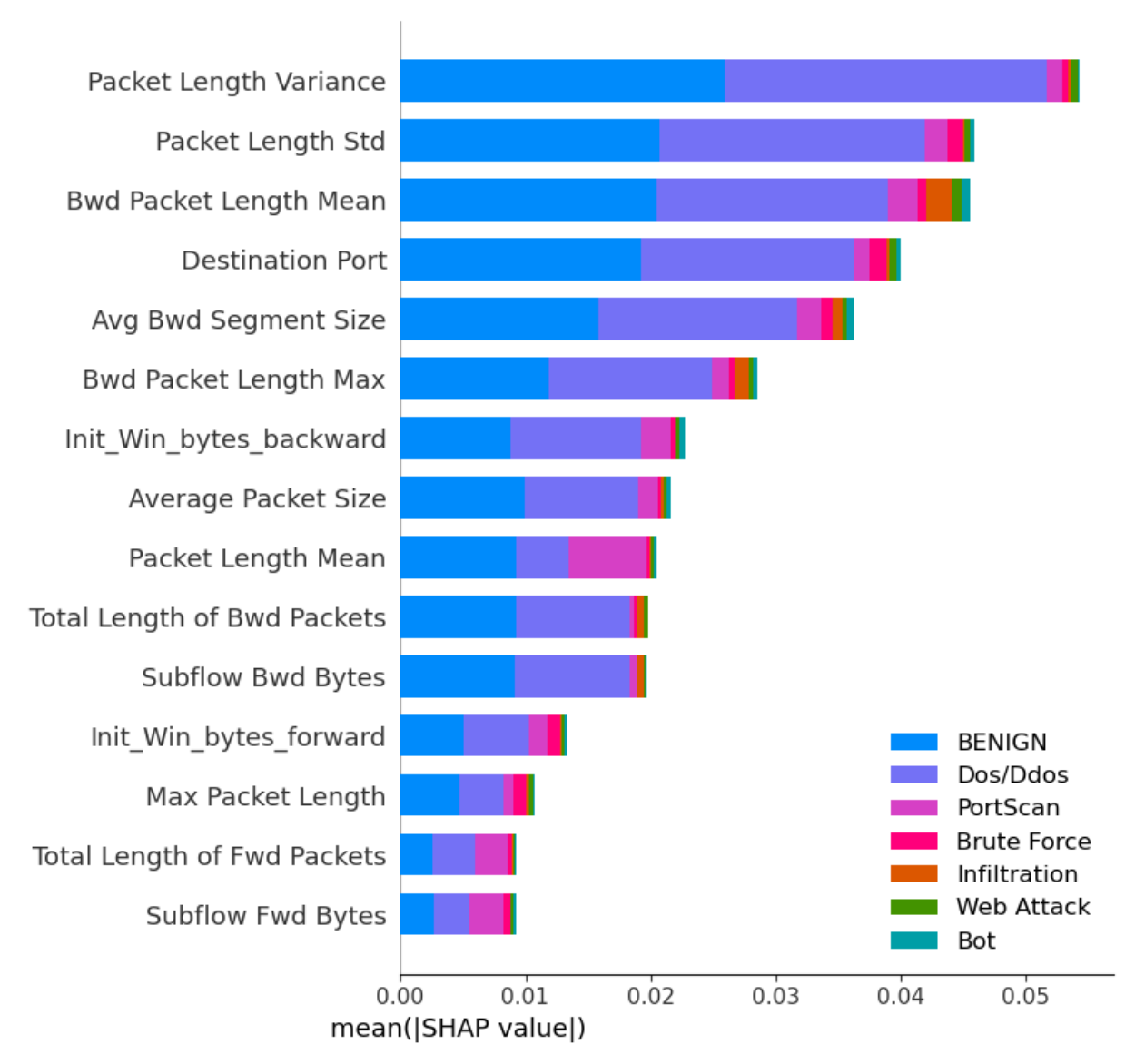}
  {(b) RF}
  \label{fig:RF_shap_15_summary_cicids}
\end{minipage}

\begin{minipage}{.3\textwidth}
  \centering
  \includegraphics[width=1\linewidth]{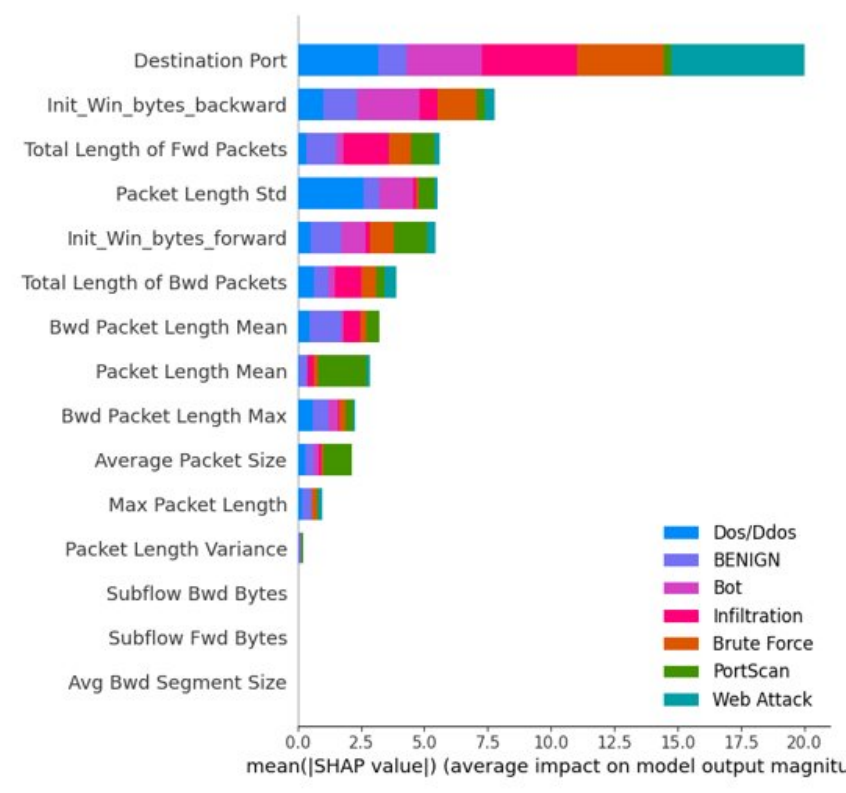}
  {(c) LightGBM}
  \label{fig:Lightbm_shap_15_summary_cicids}
\end{minipage}%
\begin{minipage}{.3\textwidth}
  \centering
  \includegraphics[width=1\linewidth]{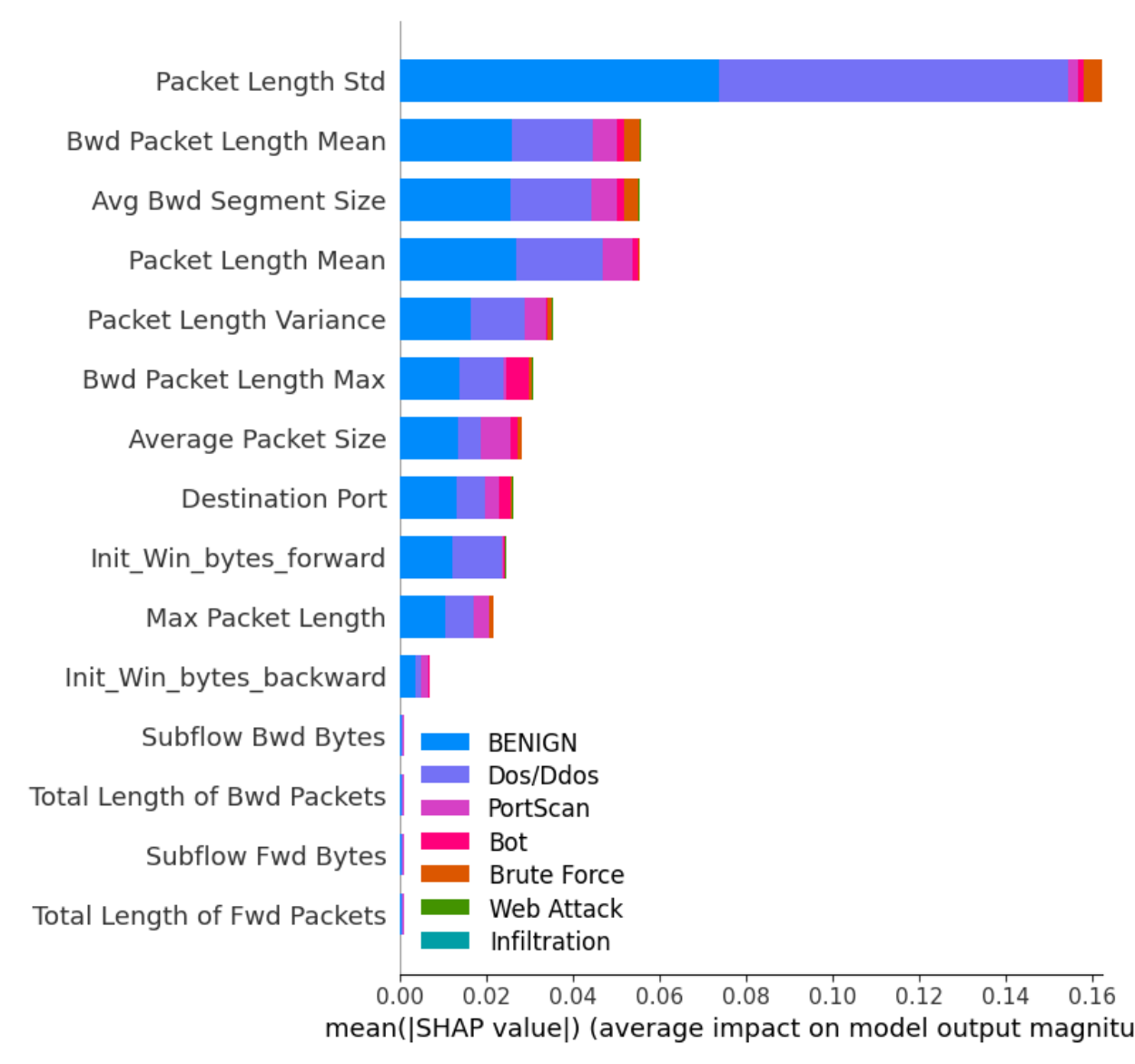}
  {(d) SVM}
  \label{fig:SVM_shap_15_summary_cicids}
\end{minipage}

\begin{minipage}{.3\textwidth}
  \centering
  \includegraphics[width=1\linewidth]{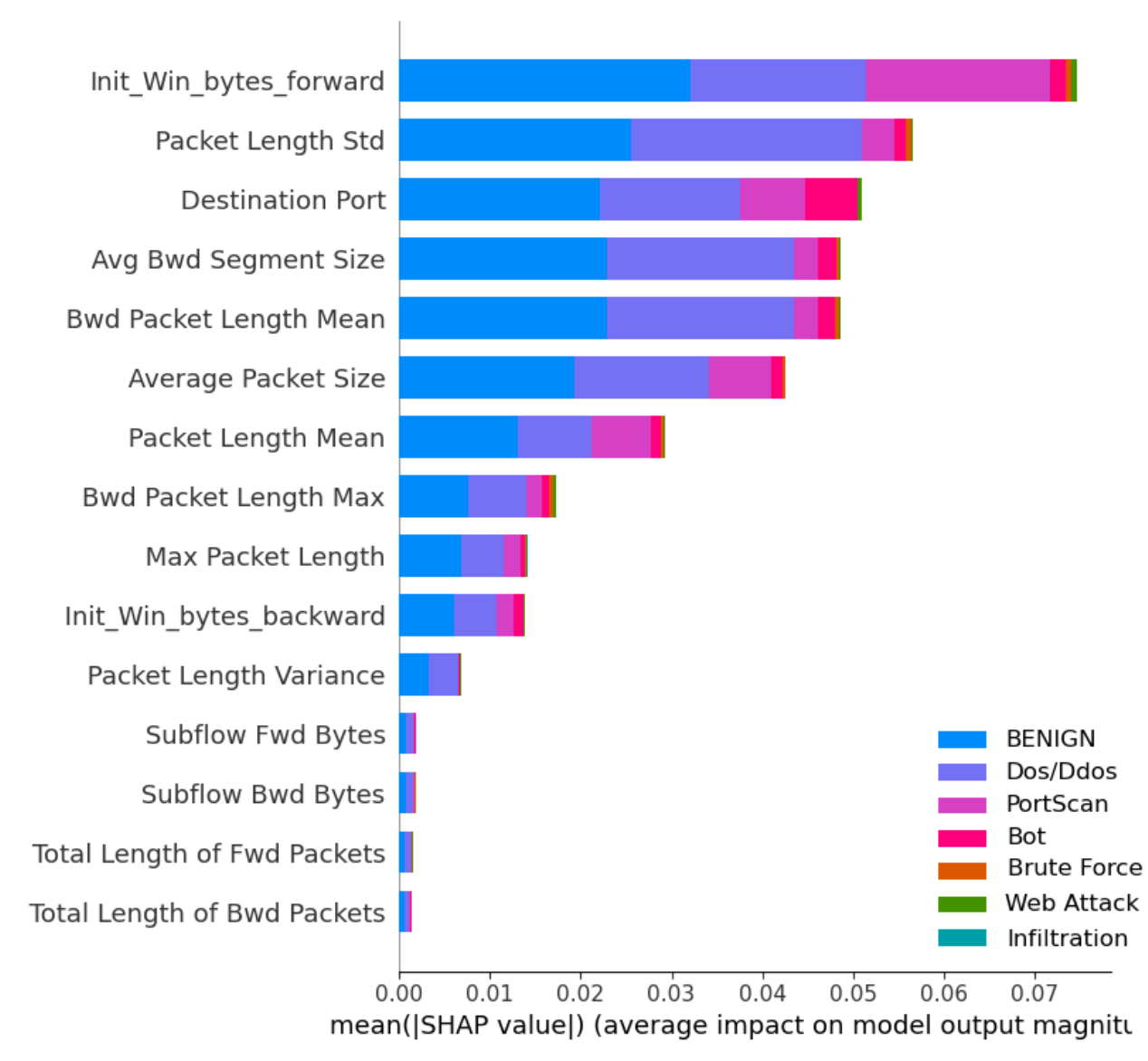}
  {(e) KNN}
  \label{fig:KNN_shap_15_summary_cicids}
\end{minipage}%
\begin{minipage}{.3\textwidth}
  \centering
  \includegraphics[width=1\linewidth]{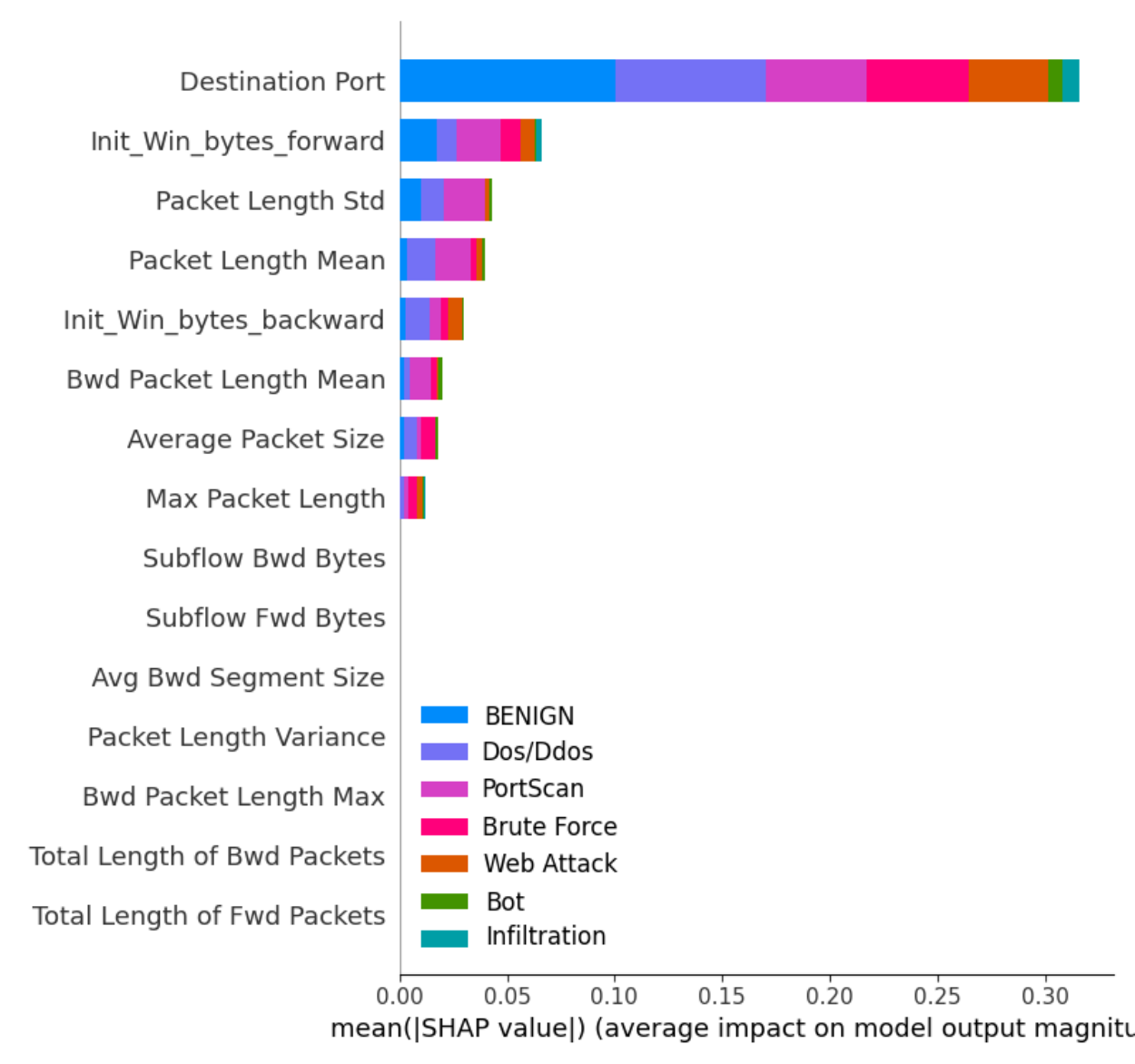}
  {(f) ADA}
  \label{fig:summary_ada_sensor_cicids}
\end{minipage}%
\begin{minipage}{.3\textwidth}
  \centering
  \includegraphics[width=1\linewidth]{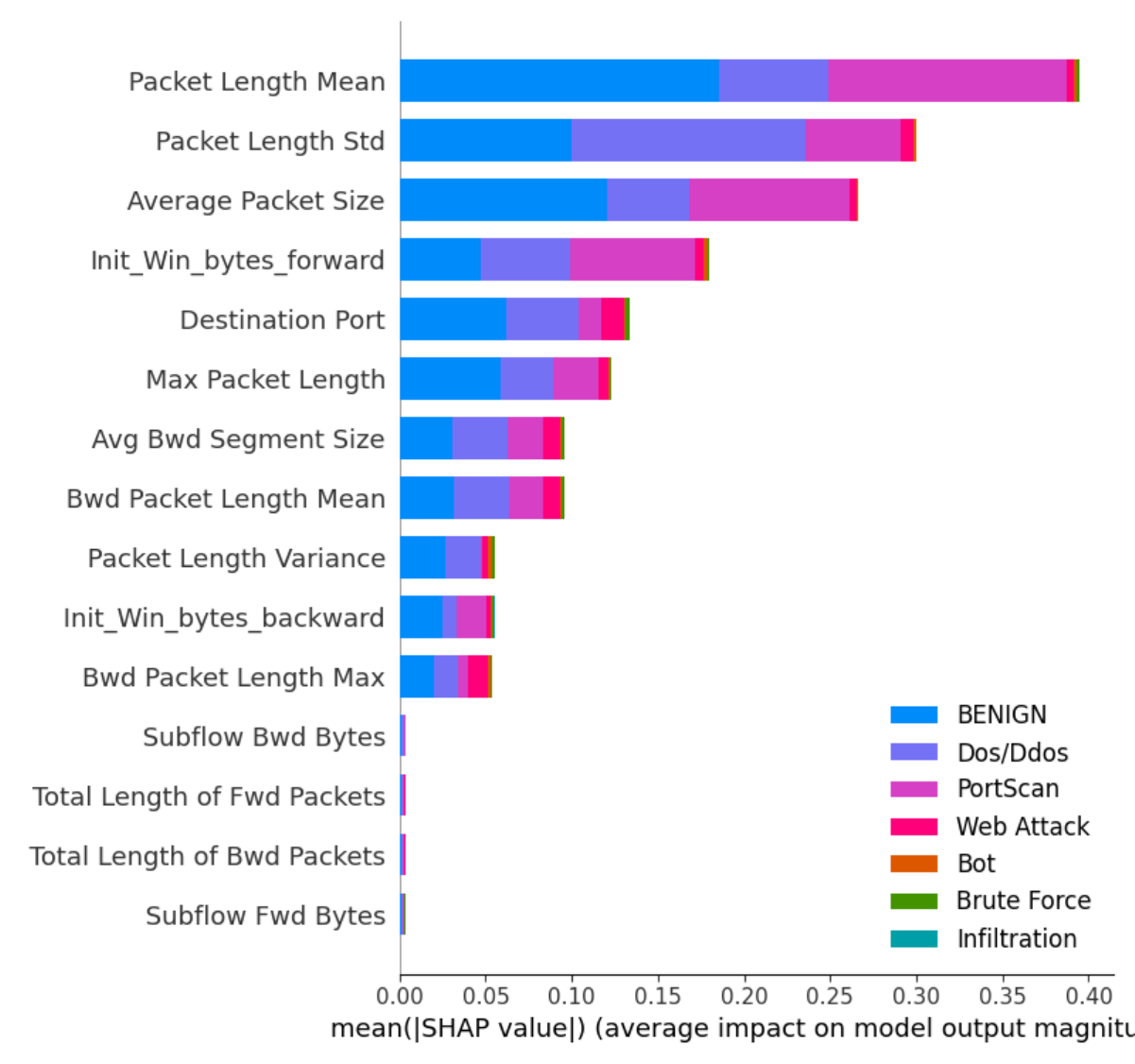}
  {(g) MLP}
  \label{fig:summary_mlp_sensor_cicids}

\end{minipage}%

 \caption{Global summary graphs for each intrusion detection AI model for the CICIDS-2017 dataset, produced by SHAP. It displays the relative weights of several attributes, with varying hues signifying the relevance of each kind of attack.}
\label{fig:summary_plots_cicids_dataset}

\end{figure*}

\textbf{Global Summary Plot:} 
First, we show the global summary charts for several AI models, which highlight the key factors that influence the algorithms' decisions. These plots give important insights into the decision hierarchy by showing the elements in declining order of relevance. For the RoEduNet-SIMARGL2021 and CICIDS-2017 datasets, respectively, Figures~\ref{fig:summary_plots_sensors_dataset} and \ref{fig:summary_plots_cicids_dataset} illustrate the feature significance across various AI models. By averaging the mean of each Shapley value, the feature significance values are obtained. Notably, by assigning different colors to each assault label, this representation also emphasizes the importance of certain attributes for each sort of attack (or intrusion).

\textbf{Feature Importance based on XAI:} 
The next stage is to extract the most important features from all intrusion detection AI models for both network intrusion datasets after analyzing the feature importance of each model. The top eight features for each dataset are shown in Table~\ref{tbl:feature_importance_all}. For instance, TCP\_WIN\_SCALE\_IN is a prominent characteristic in the RoEduNet-SIMAR GL2021 dataset that affects all AI models' judgments. Similar to this, every AI model in the CICIDS-2017 dataset relies heavily on the Destination Port attribute to make judgments. When the precise attack type is unclear, security analysts can monitor these features in network traffic thanks to this overall feature rating.

\begin{table}[hbt!]
\caption{For RoEduNet-SIMARGL2021 and other AI models, an overall feature importance (rank)
CICIDS-2017 datasets for intrusions. The order of the rank is downward.
}
\vspace{1mm}
\resizebox{\columnwidth}{!}{
\begin{tabular}{cll}
\hline
\begin{tabular}[c]{@{}c@{}}\textbf{Feature} \\ \textbf{Rank}\end{tabular} & \textbf{CICIDS-2017}                      & \textbf{RoEduNet-SIMARGL2021}                       \\ \hline
1                                                             & Destination Port            & TCP\_WIN\_MSS\_IN         \\
2                                                             & Init\_Win\_Bytes\_Backward  & TCP\_WIN\_SCALE\_IN \\
3                                                             & Packet Length Std           & TCP\_WIN\_MAX\_IN           \\
4                                                             & Bwd Packet Length Mean      & TCP\_WIN\_MIN\_IN            \\
5                                                             & Total length of Bwd Packets & TCP\_WIN\_WIN\_OUT            \\
6                                                             & Packet Length Mean          & TCP\_FLAGS           \\
7                                                             & Subflow Backward Bytes      & PROTOCOL                   \\
8                                                             & Packet Length Variance      & FLOW\_DURATION\_MILLISECONDS \\  
\hline
\end{tabular}
}
\label{tbl:feature_importance_all}
\vspace{-4mm}
\end{table}

\textbf{Outlining RoEduNet-SIMARGL2021's Principal Features:} Understanding the nature of the ``TCP\_WIN\_SCALE\_IN'' characteristic and its possible influence on network intrusion detection tasks is crucial when evaluating its significance in AI models. The inbound TCP Window Scale, a parameter in the Transmission Control Protocol (TCP) for internet communication, is referred to as the ``TCP\_WIN\_SCALE\_IN" feature. This option allows the window size that controls the data flow to be adjusted. As a result, by increasing the TCP window size, an attacker can use this parameter to attack the network and regulate data movement.

\textbf{Explaining CICIDS-2017's Top Features:}
The port number on the destination device or server in a network connection is referred to as the ``Destination Port" in the context of intrusion detection systems and network security. Remember that ports are only numbers that are used to assist in identifying which application or service on a device is receiving network data. Therefore, the port that a packet is meant to be received at is indicated by the `{Destination Port'. Monitoring and examining destination port addresses in the context of IDS might be essential for identifying specific kinds of network assaults or unwanted activity.

\textbf{Feature Importance for Each Attack Type:} 
The intrusion-specific feature importance, which lists the top five attributes for every kind of assault, is next shown. It should be noted that by going back to the Figures~\ref{fig:summary_plots_sensors_dataset} and ~\ref{fig:summary_plots_cicids_dataset}, the attack-specific characteristics are produced. Next, we use frequency analysis to examine each assault (intrusion) separately and extract the most important attributes of each attack. This data is shown for the RoEduNet-SIMARGL2021 dataset in Table~\ref{tbl:feat_importance_per_attack_sensor}, and for the CICIDS-2017 dataset in Table~\ref{tbl:feat_imp_attack_CICIDS}. 
Interestingly, we find that some common aspects across the two datasets—such as Destination Port and Init\_Win\_Bytes\_Bwd for CICIDS-2017, and TCP\_WIN\_SCALE\_IN and FLOW\_DURATION\_MS for RoEduNet-SIMARGL2021—are shared across various attack types. These results point out the most important factors that analysts have to consider while looking into certain network invasions. Furthermore, by understanding the key indicators (features) of these attacks in the network flow, this knowledge can help develop attack-specific solutions for intrusion detection systems (IDS) and adapt already-existing AI models for IDS based on these identified features for more accurate detection of specific network attacks.

\begin{table*}[hbt!]
\centering
\caption{The RoEduNet-SIMARGL2021 dataset's feature significance (the top five features) for every kind of assault.} 
% \vspace{-3mm}
\resizebox{.75\linewidth}{!}{
\begin{tabular}{clll}
\hline
\begin{tabular}[c]{@{}c@{}}\textbf{Feature}\\ \textbf{Rank}\\ \end{tabular} & \textbf{Normal}                                                                  & \textbf{DoS}                                                                     & \textbf{Port Scan}                                                                \\ \hline
1                                                                     & TCP\_WIN\_SCALE\_IN & TCP\_WIN\_SCALE\_IN&      TCP\_WIN\_SCALE\_IN     \\
2                                                                     & TCP\_WIN\_MAX\_OUT  & TCP\_WIN\_MIN\_IN            & FLOW\_DURATION\_MS \\
3                                                                     & FLOW\_DURATION\_MS & TCP\_WIN\_MAX\_IN            & TCP\_WIN\_MAX\_IN            \\
4                                                                     & TCP\_WIN\_MIN\_OUT           & FLOW\_DURATION\_MS & TCP\_WIN\_MAX\_OUT           \\
5                                                                     & TCP\_WIN\_MAX\_IN            & TCP\_FLAGS                                                              & TCP\_WIN\_MIN\_IN \\
\hline
\end{tabular}
}
\label{tbl:feat_importance_per_attack_sensor}
%\vspace{-2mm}
\end{table*}

\begin{table*}[hbt!]
% \vspace{0.5mm}
\caption{The CICIDS-2017 dataset's feature significance (the top five features) for every kind of assault.
 }
\vspace{1mm}
\resizebox{0.95\linewidth}{!}{
\begin{tabular}{cllll}
\hline
\textbf{Feature}\\ \textbf{Rank} & \textbf{Normal}                                                                  & \textbf{DoS}                                                                    & \textbf{Port Scan}   &  \textbf{Bot}                                                               \\ \hline
1                                                                     & Destination Port                                                       & Destination Port                                                      & Packet Length Mean & Destination Port                \\
2                                                                     & Bwd Packet Length Mean      & Packet Length Std           & Init\_Win\_Bytes\_Fwd &     Init\_Win\_Bytes\_Bwd \\

3                                                                     & Total Length of Bwd Packets & Init\_Win\_Bytes\_Bwd      & Packet Length Std  & Init\_Win\_Bytes\_Fwd   \\
4                                                                     & Packet Length Std           & Total Length of Bwd Packets & Total Length Fwd Packets &  Packet Length Std   \\
5                                                                     & Init\_Win\_Bytes\_Fwd      & Bwd Packet Length Max       & Average Packet Size & Packet Length Mean    \\ 
%& & & \\
\hline
% & & & \\
\textbf{Feature}\\ \textbf{Rank}      & \textbf{Web Attack}                                                          & \textbf{Brute Force}                                                            & \textbf{Infiltration}                                                              \\ \hline
1       & Destination Port       & Destination Port            & Destination Port                \\
2       &  Init\_Win\_Bytes\_Bwd  & Init\_Win\_Bytes\_Bwd    & Init\_Win\_Bytes\_Bwd   \\

3                                                                     &  Init\_Win
\_bytes\_Fwd  & Init\_Win
\_bytes\_Fwd & Init\_Win
\_bytes\_Fwd \\
4    & Bwd Packet Length Mean & Total Length of Fwd Packets & Bwd Packet Length Mean  \\
5      & Packed Length Std      & Total Length of Bwd packets & Total Length of Fwd Packets  \\ \hline  
\end{tabular}
}
\label{tbl:feat_imp_attack_CICIDS}
%\vspace{-2mm}
\end{table*}

\vspace{-2mm}

\begin{table*}[ht]
%\vspace{-0.9mm}
\centering

\caption{AI performance under the best features chosen by our framework compared to information gain~\cite{CICIDS} and K-best~\cite{SENSOR}. For the two datasets, our system performs better (bold text) in 22 out of 28 AI models.
}
% \vspace{-3mm}
\resizebox{0.99\linewidth}{!}{
\begin{tabular}{lccccccc||lccccccc}
\hline
\multicolumn{8}{c||}{\textbf{CICIDS-2017}} & \multicolumn{8}{c}{\textbf{RoEduNet-SIMARGL2021}} \\ 
\hline

%%%%%%%% first part 

\textbf{AI Model k = 15}                                 & \textbf{Acc} & \textbf{Prec} & \textbf{Rec} & \textbf{F1} & \textbf{Bacc} & \textbf{Mcc} & \textbf{AucRoc} & 
\textbf{AI Model k = 15}            
& 
\textbf{Acc} & \textbf{Prec} & \textbf{Rec} & \textbf{F1} & \textbf{Bacc} & \textbf{Mcc} & \textbf{AucRoc}

\\ \hline
RF (our framework)      & \textbf{ 0.988}         & \textbf{0.959}          & \textbf{0.959}          & \textbf{0.959}        & \textbf{0.976}          & \textbf{0.952}         & 0.385     
& RF (our framework) & \textbf{0.999}         & \textbf{0.999}          & \textbf{0.999}          & \textbf{0.999}        & \textbf{0.999}          & \textbf{0.999}         & \textbf{0.999}          \\

RF (IG~\cite{CICIDS})         & 0.979         & 0.928         & 0.928          & 0.928        & 0.958          & 0.916         & \textbf{0.678}  
& RF (K-best~\cite{SENSOR}) & 0.998         & 0.998          & 0.998          & 0.998        & 0.998          & 0.997        & \textbf{0.999}             \\

 \hline
ADA (our framework)  &  0.745         & 0.110          & 0.110          & 0.110        & 0.481          & \textbf{ -0.037}         & \textbf{0.972}   
& ADA (our framework) & \textbf{ 0.765}         & \textbf{0.647}          & \textbf{0.647}          & \textbf{0.647}        & \textbf{0.735}          & \textbf{0.471}         & 0.513                   \\

ADA (IG~\cite{CICIDS}) & \textbf{0.762}         &  \textbf{0.169}          &  \textbf{0.169}          & \textbf{0.169}       & \textbf{0.515}          & 0.030         & 0.432 
& ADA (K-best~\cite{SENSOR}) & 0.745      & 0.617         &0.617    & 0.617  &  0.713     & 0.426       & \textbf{0.566}             \\

\hline
DNN (our framework)  & 0.737         & 0.080          & 0.080          & 0.080        &  0.463          &  \textbf{-0.072}       & \textbf{0.937}
& DNN (our framework) & 0.557         & 0.336          & 0.336          & 0.336        &  0.502          & 0.004       & 0.392         \\

DNN (IG~\cite{CICIDS})  & \textbf{0.769}   & \textbf{0.194}   & \textbf{0.194}   & \textbf{0.194}   & \textbf{0.529}  &  0.059  &  0.931     
& DNN (K-best~\cite{SENSOR}) & \textbf{0.560}         & \textbf{ 0.341}          & \textbf{0.341}          & \textbf{ 0.341}        & \textbf{0.505}          & \textbf{0.011}       & \textbf{0.472}      \\

\hline
SVM (our framework)  & \textbf{0.972}         & \textbf{ 0.904}          & \textbf{ 0.904}          & \textbf{ 0.904}        &\textbf{0.944}          & \textbf{ 0.888}         & \textbf{0.795}     
& SVM (our framework) & \textbf{ 0.986}         & \textbf{ 0.979}          & \textbf{ 0.979}          & \textbf{ 0.979}        & \textbf{0.984}          & \textbf{0.969}       & \textbf{0.972}               \\

SVM (IG~\cite{CICIDS})       & 0.933        & 0.765          &  0.765          &  0.765        &  0.863  & 0.726         & 0.532 
& SVM (K-best~\cite{SENSOR}) &  0.940       &   0.910        &   0.910        &   0.910        & 0.932       & 0.865  &  0.965           \\

\hline
KNN (our framework)  & \textbf{0.999}         & \textbf{0.999}          & \textbf{0.999}          & \textbf{0.999}        & \textbf{0.999}          & \textbf{0.999}         & \textbf{0.998}  
&  KNN (our framework) & \textbf{0.998}         & \textbf{ 0.998}          & \textbf{0.998}          & \textbf{0.998}        & \textbf{ 0.998}          & \textbf{0.997}       & 0.998               \\

KNN (IG~\cite{CICIDS})  & \textbf{0.999} & 0.997  & 0.997   & 0.997 &  0.998     & 0.998 & 0.996 
& KNN (K-best~\cite{SENSOR}) & 0.997         & 0.996          & 0.996          & 0.996        & 0.997          & 0.995       & \textbf{0.999}               \\

\hline
MLP (our framework) & \textbf{ 0.766}         & \textbf{ 0.181}          & \textbf{ 0.181}          & \textbf{ 0.181}        & \textbf{0.522}          & \textbf{0.045}         & \textbf{0.991} 
& MLP (our framework) & \textbf{0.555}         & \textbf{ 0.333}          & \textbf{ 0.333}  & \textbf{0.333}    & \textbf{0.500}          & \textbf{0.000}       & \textbf{0.999}         \\

MLP (IG~\cite{CICIDS})   &  0.759      &  0.158 &  0.158   &  0.158  &  0.508  &  0.017  & 0.458  
& MLP (K-best~\cite{SENSOR}) & \textbf{0.555}         & \textbf{0.333}          & \textbf{0.333}          & \textbf{0.333}        & \textbf{0.500}          & \textbf{0.000}       & 0.995         \\

\hline
LightGBM (our framework) & \textbf{0.996}         & \textbf{0.988}          & \textbf{0.988}          & \textbf{0.988}        & \textbf{0.993}          & \textbf{0.986}         & \textbf{0.991} 
& LightGBM (our framework) &  \textbf{0.999} &  \textbf{0.999} &  \textbf{0.999} &  \textbf{0.999} &  \textbf{0.999} &  \textbf{0.999} &  \textbf{0.999}          \\

LightGBM (IG~\cite{CICIDS})   & 0.872        &  0.552          &  0.552 &  0.552    &  0.738   & 0.477  & 0.981 
&  LightGBM (K-best~\cite{SENSOR}) &  0.872         & 0.552          & 0.552           & 0.552         & 0.738         & 0.477         & 0.981

%%%% end of the first part
\\
\hline 
%%%% Second part k = 10

\textbf{AI Model k = 10}                                 & \textbf{Acc} & \textbf{Prec} & \textbf{Rec} & \textbf{F1} & \textbf{Bacc} & \textbf{Mcc} & \textbf{AucRoc} & 
\textbf{AI Model k = 10}                               & 
\textbf{Acc} & \textbf{Prec} & \textbf{Rec} & \textbf{F1} & \textbf{Bacc} & \textbf{Mcc} & \textbf{AucRoc}

\\ \hline
RF (our framework) & \textbf{0.992} &  \textbf{0.971} & \textbf{0.971} & \textbf{0.971} &  \textbf{0.983} &  \textbf{0.967} & \textbf{0.679}

& RF (our framework) & \textbf{0.998} &  \textbf{0.997} &  \textbf{0.997}    &  \textbf{0.997} &  \textbf{0.998}   &  0.995  &  \textbf{0.999}          \\

RF (IG~\cite{CICIDS}) & 0.973 & 0.906 &  0.906  & 0.906 &  0.945 & 0.891 & 0.572 

& RF (K-best~\cite{SENSOR}) & \textbf{0.998} & \textbf{0.997} & \textbf{0.997}  & \textbf{0.997} &   \textbf{0.998}   &  \textbf{0.996}  & \textbf{0.999}          \\

 \hline
 
ADA (our framework)  & 0.752 &  0.132 & 0.132 & 0.132 &  0.494 &  \textbf{-0.012} & \textbf{0.960}

& ADA (our framework) & \textbf{0.999} & \textbf{0.999} &  \textbf{0.999}  & \textbf{0.999} & \textbf{0.999} & \textbf{0.999} & \textbf{0.579} 
\\
ADA (IG~\cite{CICIDS}) &  \textbf{0.753} &   \textbf{0.135} &  \textbf{0.135} & \textbf{0.135} &  \textbf{0.495}  &  -0.009  & 0.959 

& ADA (K-best~\cite{SENSOR}) & \textbf{0.999} & \textbf{0.999} &  \textbf{0.999}  & \textbf{0.999} & \textbf{0.999} & \textbf{0.999} & 0.567     \\

\hline

DNN (our framework) & \textbf{0.759} & \textbf{0.156} & \textbf{0.156} & \textbf{0.156} &  \textbf{0.508} & \textbf{0.015} &  0.929

& DNN (our framework) & 0.560 &   0.341 &   0.341 &   0.341 &  0.506 & 0.012 & \textbf{0.988}         \\

DNN (IG~\cite{CICIDS}) & 0.756 & 0.147 & 0.147 & 0.147 & 0.502 &  0.005 & \textbf{0.931}

& DNN (K-best~\cite{SENSOR})& \textbf{0.563} & \textbf{0.345} & \textbf{0.345} & \textbf{0.345} & \textbf{0.509} & \textbf{0.018} & 0.411      \\      
\hline

SVM (our framework) & \textbf{0.929} &   \textbf{0.751} &  \textbf{0.751}    &  \textbf{0.751} &  \textbf{0.854}   &  \textbf{0.709}  &  \textbf{0.626} 

& SVM (our framework) &  \textbf{0.956} & \textbf{0.934} &  \textbf{0.934} & \textbf{0.934} & \textbf{0.951} &   \textbf{0.902} &  \textbf{0.921}   \\

SVM (IG~\cite{CICIDS})     &  0.917 &  0.709 & 0.709 & 0.709 &  0.830 & 0.661 & 0.615 

& SVM (K-best~\cite{SENSOR}) &  0.940 & 0.910 & 0.910 & 0.910 &  0.933 & 0.866 & 0.916     \\
\hline

KNN (our framework)  & \textbf{0.999} &  \textbf{0.997} &  \textbf{0.997} &  \textbf{0.997} &  \textbf{0.998} &  \textbf{0.997} &  \textbf{0.998} 

&  KNN (our framework) &  \textbf{0.999} &   \textbf{0.999} &   \textbf{0.999}    &   \textbf{0.999} &  \textbf{0.999}   &   \textbf{0.998} &  0.996  \\

KNN (IG~\cite{CICIDS}) &\textbf{0.999} &  0.996 &  0.996 &  0.996 &  \textbf{0.998} &  0.995 & 0.996

& KNN (K-best~\cite{SENSOR})&  \textbf{0.999} &   \textbf{0.999} & \textbf{0.999} &  \textbf{0.999} & \textbf{0.999} &  \textbf{0.998} &  \textbf{0.998}   \\
\hline

MLP (our framework) & \textbf{0.759} &  \textbf{0.158} & \textbf{0.158}  & \textbf{0.158} &  \textbf{0.509} &  \textbf{0.018} &  \textbf{0.987}

& MLP (our framework) & \textbf{0.555} &  \textbf{0.333}&  \textbf{0.333}  &  \textbf{0.333}&  \textbf{0.500}  &  \textbf{0.000 } & \textbf{ 0.995 }  \\

MLP (IG~\cite{CICIDS})  & 0.758 & 0.153 &  0.153 & 0.153 &  0.506 &  0.012  &  0.976 

& MLP (K-best~\cite{SENSOR})&  \textbf{0.555} &  \textbf{0.333} & \textbf{0.333} & \textbf{0.333} & \textbf{ 0.500} &  \textbf{0.000}  &  0.993 \\

\hline

LightGBM (our framework)  &  0.872 &  0.552 &  0.552 & 0.552 &  0.738 &  0.477 &   0.986

& LightGBM (our framework) & \textbf{0.999} &  \textbf{0.999} &  \textbf{0.999} &  \textbf{0.999} &  \textbf{0.999}   &  \textbf{0.998}  & \textbf{0.999}   \\

LightGBM (IG~\cite{CICIDS})  & \textbf{0.910} &  \textbf{0.685} &  \textbf{0.685} & \textbf{0.685} &  \textbf{0.817} &  \textbf{0.633} & \textbf{0.999}

&  LightGBM (K-best~\cite{SENSOR})&  \textbf{0.999} &  \textbf{0.999} &  \textbf{0.999} &  \textbf{0.999} &  \textbf{0.999} &   \textbf{0.998}  &  \textbf{0.999} \\
\end{tabular}
}
\label{tbl:comparison_with_baselines_top features effect on AI}
%\vspace{-2mm}
\end{table*}

\begin{table*}[ht]
\centering

\caption{Results of feature selection (top incursion features using various approaches) for the datasets RoEduNet-SIMARGL2021 and CICIDS-2017. For both datasets, we extract the top-5 (k = 5) and top-10 (k = 10) features, respectively.}
% \vspace{-3mm}
\resizebox{1.0\linewidth}{!}{
\begin{tabular}{|ccccc|}
\hline
\textbf{CICIDS-2017}  & \textbf{Common Features by Overall Rank}           & \textbf{Chi-square}                               & \textbf{Feature Correlation}                                 & \textbf{Feature Importance}             \\ \hline
\textbf{1}              & {Packet Length Std}           & {Destination Port}          & Init\_Win\_bytes\_forward                                 & Avg Bwd Segment Size                \\
\textbf{2}              & Destination Port         & Bwd Packet Length Max                          & Init\_Win\_bytes\_backward                               & Bwd Packet Length Mean              \\
\textbf{3}              & {Init\_Win\_bytes\_forward}    & Bwd Packet Length Mean                        & Packet Length Variance                                   & Destination Port\\
\textbf{4}              & {Packet Length Mean}          & Max Packet Length                            & Bwd Packet Length Max                                     & Max Packet Length                   \\
\textbf{5}              & {Bwd Packet Length Mean}      & Packet Length Mean                            & Packet Length Std                                        & Packet Length Mean                  \\
\hline
\textbf{6}              & {Average Packet Size}         & Packet Length Std                             & Destination Port                    & Bwd Packet Length Max                \\
\textbf{7}              & {Init\_Win\_bytes\_backward}  & Packet Length Variance                        & Bwd Packet Length Mean                                  & Init\_Win\_bytes\_forward            \\
\textbf{8}              & {Avg Bwd Segment Size}        & Average Packet Size                           & Avg Bwd Segment Size                                     & Average Packet Size                 \\
\textbf{9}              & {Bwd Packet Length Max}        & Avg Bwd Segment Size                          & Max Packet Length                                        & Packet Length Std                   \\
\textbf{10}             & {Packet Length Variance}     & Init\_Win\_bytes\_forward                      & Subflow Fwd Bytes                                            & Init\_Win\_bytes\_backward          \\ \hline
\textbf{CICIDS-2017}  & \textbf{Models + Attacks Ranking Score}            & \textbf{Common Features by Overall Weighted Rank} & \textbf{Common Features by Overall Normalized Weighted Rank} & \textbf{Combined Selection}                                       \\ \hline
\textbf{1}              & Destination Port           & Average Packet Size                           & Average Packet Size                                      & Packet Length Std                                      \\
\textbf{2}              & Packet Length Std                              & Init\_Win\_bytes\_backward                    & Destination Port                  & Init\_Win\_bytes\_forward                                     \\
\textbf{3}             & Init\_Win\_bytes\_forward                      & Init\_Win\_bytes\_forward                      & Packet Length Mean                                       & Bwd Packet Length Mean                                     \\
\textbf{4}              & Avg Bwd Segment Size                          & Total Length of Fwd Packets                   & Init\_Win\_bytes\_forward                                & Bwd Packet Length Max                                       \\
\textbf{5}              & Bwd Packet Length Mean                        & Bwd Packet Length Max                          & Init\_Win\_bytes\_backward                               & Destination Port                                      \\ \hline
\textbf{6}              & Packet Length Mean                             & Packet Length Std                             & Packet Length Std                                        & Packet Length Mean                                      \\
\textbf{7}              & Average Packet Size                            & Packet Length Mean                            & Avg Bwd Segment Size                                     & Average Packet Size                                      \\
\textbf{8}              & Init\_Win\_bytes\_backward                     & Max Packet Length                            & Bwd Packet Length Max                                     & Init\_Win\_bytes\_backward                                       \\
\textbf{9}             & Bwd Packet Length Max                           & Total Length of Bwd Packets                   & Bwd Packet Length Mean                                   & Avg Bwd Segment Size                                      \\
\textbf{10}             & Max Packet Length                              & Bwd Packet Length Mean                        & Max Packet Length                                        &  Init\_Win\_bytes\_forward                                       \\ \hline \hline
\textbf{SIMARGL2021} & \textbf{Common Features by Overall Rank}           & \textbf{Chi-square}                               & \textbf{Feature Correlation}                                 & \textbf{Feature Importance}             \\ \hline
\textbf{1}              & {TCP\_WIN\_SCALE\_IN}         & FLOW\_DURATION\_MILLISECONDS                  & TCP\_WIN\_MSS\_IN                                        & TCP\_WIN\_MSS\_IN                    \\
\textbf{2}              & {TCP\_WIN\_MIN\_IN}           & PROTOCOL                                      & TCP\_WIN\_MIN\_IN                                        & TCP\_WIN\_MAX\_IN                    \\
\textbf{3}              & {TCP\_WIN\_MAX\_IN}           & TCP\_WIN\_MAX\_IN                             & TCP\_WIN\_MAX\_IN                                        & TCP\_WIN\_MIN\_IN                    \\
\textbf{4}              & {TCP\_WIN\_MSS\_IN}           & TCP\_WIN\_MAX\_OUT                            & TCP\_WIN\_SCALE\_IN                                      & TCP\_WIN\_SCALE\_IN                  \\
\textbf{5}              & {TCP\_FLAGS}                  & TCP\_WIN\_MIN\_IN                             & PROTOCOL                                                 & TCP\_FLAGS                           \\\hline
\textbf{6}              & {FLOW\_DURATION\_MILLISECONDS}& TCP\_WIN\_MIN\_OUT                            & TCP\_WIN\_MAX\_OUT                                       & FLOW\_DURATION\_MILLISECONDS         \\
\textbf{7}              & {TCP\_WIN\_MAX\_OUT}          & TCP\_WIN\_SCALE\_IN                           & TCP\_WIN\_MIN\_OUT                                       & TCP\_WIN\_MAX\_OUT                   \\
\textbf{8}              & {TCP\_WIN\_MIN\_OUT}          & TCP\_WIN\_SCALE\_OUT                          & TCP\_WIN\_SCALE\_OUT                                     & TCP\_WIN\_MIN\_OUT                   \\
\textbf{9}              & {SRC\_TOS}                    & SRC\_TOS                                      & FIRST\_SWITCHED                                          & TOTAL\_FLOWS\_EXP                    \\
\textbf{10}             & {DST\_TOS}                    & DST\_TOS                                      & LAST\_SWITCHED                                           & LAST\_SWITCHED                       \\ \hline
\textbf{SIMARGL2021} & \textbf{Models + Attacks Ranking Score}            & \textbf{Common Features by Overall Weighted Rank} & \textbf{Common Features by Overall Normalized Weighted Rank} & \textbf{Combined Selection}                                       \\ \hline
\textbf{1}             & TCP\_WIN\_MSS\_IN                              & FLOW\_DURATION\_MILLISECONDS                  & TCP\_WIN\_SCALE\_IN                                      & TCP\_WIN\_SCALE\_IN                                     \\
\textbf{2}             & TCP\_WIN\_MAX\_IN                              & FIRST\_SWITCHED                               & FLOW\_DURATION\_MILLISECONDS                             & TCP\_WIN\_MIN\_IN                                     \\
\textbf{3}             & TCP\_WIN\_SCALE\_IN                            & TOTAL\_FLOWS\_EXP                             & TCP\_WIN\_MSS\_IN                                        & TCP\_WIN\_MAX\_IN                                     \\
\textbf{4}             & TCP\_WIN\_MIN\_IN                              & LAST\_SWITCHED                                & TCP\_WIN\_MAX\_IN                                        & TCP\_WIN\_MSS\_IN                                       \\
\textbf{5}             & FLOW\_DURATION\_MILLISECONDS                   & TCP\_WIN\_SCALE\_IN                           & TCP\_WIN\_MIN\_IN                                        & FLOW\_DURATION\_MILLISECONDS                                      \\\hline
\textbf{6}             & TOTAL\_FLOWS\_EXP                              & TCP\_WIN\_MSS\_IN                             & TOTAL\_FLOWS\_EXP                                        & TCP\_WIN\_MAX\_OUT                                      \\
\textbf{7}             & TCP\_FLAGS                                     & TCP\_WIN\_MAX\_IN                             & FIRST\_SWITCHED                                          & TCP\_FLAGS                                     \\
\textbf{8}             & PROTOCOL                                       & TCP\_WIN\_MIN\_IN                             & TCP\_FLAGS                                               & PROTOCOL                                      \\
\textbf{9}             & TCP\_WIN\_MAX\_OUT                             & PROTOCOL                                      & TCP\_WIN\_MAX\_OUT                                        & TCP\_WIN\_MIN\_OUT                                     \\
\textbf{10}             & SRC\_TOS                                        & TCP\_FLAGS                                     & PROTOCOL                                                  & TOTAL\_FLOWS\_EXP      \\ \hline                             
\end{tabular}
}
\label{tbl:top_features_baselines}
\vspace{-3mm}
\end{table*}

\begin{table*}[hbt!]
\caption{The optimal feature selection strategies for various configurations of RoEduNet-SIMARGL2021 and CICIDS-2017. When compared to baseline approaches, our suggested methods perform better at choosing the optimal features for various AI models.}
% \vspace{-3mm}
\resizebox{\linewidth}{!}{
\begin{tabular}{|l|l|c c c c c c c|}
\hline

\multicolumn{1}{|l|}{\textbf{AI Model}}  & \textbf{Best Feature Selection Method}                                           & \textbf{Accuracy} & \textbf{Precision} & \textbf{Recall} & \textbf{F1}    & \textbf{BACC}  & \textbf{MCC}   & \textbf{AUC\_ROC} \\ \hline
\multicolumn{9}{|c|}{CICIDS-2017  ($k$ = 5)} \\ \hline
\textbf{RF}   & \textbf{Common Features by Overall Weighted Rank (Ours)}            & \textbf{0.999}    & \textbf{0.999}     & \textbf{0.999}  & \textbf{0.999} & \textbf{0.999} & \textbf{0.998} & \textbf{0.828}    \\ % 
                               
\textbf{ADA}  &  \textbf{Feature Correlation}                                 & \textbf{0.943}    & \textbf{0.802}     & \textbf{0.802}  & \textbf{0.802} & \textbf{0.885} & \textbf{0.769} & \textbf{0.869}    \\  
                                                    
\textbf{DNN}  & \textbf{Common Features by Overall Rank (Ours)}                     & \textbf{0.949}    & \textbf{0.824}     & \textbf{0.824}  & \textbf{0.824} & \textbf{0.897} & \textbf{0.795} & \textbf{0.482}    \\  
                              
\textbf{LightGBM} & \textbf{Common Features by Overall Rank (Ours)}                     & \textbf{0.987}    & \textbf{0.954}     & \textbf{0.954}  & \textbf{0.954} & \textbf{0.973} & \textbf{0.946} & \textbf{0.999}    \\  
                           
\textbf{MLP}  &  \textbf{Common Features by Overall Normalized Weighted Rank (Ours)} & \textbf{0.980}    & \textbf{0.931}     & \textbf{0.931}  & \textbf{0.931} & \textbf{0.959} & \textbf{0.919} & \textbf{0.996}    \\ 
                        
\textbf{KNN} &  \textbf{Common Features by Overall Weighted Rank (Ours)}            & \textbf{0.999}    & \textbf{0.996}     & \textbf{0.996}  & \textbf{0.996} & \textbf{0.997} & \textbf{0.995} & \textbf{0.656}    \\
                               
\textbf{SVM}  &  \textbf{Common Features by Overall Normalized Weighted Rank (Ours)} & \textbf{0.993}    & \textbf{0.977}     & \textbf{0.977}  & \textbf{0.977} & \textbf{0.987} & \textbf{0.973} & \textbf{0.759}    \\ 
\hline
\multicolumn{9}{|c|}{CICIDS-2017  ($k$ = 10)} \\ \hline

\textbf{RF}   &  \textbf{Common Features by Overall Weighted Rank (Ours)}            & \textbf{0.999} & \textbf{0.999}                      & \textbf{0.999} & \textbf{0.999} & \textbf{0.999}                 & \textbf{0.998} & \textbf{0.546}  \\

\textbf{ADA}      & \textbf{Chi-square}                                                   & \textbf{0.941}          & \textbf{0.793}                               & \textbf{0.793}          & \textbf{0.793}          & \textbf{0.879}                          & \textbf{0.758}          & \textbf{0.940}             \\ 

\textbf{DNN}   & \textbf{Model Specific Features (Ours)}                                     & \textbf{0.944}          & \textbf{0.803}                               & \textbf{0.803}          & \textbf{0.803}          & \textbf{0.885}                          & \textbf{0.770}          & \textbf{0.589}             \\ 
                              
\textbf{LightGBM}  & \textbf{Chi-square}                                                   & \textbf{0.983}          & \textbf{0.940}                              & \textbf{0.940}          & \textbf{0.940}          & \textbf{0.965}                          & \textbf{0.930}          & \textbf{0.997}             \\ 
                                
\textbf{MLP}  & \textbf{Models + Attacks Ranking Score (Ours)}                               & \textbf{0.997}          & \textbf{0.988}                               & \textbf{0.988}          & \textbf{0.988}          & \textbf{0.993}                          & \textbf{0.986}          & \textbf{0.997}             \\ 
                              
							  \textbf{KNN}  & \textbf{Common Features by Overall Rank (Ours)}                              & \textbf{0.976}         & \textbf{0.917}                               & \textbf{0.917}          & \textbf{0.917}          & \textbf{0.952}                          & \textbf{0.903}          & \textbf{0.633}             \\ 
                              
                                & \textbf{Chi-square}                                                   & \textbf{0.976}          & \textbf{0.917}                               & \textbf{0.917}          & \textbf{0.917}          & \textbf{0.952}                          & \textbf{0.903}          & \textbf{0.633}             \\ 
                                & \textbf{Feature Correlation}                                          & \textbf{0.976}          & \textbf{0.917}                               & \textbf{0.917}          & \textbf{0.917}          & \textbf{0.952}                          & \textbf{0.903}          & \textbf{0.633}             \\ 
                                & \textbf{Feature Importance}                                           & \textbf{0.976}          & \textbf{0.917}                               & \textbf{0.917}          & \textbf{0.917}          & \textbf{0.952}                          & \textbf{0.903}          & \textbf{0.633}             \\ 
                                & \textbf{Models + Attacks Ranking Score (Ours)}                               & \textbf{0.976}          & \textbf{0.917}                               & \textbf{0.917}          & \textbf{0.917}          & \textbf{0.952}                          & \textbf{0.903}          & \textbf{0.633}             \\ 
                                & \textbf{Common Features by Overall Normalized Weighted Rank (Ours)}          & \textbf{0.976}          & \textbf{0.917}                               & \textbf{0.917}          & \textbf{0.917}          & \textbf{0.952}                          & \textbf{0.903}          & \textbf{0.633}             \\

\textbf{SVM}   & \textbf{Common Features by Overall Rank (Ours)}                              & \textbf{0.976}          & \textbf{0.915}                               & \textbf{0.915}          & \textbf{0.915}          & \textbf{0.950}                          & \textbf{0.901}         & \textbf{0.598}             \\ 
                                & \textbf{Feature Correlation}                                          & \textbf{0.976}          & \textbf{0.915}                               & \textbf{0.915}          & \textbf{0.915}          & \textbf{0.951}                          & \textbf{0.901}          & \textbf{0.609}             \\ 
                                & \textbf{Feature Importance}                                           & \textbf{0.976}          & \textbf{0.915}                               & \textbf{0.915}          & \textbf{0.915}          & \textbf{0.951}                          & \textbf{0.901}          & \textbf{0.592}             \\ 
                                & \textbf{Models + Attacks Ranking Score (Ours)}                               & \textbf{0.976}          & \textbf{0.915}                               & \textbf{0.915}          & \textbf{0.915}          & \textbf{0.951}                          & \textbf{0.901}          & \textbf{0.592}             \\ 
                                & \textbf{Common Features by Overall Normalized Weighted Rank (Ours)} & \textbf{0.976} & \textbf{0.915}                      & \textbf{0.915} & \textbf{0.915} & \textbf{0.951}                 & \textbf{0.901} & \textbf{0.592}    \\
                                \hline
                                \multicolumn{9}{|c|}{RoEduNet-SIMARGL2021  ($k$ = 5)} \\ 
                                \hline

                                \textbf{RF}   & \textbf{Models + Attacks Ranking Score (Ours)}                      & \textbf{0.944}    & \textbf{0.917}     & \textbf{0.917}  & \textbf{0.917} & \textbf{0.938} & \textbf{0.875} & \textbf{0.987}    \\  
                               & \textbf{Common Features by Overall Weighted Rank (Ours)}            & \textbf{0.944}    & \textbf{0.917}     & \textbf{0.917}  & \textbf{0.917} & \textbf{0.938} & \textbf{0.875} & \textbf{0.987}    \\  
                               & \textbf{Common Features by Overall Normalized Weighted Rank (Ours)} & \textbf{0.944}    & \textbf{0.917}     & \textbf{0.917}  & \textbf{0.917} & \textbf{0.938} & \textbf{0.875} & \textbf{0.987}    \\

                               \textbf{ADA}  & \textbf{Combined Selection (Ours)}      & \textbf{0.999}              & \textbf{0.999}             & \textbf{0.999}           & \textbf{0.999}         & \textbf{0.999}          & \textbf{0.999}         & \textbf{0.487}           \\

\textbf{DNN}  & \textbf{Common Features by Overall Rank (Ours)}                     & \textbf{0.750}    & \textbf{0.625}     & \textbf{0.625}  & \textbf{0.625} & \textbf{0.719} & \textbf{0.438} & \textbf{0.729}      \\  

 \textbf{LightGBM} & \textbf{Models + Attacks Ranking Score (Ours)}                      & \textbf{0.967}    & \textbf{0.950}     & \textbf{0.950}  & \textbf{0.950} & \textbf{0.963} & \textbf{0.926} & \textbf{0.997}    \\  
                               & \textbf{Common Features by Overall Weighted Rank (Ours)}            & \textbf{0.967}    & \textbf{0.950}     & \textbf{0.950}  & \textbf{0.950} & \textbf{0.963} & \textbf{0.926} & \textbf{0.997}    \\  
                               & \textbf{Common Features by Overall Normalized Weighted Rank (Ours)} & \textbf{0.967}    & \textbf{0.950}     & \textbf{0.950}  & \textbf{0.950} & \textbf{0.963} & \textbf{0.926} & \textbf{0.997}    \\ 
                               & \textbf{Combined Selection (Ours)}      & \textbf{0.999}             & \textbf{0.999}            & \textbf{0.999}          & \textbf{0.999}      & \textbf{0.999}        & \textbf{0.999}         & \textbf{0.487}              \\

                               \textbf{MLP}           & \textbf{Combined Selection (Ours)}      & \textbf{0.996}             & \textbf{0.995} &  \textbf{0.995}  &  \textbf{0.995} &  \textbf{0.996}      & \textbf{0.993}        & \textbf{0.952}           \\

                                          \textbf{KNN}  &  \textbf{Feature Importance}                                  & \textbf{0.760}    & \textbf{0.640}     & \textbf{0.640}  & \textbf{0.640} & \textbf{0.730} & \textbf{0.461} & \textbf{0.592}    \\  
                               & \textbf{Models + Attacks Ranking Score (Ours)}                      & \textbf{0.760}    & \textbf{0.640}     & \textbf{0.640}  & \textbf{0.640} & \textbf{0.730} & \textbf{0.460} & \textbf{0.593}    \\  
                               & \textbf{Common Features by Overall Weighted Rank (Ours)}            & \textbf{0.760}    & \textbf{0.640}     & \textbf{0.640}  & \textbf{0.640} & \textbf{0.730} & \textbf{0.460} & \textbf{0.593}    \\  
                               & \textbf{Common Features by Overall Normalized Weighted Rank (Ours)} & \textbf{0.760}    & \textbf{0.640}     & \textbf{0.640}  & \textbf{0.640} & \textbf{0.730} & \textbf{0.460} & \textbf{0.593}    \\ 

                               \textbf{SVM}  & \textbf{Feature Importance}                                  & \textbf{0.758}    & \textbf{0.636}     & \textbf{0.636}  & \textbf{0.636} & \textbf{0.727} & \textbf{0.455} & \textbf{0.373}    \\  
                               \hline
 \multicolumn{9}{|c|}{RoEduNet-SIMARGL2021  ($k$ = 10)} \\ 
                                \hline

\textbf{RF}        & \textbf{Feature Correlation}                                 & \textbf{0.999}      & \textbf{0.998}     & \textbf{0.998}     & \textbf{0.998}     & \textbf{0.998} & \textbf{0.997}    & \textbf{0.999} \\ 
                                & \textbf{Common Features by Overall Weighted Rank (Ours)}                 & \textbf{0.999}            & \textbf{0.998}              & \textbf{0.998}              & \textbf{0.998}              & \textbf{0.998}          & \textbf{0.997}             & \textbf{0.999}\\ 
                                & \textbf{Common Features by Overall Normalized Weighted Rank (Ours)}         & \textbf{0.999}      & \textbf{0.998}     & \textbf{0.998}     & \textbf{0.998}     & \textbf{0.998} & \textbf{0.997}    & \textbf{0.999} \\ 

\textbf{ADA}                                    & \textbf{Combined Selection (Ours)}  & \textbf{0.999} & \textbf{0.999} & \textbf{0.999}  & \textbf{0.999}    & \textbf{0.999}  & \textbf{0.999}  & \textbf{0.999}      \\

\textbf{DNN}      & \textbf{Chi-square}                                          & \textbf{0.572}      & \textbf{0.358}     & \textbf{0.358}     & \textbf{0.358}     & \textbf{0.519} & \textbf{0.374}    & \textbf{0.541}   \\ 

							 \textbf{LightGBM}  & \textbf{Feature Importance}                                  & \textbf{0.999}      & \textbf{0.999}     & \textbf{0.999}     & \textbf{0.999}     & \textbf{0.999} & \textbf{0.999}    & \textbf{0.999} \\ 
                                & \textbf{Models + Attacks Ranking Score (Ours)}              & \textbf{0.999}      & \textbf{0.999}     & \textbf{0.999}     & \textbf{0.999}     & \textbf{0.999} & \textbf{0.999}    & \textbf{0.999} \\ 
                                
                                & \textbf{Common Features by Overall Weighted Rank (Ours)}       & \textbf{0.999}      & \textbf{0.999}     & \textbf{0.999}     & \textbf{0.999}     & \textbf{0.999} & \textbf{0.999}    & \textbf{0.999} \\                
                                
                                & \textbf{Common Features by Overall Normalized Rank (Ours)} & \textbf{0.999}      & \textbf{0.999}     & \textbf{0.999}     & \textbf{0.999}     & \textbf{0.999} & \textbf{0.999}    & \textbf{0.999} \\   

                                \textbf{MLP}           & \textbf{Common Features by Overall Rank (Ours)}                     & \textbf{0.995}      & \textbf{0.981}     & \textbf{0.981}     & \textbf{0.981}     & \textbf{0.989} & \textbf{0.978}    & \textbf{0.996} \\ 
                          
						  & \textbf{Combined Selection (Ours)}      & \textbf{0.999}    & \textbf{0.999}        & \textbf{0.999}           & \textbf{0.999}    & \textbf{0.999}      & \textbf{0.999}      & \textbf{0.999}          \\

\textbf{KNN}             & \textbf{Combined Selection} (Ours)      & \textbf{0.999}             & \textbf{0.999}   & \textbf{0.999}           & \textbf{0.999}        & \textbf{0.999}        & \textbf{0.999}          & \textbf{0.648}   \\

       \textbf{SVM}            & \textbf{Combined Selection (Ours)} & \textbf{0.956} & \textbf{0.934} & \textbf{0.934} & \textbf{0.934} & \textbf{0.951}  & \textbf{0.902} &  \textbf{0.795}     \\
                                \hline

\end{tabular}
}
\label{tbl:comparison_best_features}
\vspace{-1mm}
\end{table*}

\subsection{Baseline Systems}

Our approach to feature extraction is compared to five baseline methods: information gain~\cite{CICIDS}, feature importance~\cite{dutta2018analysing}, feature correlation~\cite{feat_corr_citation}, K-best~\cite{SENSOR}, and Chi-square~\cite{gajawada2019chi}. The entire list of the top features retrieved by our suggested approaches is shown along with these various baselines in Table~\ref{tbl:top_features_baselines}. The comparison shows that our methodologies and the baselines for both datasets do not agree on the feature significance rankings. It is important to remember, nevertheless, that several feature extraction techniques have discovered similar top incursion characteristics.

\textbf{Effect of Different Top Features:}
We now juxtapose the AI models' performance under our framework's top five significant characteristics with those chosen by information gain~\cite{CICIDS} and K-best~\cite{SENSOR}. Table~\ref{tbl:comparison_with_baselines_top features effect on AI} demonstrates that, for the two datasets, our framework performs better in 22 out of 28 AI models (bold text indicates superior performance). Additionally, we present the optimal feature selection strategies for various configurations of the CICIDS-2017 and RoEduNet-SIMARGL202 datasets. Table~\ref{tbl:comparison_best_features} demonstrates that, when compared to baselines, our suggested techniques perform better at choosing the best features for various AI models.

\textbf{Xplique Experiment:} 
For this experiment, we do feature selection using nine alternative XAI approaches for the DNN model, utilizing the Xplique toolbox. Table~\ref{tbl:Xplique_methods} (\ref{app:feat_selec_details}) displays the experiment's findings. In Table~\ref{tbl:Xplique_methods}, the 'Voting' column aggregates all nine Xplique methods to produce an overall feature ranking result for feature selection, similarly to Table~\ref{tbl:top_features_baselines} `Combined Selection' column does. Subsequently, we observed that `Combined Selection' and `Voting' selected six features in common out of ten for the CICIDS-2017 dataset and eight features in common out of ten for the RoEduNet-SIMARGL2021 dataset. These results were obtained by comparing both approaches from Table~\ref{tbl:top_features_baselines} and Table~\ref{tbl:Xplique_methods}. 
This shows some convergence when taking into account both approaches. However, since all of Xplique's XAI techniques attempt to explain such design, Xplique is limited to only taking neural network-based AI models (DNNs in this case). However, as our framework demonstrates, our approaches are independent of any particular model, and every AI model that is taken into consideration for this study has been tested using our suggested selection techniques.

\begin{table*}[ht]
\centering
\caption{Runtimes for the CICIDS-2017 and RoEduNet-SIMARGL2021 datasets (training and testing).}
\vspace{1mm}
\resizebox{1\linewidth}{!}{
\begin{tabular}{|c|cc|cc|cc|cc|}
\hline
\textbf{RoEduNet-SIMARGL2021}            & 
\multicolumn{2}{c|}{\textbf{k = 5}}  & \multicolumn{2}{c|}{\textbf{k = 10}} & \multicolumn{2}{c|}{\textbf{k = 15}} & \multicolumn{2}{c|}{\textbf{k = all}} \\
\hline
\textbf{Running Time (min)} & \textbf{Training} & \textbf{Testing} & \textbf{Training} & \textbf{Testing} & \textbf{Training} & \textbf{Testing} & \textbf{Training} & \textbf{Testing} \\
\hline
\textbf{RF}                 & 0.002             & 0.148            & 0.504             & 0.019            & 11.572            & 0.697            & 0.793                & 0.019                \\
\textbf{DNN}                & 0.876             & 0.387            & 0.526             & 0.632            & 5.993             & 5.667            &  5.613                & 3.126                \\
\textbf{ADA}                & 0.696             & 1.555            & 9.507             & 0.769            & 10.913            & 1.472            & 4.696                 & 0.050                \\
\textbf{SVM}                & 0.401             & 0.193            & 3.226             & 0.444            & 828.511           & 3.409            & 3.348                & 0.192               \\
\textbf{MLP}                & 0.012             & 23.467           & 5.267             & 0.016            & 7.851             & 1.683            & 0.535                & 4.165               \\
\textbf{LightGBM}               & 0.002             & 16.828           & 21.765            & 0.025            & 9.012             & 0.041            & 2.346                & 0.001               \\
\textbf{KNN}                & 34.057            & 0.409            & 0.001             & 0.008            & 1.243             & 24.842           & 0.009                & 50.849               \\ \hline \hline
\textbf{CICIDS-2017}             & \multicolumn{2}{c|}{\textbf{k = 5}}  & \multicolumn{2}{c|}{\textbf{k = 10}} & \multicolumn{2}{c|}{\textbf{k = 15}} & \multicolumn{2}{c|}{\textbf{k = all}} \\
\hline
\textbf{Running Time (min)} & \textbf{Training} & \textbf{Testing} & \textbf{Training} & \textbf{Testing} & \textbf{Training} & \textbf{Testing} & \textbf{Training} & \textbf{Testing} \\
\hline
\textbf{RF}                 & 0.148             & 0.011            & 0.162             & 0.009            & 0.193             & 0.024            & 0.322             & 0.021            \\
\textbf{DNN}                & 0.387             & 0.230            & 0.492             & 0.413            & 0.588             & 0.111            & 1.042             & 0.299            \\
\textbf{ADA}                & 1.555             & 0.102            & 1.766             & 0.103            & 1.922             & 0.154            & 5.054             & 0.171            \\
\textbf{SVM}                & 0.193             & 0.010            & 0.172             & 0.011            & 0.202             & 0.011            & 0.758             & 0.033            \\
\textbf{MLP}                & 23.467            & 0.020            & 2.919             & 0.001            & 0.812             & 0.113            & 1.444             & 0.111            \\
\textbf{LightGBM}               & 16.828            & 0.020            & 1.236             & 0.002            & 7.588             & 0.022            & 13.352            & 0.111            \\
\textbf{KNN}                & 0.409             & 0.168            & 0.372             & 1.201            & 0.094             & 1.688            & 0.145             & 2.711  \\         \hline
\end{tabular}
}
\label{tbl: run_time}
\end{table*}

\subsection{RunTime Analysis}
Lastly, we display the computational effectiveness of the various AI models that our system makes use of. We also demonstrate how runtime metrics are impacted by the quantity of features. Now, we get into the setup and key results for the runtime analysis.

\textbf{Computing Resources:} 
A renowned public university's high-performance computer (HPC) was used for the tests. Four NVIDIA A100 GPUs, 64 GPU-accelerated nodes with 256 GB of RAM apiece, and a single 64-core AMD EPYC 7713 CPU with a clock speed of 2.0 GHz and 225 watts are all included in the HPC system. Maximum performance with this arrangement is about 7 petaFLOPs. This supercomputer is meant to support academics in carrying out complex AI and machine learning tasks at a high level.

\textbf{(a) AI Model Training Runtime:} 
The training runtimes of the AI models on the CICIDS-2017 dataset are shown in Table~\ref{tbl: run_time}. As can be shown, most of the AI models in our study simply require a few minutes to train these models on millions of data. The chart also shows that using all characteristics for training, as opposed to just the top 15, leads to longer training periods, with an average increase of 2.1X for all AI models. For this dataset, the individual models with the quickest training runtimes are KNN and RF. In contrast, out of all the AI models under consideration, LightGBM has the slowest training duration. Table~\ref{tbl: run_time} shows that SVM has the slowest training runtime while RF and MLP have the fastest training runtimes for the RoEduNet-SIMARGL2021 dataset.

\textbf{(b) AI Classification Runtime:} 
The prediction runtimes of the AI models are displayed in Table~\ref{tbl: run_time}. The traffic test data from the CICIDS-2017 dataset shows that most AI models have quick prediction speeds, finishing the task in less than two minutes. RF has the quickest prediction runtime of all the AI models for all feature combinations. Table~\ref{tbl: run_time} compares the two datasets and shows that the RoEduNet-SIMARGL2021 dataset has a longer classification runtime than the CICIDS-2017 dataset. The RoEduNet-SIMARGL2021 dataset is around 15 times bigger than the CICIDS-2017 dataset, which is the cause of this discrepancy. 
Furthermore, Table~\ref{tbl: run_time} illustrates that, despite its effective training, KNN has the slowest prediction runtime, whereas RF exhibits the quickest prediction runtime. It is significant to remember that these timeframes might change based on the computer power used.

\textbf{(c) SHAP Feature Importance Runtime:} 
We also present the efficiency (duration in hours) of SHAP explanation generation for various AI models and sample counts. The Appendix's Table~\ref{tab:efficiency} illustrates the time required to generate the importance of the SHAP feature. In SHAP, the global feature significance is generated quickly by the RF, DNN, and LightGBM models, regardless of the quantity of data used for the explanation. 
For instance, generating feature significance using SHAP with 10,000 samples takes 18.36 minutes with DNN, 2.76 minutes with RF, and 0.36 minutes with LightGBM. However, ADA and KNN models take many hours to run and are not scalable for a higher number of samples. All things considered, SHAP outperforms the other models in terms of time performance when using RF, DNN, and LightGBM.

\begin{table}[hbt!]
\centering
\vspace{1.5mm}
\caption{The efficiency, measured in hours, for producing SHAP explanations for various AI models and sample counts.}

\resizebox{\columnwidth}{!}
{
\begin{tabular}{l|lllllll}
\hline
\textbf{SHAP Efficiency (Samples x Hours)} & \textbf{RF}     & \textbf{DNN}    & \textbf{LGBM}   & \textbf{MLP}    & \textbf{ADA}    & \textbf{SVM}    & \textbf{KNN}    \\
\hline
100                        & 0.001  & 0.001  & 0.001  & 0.014  & 0.129  & 0.041  & 1.104  \\
500                        & 0.002  & 0.004  & 0.001  & 0.925  & 3.557  & 0.841  & 28.33  \\
2500                       & 0.011  & 0.029  & 0.002  & 21.18  & 99.00  & 20.87  & 706.0  \\
10000                      & 0.046  & 0.306  & 0.006  & 385.0  & 1512   & 343.0  & 1654   \\
\hline
\textbf{SHAP Local Efficiency (Minutes)} & 0.001  & 0.001  & 0.001  & 0.001  & 0.001  & 0.001  & 0.001  \\
\hline
%\textbf{LIME Local Efficiency (Minutes)} & 0.351  & 0.311  & 0.294  & 0.284  & 0.294  & 0.306  & 0.357  \\
%\hline
\end{tabular}
}
\label{tab:efficiency}
\vspace{-4mm}
\end{table}

\begin{table*}[t]
\centering
%\vspace{-3mm}
\caption{a comparative analysis of several facets of our own research and previous pertinent studies on XAI for network intrusion detection (including datasets, AI models, and explanation techniques).} 
\vspace{-3mm}
\resizebox{1\textwidth}{!}{
\begin{tabular}{l|c|c|c|c}
%header
\textbf{Paper} & \textbf{Dataset}  & \textbf{Model} & \textbf{Explainer} & \textbf{Evaluating XAI Feature Selection} % & %\textbf{Contribution} & \textbf{Limitation}  
\\
\hline
\hline
%first row
Our  Work  & \makecell{CICIDS-2017,  \\ RoEduNet-SIMARGL2021} & \makecell{LGBM,   DNN, MLP,\\ RF, SVM, ADA, KNN}   & SHAP, Xplique & Yes\\
\hline
%second row
DeepAID~\cite{deepaid} & \makecell{Kitsune-Mirai, HDFS, \\ CICIDS-2017, LANL-CMSCSE }  & \makecell{DNN,   LSTM, GNN \\ Autoencoders, Kitsune } & GLGV,   DeepLog,   DeepAID & No \\
\hline
%third row
Kitsune~\cite{kitsune} & \makecell{Kitsune   (OS Scan, \\Fuzzing, and Mirai)}  & \makecell{Kitnet, GMM, SVM,\\ DNN, Autoencoders}  & Kitsune & No \\
\hline
%fourth row
OWAD~\cite{owad}  & \makecell{Kyoto 2006+, BGL, \\ LANL-CMSCSE}  & \makecell{DNN, Autoencoder, DeepLog, \\ APT (GLGV), GAN, LSTM} & OWAD, CADE, TRANS & No \\
\hline
Feature-oriented   Design~\cite{Feature_oriented_Design} & NSL-KDD  & CNN,   DNN  & \makecell{LIME, and Saliency View} & No  \\
\hline
Explainable ML Framework~\cite{Explainable_Machine_Learning_Framework} & NSL-KDD &\makecell{ DNN,   RF, SVM, KNN,\\ ResNet50}  & SHAP & No  \\
\hline
Fooling LIME and SHAP~\cite{adversary} &  \makecell{COMPAS,  German Credit, \\ Communities and Crime}  & RF &  SHAP, LIME& No \\
\hline
\makecell{Evaluating XAI
in Security~\cite{evaluating6metrics}} & \makecell{Malware Genome Project,\\CWE-119} & CNN, MLP, RNN & \makecell{SHAP, LIME , LEMNA,\\ IG, LRP, Gradients} & No \\
\hline
%seventh row
Survey of Current Methods~\cite{survey}  & \makecell{NSL-KDD \\ CICIDS-2017\\ UGR} &\makecell{ SVM,   CNN, RF, MLP,\\ LSTM, GAN }& \makecell{IHMC, LIME, SHAP, \\PDP, ALE, H-statistic, \\Anchors,LOCO, IG, \\Gradient Input, Grad-CAM, \\DeepLIFT, DTD, LRP} & Yes \\
\hline
\end{tabular}
}
\vspace{-4mm}
\label{tbl:related_works_XAIEval}
\end{table*}

\vspace{-1mm}
\section{Related Work}\label{sec:related_work}

\textbf{Existing Efforts for Leveraging XAI for IDS:} 
The use of XAI in IDS has been the subject of many recent research. A thorough summary of these efforts to provide a framework for the creation of Explainable AI-based intrusion detection systems can be found in the survey~\cite{survey}. To standardize notions, this survey presents a taxonomy, highlighting the differences between opaque and transparent models, sometimes known as black-box and white-box models, and examining the trade-offs between explainability and performance. Utilizing Explainable Artificial Intelligence (XAI) for intrusion detection systems (IDS) has been the focus of recent research~\cite{mahbooba2021explainable,patil2022explainable,islam2019domain,roponena2022towards}.

For example, \cite{mahbooba2021explainable} used the KDD dataset and decision tree methods to extract rules in IDS. Using the CICIDS-2017 dataset, [42] integrated the CIA principle to improve the explainability of AI models, whereas the work~\cite{patil2022explainable} integrated LIME to offer local explanations for an SVM model in IDS. Most of these attempts are compiled in~\cite{survey}, which also proposes an IDS framework and maps the state-of-the-art works in the field as of right now. 

This framework includes a ``pre-processing phase" that turns a raw dataset into a high-quality dataset; a ``modeling phase" that uses the high-quality dataset to produce explanations for results; and a ``post-modeling explainability phase" that includes an audience-specific user interface. Some notable works are~\cite{deepaid}, which contributes the DeepAID framework to overcome constraints in unsupervised deep learning-based anomaly detection systems, and~\cite{owad}, which introduces the OWAD framework, addressing typical concept drift in deep learning-based models. ~\cite{kitsune} introduces Kitsune, an online anomaly detection system for network intrusion detection, focusing on assaults in camera network systems.

Our current work, on the other hand, addresses several shortcomings in previous works by investigating the impact of each feature on the AI's decision, supplying attack-specific and model-specific feature importance, and taking into account various XAI methods for producing feature importance for various network intrusion datasets. Our methodology provides a more thorough approach to XAI for feature analysis in IDS by taking these factors into account. Moreover, our work considers a diverse XAI-based set of feature selection methods for IDS, testing their performance on different classes of AI models. Table~\ref{tbl:related_works_XAIEval} lists the salient features that distinguish our work from these earlier attempts.

\textbf{Feature Selection for Intrusion Detection:} 
Effective feature selection techniques in the context of intrusion detection have been investigated in several studies~\cite{alazab2012using,MEBAWONDU2020e00497,CICIDS,SENSOR}. These techniques include the use of the K-best method~\cite{SENSOR,li2021lnnls}, correlation-based approaches~\cite{kamalov2020feature,feat_corr_citation}, and information gain~\cite{alazab2012using,CICIDS,MEBAWONDU2020e00497}. However none of these approaches considers the unique properties of the AI models themselves; instead, they are all primarily concerned with the attributes found in the dataset. Our XAI-based approach, on the other hand, takes into account how characteristics affect AI model performance. Our research has shown that every AI model has a unique collection of key characteristics that frequently set it apart from other models. This discovery enhances our comprehension of the AI models' decision-making process within the framework of IDS.

\textbf{Differentiation from Our Previous Related Works:} 
We stress that this current work tackles a different area (i.e., XAI-based Feature Selection) of the research in the context of XAI-IDS for network security compared to our previous works~\cite{osvaldo_1st} and~\cite{osvaldo_3rd}. In particular, the recent work~\cite{osvaldo_1st} proposes an XAI-IDS framework using SHAP and LIME, focusing on plot generation, global and local explanations, and proof of concept for a user-friendly interface. Although it displays a couple of feature selection experiments, it does not propose any new XAI-based feature selection methods and does not provide an in-depth analysis focused on the feature selection of intrusion features subject compared to this current work. Meanwhile, the paper~\cite{osvaldo_3rd} evaluates the quality of the XAI method itself when applied to IDS by applying six metrics described as Descriptive Accuracy, Sparsity, Efficiency, Stability, Completeness, and Robustness. Then, it discusses points of improvement and strong suits of the analyzed XAI methods based on these metrics. However, it does not consider any XAI-based feature selection and does not provide comprehensive evaluations for the different feature selection methods considered in this current work.}

\vspace{-3mm}

\section{Limitations and Discussion}\label{sec: Discussion}

\textbf{(1) Importance of XAI and Feature Selection for IDS:} 
Network attack frequency is predicted to rise in the present era of fast information expansion~\cite{cyber-attacks-decade}. Consequently, it becomes imperative to have accurate intrusion detection systems that are backed by explainable artificial intelligence (XAI). We can obtain a more profound understanding of the particular characteristics that are important for both AI models and network invasions by extracting complete intrusion features, and to effectively guide security analysts by gaining a thorough grasp of their unique traits and behaviors thanks to this in-depth examination. By examining the impact of XAI-based feature selection on the effectiveness of various AI intrusion detection models and taking into account various XAI-based techniques for extracting the best intrusion features on various levels (intrusion-specific level, model-specific level, and overall level), our work offered a crucial step in this research direction.

\textbf{(2) Exploring Our Framework on Other Benchmark Datasets:}
We want to present a thorough examination of our system with the RoEduNet-SIMARGL2021 and CICIDS-2017 datasets in this article. Prospective directions for future study, however, include testing our XAI-based feature selection framework's efficacy on other benchmark datasets for network intrusion detection. NSL-KDD~\cite{dhanabal2015study}, UNSW-NB15~\cite{moustafa2015unsw}, CICIDS-2018~\cite{CSE-CIC-IDS2018}, CTU-13~\cite{sebastian_garcia_martin_grill_jan_stiborek_alejandro_zunino_2022}, and UMass~\cite{UMASS_IDS} are a few possible datasets that may be of interest.
This would increase the possibility of our approach being used in improving feature selection for real-world network intrusion detection applications.

\textbf{(3) Impact of Bias in Top Intrusion Features:} 
Some characteristics could have been biased during the dataset production procedure, which would have an impact on the incursion features' overall relevance. In particular, it has been noted that the feature ``Destination Port" may be biased~\cite{d2022establishing} and that adding it to AI network intrusion detection models might have contaminating consequences. In the RoEduNet-SIMARGL2021 dataset, it can be deduced that ``{TCP WIN SCALE IN'' may have a comparable contaminating impact. We emphasize that our XAI-based feature significance decision is not dependent on a single feature, which reduces the possibility of bias influencing the model as a whole. 
As a result, even when certain features exhibit bias, the bulk of features are made to offset it. A different strategy would be to include a warning in addition to the XAI explanation if one of the top features is generated from these two possibly biased characteristics.  By drawing attention to possible biases and their effects on the model's predictions or classifications, this method promotes more informed decision-making.

\textbf{(4) Combining Different XAI Frameworks for Feature Selection:} 
To more effectively pick the top intrusion features for computer network security, our study demonstrates the possibility of integrating the XAI techniques (SHAP and Xplique) that were taken into consideration in this work with additional XAI methods (such LEMNA~\cite{LEMNA} and LRP~\cite{survey}). We leave full investigation of this idea for future works.

\section{Conclusion}\label{sec: conclusion}

This research proposed an XAI-based framework to improve the feature selection of AI models for network intrusion detection tasks. It provided a variety of novel global feature selection techniques based on SHAP,  identifying key characteristics unique to various AI models and forms of network penetration to comprehend the primary aspects that intrusion detection AI models rely on. To rank the top incursion features, we utilized seven black-box AI models, benchmarking and evaluating our framework on two invaluable network intrusion datasets: the CICIDS-2017 benchmark dataset and the recently released RoEduNet-SIMARGL2021 dataset. As a holistic analysis, we considered the top features for each type of intrusion, the overlapping features across models, the overall feature importance, and the contribution of each feature to the model's decision. Additionally, we performed a benchmark comparison using five state-of-the-art feature selection methods for network intrusion detection. The outcomes show that compared to these baselines, the features chosen by our feature selection techniques provide a more accurate depiction of the AI models. 
Moreover, we contrasted our feature selection techniques with those produced by the various XAI techniques found in the Xplique toolkit
to further our comprehension of the significant components of deep neural network models. We have released our source codes so the research community can use them and expand upon them with other techniques and datasets.

% \vspace{1mm}

\section*{Declarations}

\noindent \textbf{Conflict of interest} The authors declare full compliance with ethical
standards. This paper does not contain any studies involving humans or
animals performed by any of the authors. The authors have no conflict of
interest that might be perceived
to influence the results and/or discussion reported in this paper.

\bibliographystyle{splncs04}
\bibliography{ref}

\begin{thebibliography}{10}
\providecommand{\url}[1]{\texttt{#1}}
\providecommand{\urlprefix}{URL }
\providecommand{\doi}[1]{https://doi.org/#1}

\bibitem{flow1234}
Flow information elements - nprobe 10.1 documentation, \url{https://www.ntop.org/guides/nprobe/flow_information_elements.html}

\bibitem{vargrad}
Adebayo, J., Gilmer, J., Muelly, M., Goodfellow, I., Hardt, M., Kim, B.: Sanity checks for saliency maps (2020)

\bibitem{ahlashkari_2021}
Ahlashkari: Cicflowmeter/readme.txt at master · ahlashkari/cicflowmeter, https://shorturl.at/stji5 (Jun 2021), \url{https://shorturl.at/BM9r8}

\bibitem{al2021intelligent}
Al-Omari, M., Rawashdeh, M., Qutaishat, F., Alshira’H, M., Ababneh, N.: An intelligent tree-based intrusion detection model for cyber security. Journal of Network and Systems Management  \textbf{29}(2),  1--18 (2021)

\bibitem{alazab2012using}
Alazab, A., Hobbs, M., Abawajy, J., Alazab, M.: Using feature selection for intrusion detection system. In: 2012 international symposium on communications and information technologies (ISCIT). pp. 296--301. IEEE (2012)

\bibitem{amor2004naive}
Amor, N.B., Benferhat, S., Elouedi, Z.: Naive bayes vs decision trees in intrusion detection systems. In: Proceedings of the 2004 ACM symposium on Applied computing. pp. 420--424 (2004)

\bibitem{arisdakessian2022survey}
Arisdakessian, S., Wahab, O.A., Mourad, A., Otrok, H., Guizani, M.: A survey on iot intrusion detection: Federated learning, game theory, social psychology and explainable ai as future directions. IEEE Internet of Things Journal  (2022)

\bibitem{osvaldo_1st}
Arreche, O., Guntur, T., Abdallah, M.: Xai-ids: Toward proposing an explainable artificial intelligence framework for enhancing network intrusion detection systems. Applied Sciences  \textbf{14}(10) (2024). \doi{10.3390/app14104170}, \url{https://www.mdpi.com/2076-3417/14/10/4170}

\bibitem{osvaldo_3rd}
Arreche, O., Guntur, T.R., Roberts, J.W., Abdallah, M.: E-xai: Evaluating black-box explainable ai frameworks for network intrusion detection. IEEE Access  \textbf{12},  23954--23988 (2024). \doi{10.1109/ACCESS.2024.3365140}

\bibitem{asad2022dynamical}
Asad, H., Gashi, I.: Dynamical analysis of diversity in rule-based open source network intrusion detection systems. Empirical Software Engineering  \textbf{27}(1),  1--30 (2022)

\bibitem{balyan2022hybrid}
Balyan, A.K., Ahuja, S., Lilhore, U.K., Sharma, S.K., Manoharan, P., Algarni, A.D., Elmannai, H., Raahemifar, K.: A hybrid intrusion detection model using ega-pso and improved random forest method. Sensors  \textbf{22}(16), ~5986 (2022)

\bibitem{barradas2021flowlens}
Barradas, D., Santos, N., Signorello, S., Ramos, F.M., Madeira, A.: Flowlens: Enabling efficient flow classification for ml-based network security applications.

\bibitem{botacin2021challenges}
Botacin, M., Ceschin, F., Sun, R., Oliveira, D., Gr{\'e}gio, A.: Challenges and pitfalls in malware research. Computers \& Security  \textbf{106},  102287 (2021)

\bibitem{brownlee2019choose}
Brownlee, J.: How to choose a feature selection method for machine learning. Machine Learning Mastery  \textbf{10} (2019)

\bibitem{buczak2015survey}
Buczak, A.L., Guven, E.: A survey of data mining and machine learning methods for cyber security intrusion detection. IEEE Communications surveys \& tutorials  \textbf{18}(2),  1153--1176 (2015)

\bibitem{claise2004cisco}
Claise, B.: Cisco systems netflow services export version 9. Tech. rep. (2004)

\bibitem{Web_attack_Ref2}
by~Comprmoise, D.: {Drive-by Compromise}. \url{https://attack.mitre.org/techniques/T1189/} (2023), [Online; accessed 21-October-2023]

\bibitem{CSE-CIC-IDS2018}
CSE-CIC-IDS2018: {CSE-CIC-IDS2018 on AWS}. \url{https://www.unb.ca/cic/datasets/ids-2018.html/} (2018), [Online; accessed 10-November-2023]

\bibitem{das2020opportunities}
Das, A., Rad, P.: Opportunities and challenges in explainable artificial intelligence (xai): A survey. arXiv preprint arXiv:2006.11371  (2020)

\bibitem{deng2003svm}
Deng, H., Zeng, Q.A., Agrawal, D.P.: Svm-based intrusion detection system for wireless ad hoc networks. In: 2003 IEEE 58th Vehicular Technology Conference. VTC 2003-Fall (IEEE Cat. No. 03CH37484). vol.~3, pp. 2147--2151. IEEE (2003)

\bibitem{dhanabal2015study}
Dhanabal, L., Shantharajah, S.: A study on nsl-kdd dataset for intrusion detection system based on classification algorithms. International journal of advanced research in computer and communication engineering  \textbf{4}(6),  446--452 (2015)

\bibitem{dina2021intrusion}
Dina, A.S., Manivannan, D.: Intrusion detection based on machine learning techniques in computer networks. Internet of Things  \textbf{16},  100462 (2021)

\bibitem{dutta2018analysing}
Dutta, D., Paul, D., Ghosh, P.: Analysing feature importances for diabetes prediction using machine learning. In: 2018 IEEE 9th Annual Information Technology, Electronics and Mobile Communication Conference (IEMCON). pp. 924--928. IEEE (2018)

\bibitem{d2022establishing}
D’hooge, L., Verkerken, M., Volckaert, B., Wauters, T., De~Turck, F.: Establishing the contaminating effect of metadata feature inclusion in machine-learned network intrusion detection models. In: International Conference on Detection of Intrusions and Malware, and Vulnerability Assessment. pp. 23--41. Springer (2022)

\bibitem{xplique}
Fel, T., Hervier, L., Vigouroux, D., Poche, A., Plakoo, J., Cadene, R., Chalvidal, M., Colin, J., Boissin, T., Bethune, L., Picard, A., Nicodeme, C., Gardes, L., Flandin, G., Serre, T.: Xplique: A deep learning explainability toolbox. Workshop on Explainable Artificial Intelligence for Computer Vision (CVPR)  (2022)

\bibitem{ferrag2020rdtids}
Ferrag, M.A., Maglaras, L., Ahmim, A., Derdour, M., Janicke, H.: Rdtids: Rules and decision tree-based intrusion detection system for internet-of-things networks. Future internet  \textbf{12}(3), ~44 (2020)

\bibitem{gajawada2019chi}
Gajawada, S.K.: Chi-square test for feature selection in machine learning. Towards Data Science: Toronto, ON, Canada  (2019)

\bibitem{sebastian_garcia_martin_grill_jan_stiborek_alejandro_zunino_2022}
Garcia, S., Grill, M., Stiborek, J., Zunino, A.: Ctu-13 (2022). \doi{10.34740/KAGGLE/DSV/4062076}, \url{https://www.kaggle.com/dsv/4062076}

\bibitem{LEMNA}
Guo, W., Mu, D., Xu, J., Su, P., Wang, G., Xing, X.: Lemna: Explaining deep learning based security applications. In: Proceedings of the 2018 ACM SIGSAC Conference on Computer and Communications Security. p. 364–379. CCS '18, Association for Computing Machinery, New York, NY, USA (2018). \doi{10.1145/3243734.3243792}

\bibitem{owad}
Han, D., Wang, Z., Chen, W., Wang, K., Yu, R., Wang, S., Zhang, H., Wang, Z., Jin, M., Yang, J., et~al.: Anomaly detection in the open world: Normality shift detection, explanation, and adaptation. In: 30th Annual Network and Distributed System Security Symposium (NDSS) (2023)

\bibitem{deepaid}
Han, D., Wang, Z., Chen, W., Zhong, Y., Wang, S., Zhang, H., Yang, J., Shi, X., Yin, X.: Deepaid: Interpreting and improving deep learning-based anomaly detection in security applications (09 2021)

\bibitem{squaregrad}
Hooker, S., Erhan, D., Kindermans, P.J., Kim, B.: A benchmark for interpretability methods in deep neural networks (2019)

\bibitem{hussain2021noncompliance}
Hussain, S.R., Karim, I., Ishtiaq, A.A., Chowdhury, O., Bertino, E.: Noncompliance as deviant behavior: An automated black-box noncompliance checker for 4g lte cellular devices. In: Proceedings of the 2021 ACM SIGSAC Conference on Computer and Communications Security. pp. 1082--1099 (2021)

\bibitem{ilgun1995state}
Ilgun, K., Kemmerer, R.A., Porras, P.A.: State transition analysis: A rule-based intrusion detection approach. IEEE transactions on software engineering  \textbf{21}(3),  181--199 (1995)

\bibitem{ingre2018decision}
Ingre, B., Yadav, A., Soni, A.K.: Decision tree based intrusion detection system for nsl-kdd dataset. In: Information and Communication Technology for Intelligent Systems (ICTIS 2017)-Volume 2 2. pp. 207--218. Springer (2018)

\bibitem{cyber-attacks-decade}
Insights, I.: {42 Cyber Attack Statistics by Year: A Look at the Last Decade}. \url{https://shorturl.at/7g90R} (February 2020), [Online; accessed 10-March-2023]

\bibitem{islam2019domain}
Islam, S.R., Eberle, W., Ghafoor, S.K., Siraj, A., Rogers, M.: Domain knowledge aided explainable artificial intelligence for intrusion detection and response. arXiv preprint arXiv:1911.09853  (2019)

\bibitem{jackson1991expert}
Jackson, K.A., DuBois, D.H., Stallings, C.A.: An expert system application for network intrusion detection. Tech. rep., Los Alamos National Lab.(LANL), Los Alamos, NM (United States) (1991)

\bibitem{jin2020swiftids}
Jin, D., Lu, Y., Qin, J., Cheng, Z., Mao, Z.: Swiftids: Real-time intrusion detection system based on lightgbm and parallel intrusion detection mechanism. Computers \& Security  \textbf{97},  101984 (2020)

\bibitem{kamalov2020feature}
Kamalov, F., Moussa, S., Zgheib, R., Mashaal, O.: Feature selection for intrusion detection systems. In: 2020 13th international symposium on computational intelligence and design (ISCID). pp. 265--269. IEEE (2020)

\bibitem{khan2021m2mon}
Khan, A., Kim, H., Lee, B.: M2mon: Building an mmio-based security reference monitor for unmanned vehicles. (2021)

\bibitem{kim2017method}
Kim, J., Shin, N., Jo, S.Y., Kim, S.H.: Method of intrusion detection using deep neural network. In: 2017 IEEE international conference on big data and smart computing (BigComp). pp. 313--316. IEEE (2017)

\bibitem{CICIDS}
Kurniabudi, Stiawan, D., Darmawijoyo, Bin~Idris, M.Y., Bamhdi, A.M., Budiarto, R.: Cicids-2017 dataset feature analysis with information gain for anomaly detection. IEEE Access  \textbf{8},  132911--132921 (2020). \doi{10.1109/ACCESS.2020.3009843}

\bibitem{kwon2018understand}
Kwon, B.J.: Understand, Detect, and Block Malware Distribution from a Global Viewpoint. Ph.D. thesis, University of Maryland, College Park (2018)

\bibitem{lee2003detection}
Lee, C.B., Roedel, C., Silenok, E.: Detection and characterization of port scan attacks. Univeristy of California, Department of Computer Science and Engineering  (2003)

\bibitem{li2010novel}
Li, L., Yang, D.Z., Shen, F.C.: A novel rule-based intrusion detection system using data mining. In: 2010 3rd International Conference on Computer Science and Information Technology. vol.~6, pp. 169--172. IEEE (2010)

\bibitem{li2014new}
Li, W., Yi, P., Wu, Y., Pan, L., Li, J.: A new intrusion detection system based on knn classification algorithm in wireless sensor network. Journal of Electrical and Computer Engineering  \textbf{2014} (2014)

\bibitem{li2021lnnls}
Li, X., Yi, P., Wei, W., Jiang, Y., Tian, L.: Lnnls-kh: a feature selection method for network intrusion detection. Security and Communication Networks  \textbf{2021},  1--22 (2021)

\bibitem{lukacs2015strongly}
Lukacs, S., Lutas, D.H., COLESA, A.V., et~al.: Strongly isolated malware scanning using secure virtual containers (Aug~25 2015), uS Patent 9,117,081

\bibitem{lundberg2017unified}
Lundberg, S.M., Lee, S.I.: A unified approach to interpreting model predictions. Advances in neural information processing systems  \textbf{30} (2017)

\bibitem{mahbooba2021explainable}
Mahbooba, B., Timilsina, M., Sahal, R., Serrano, M.: Explainable artificial intelligence (xai) to enhance trust management in intrusion detection systems using decision tree model. Complexity  \textbf{2021} (2021)

\bibitem{MEBAWONDU2020e00497}
Mebawondu, J.O., Alowolodu, O.D., Mebawondu, J.O., Adetunmbi, A.O.: Network intrusion detection system using supervised learning paradigm. Scientific African  \textbf{9},  e00497 (2020). \doi{https://doi.org/10.1016/j.sciaf.2020.e00497}, \url{https://www.sciencedirect.com/science/article/pii/S2468227620302350}

\bibitem{mihailescu2021proposition}
Mihailescu, M.E., Mihai, D., Carabas, M., Komisarek, M., Pawlicki, M., Ho{\l}ubowicz, W., Kozik, R.: The proposition and evaluation of the roedunet-simargl2021 network intrusion detection dataset. Sensors  \textbf{21}(13), ~4319 (2021)

\bibitem{SENSOR}
Mihailescu, M.E., Mihai, D., Carabas, M., Komisarek, M., Pawlicki, M., Hołubowicz, W., Kozik, R.: The proposition and evaluation of the roedunet-simargl2021 network intrusion detection dataset. Sensors  \textbf{21}(13) (2021). \doi{10.3390/s21134319}, \url{https://www.mdpi.com/1424-8220/21/13/4319}

\bibitem{kitsune}
Mirsky, Y., Doitshman, T., Elovici, Y., Shabtai, A.: Kitsune: An ensemble of autoencoders for online network intrusion detection (2018)

\bibitem{mirzaei2021scrutinizer}
Mirzaei, O., Vasilenko, R., Kirda, E., Lu, L., Kharraz, A.: Scrutinizer: Detecting code reuse in malware via decompilation and machine learning. In: Detection of Intrusions and Malware, and Vulnerability Assessment: 18th International Conference, DIMVA 2021, Virtual Event, July 14--16, 2021, Proceedings 18. pp. 130--150. Springer (2021)

\bibitem{moreo2016distributional}
Moreo, A., Esuli, A., Sebastiani, F.: Distributional random oversampling for imbalanced text classification. In: Proceedings of the 39th International ACM SIGIR conference on Research and Development in Information Retrieval. pp. 805--808 (2016)

\bibitem{moustafa2015unsw}
Moustafa, N., Slay, J.: Unsw-nb15: a comprehensive data set for network intrusion detection systems (unsw-nb15 network data set). In: 2015 military communications and information systems conference (MilCIS). pp.~1--6. IEEE (2015)

\bibitem{muhammad2022intelligent}
Muhammad, M., Saleem, A.: Intelligent intrusion detection system for apache web server empowered with machine learning approaches. International Journal of Computational and Innovative Sciences  \textbf{1}(1), ~1--8 (2022)

\bibitem{negandhi2019intrusion}
Negandhi, P., Trivedi, Y., Mangrulkar, R.: Intrusion detection system using random forest on the nsl-kdd dataset. In: Emerging Research in Computing, Information, Communication and Applications, pp. 519--531. Springer (2019)

\bibitem{survey}
Neupane, S., Ables, J., Anderson, W., Mittal, S., Rahimi, S., Banicescu, I., Seale, M.: Explainable intrusion detection systems (x-ids): A survey of current methods, challenges, and opportunities (2022)

\bibitem{northcutt2002network}
Northcutt, S., Novak, J.: Network intrusion detection. Sams Publishing (2002)

\bibitem{panigrahi2018detailed}
Panigrahi, R., Borah, S.: A detailed analysis of cicids2017 dataset for designing intrusion detection systems. International Journal of Engineering \& Technology  \textbf{7}(3.24),  479--482 (2018)

\bibitem{panigrahi2022intrusion}
Panigrahi, R., Borah, S., Pramanik, M., Bhoi, A.K., Barsocchi, P., Nayak, S.R., Alnumay, W.: Intrusion detection in cyber--physical environment using hybrid na{\"\i}ve bayes—decision table and multi-objective evolutionary feature selection. Computer Communications  \textbf{188},  133--144 (2022)

\bibitem{patcha2007overview}
Patcha, A., Park, J.M.: An overview of anomaly detection techniques: Existing solutions and latest technological trends. Computer networks  \textbf{51}(12),  3448--3470 (2007)

\bibitem{patil2022explainable}
Patil, S., Varadarajan, V., Mazhar, S.M., Sahibzada, A., Ahmed, N., Sinha, O., Kumar, S., Shaw, K., Kotecha, K.: Explainable artificial intelligence for intrusion detection system. Electronics  \textbf{11}(19), ~3079 (2022)

\bibitem{UMASS_IDS}
Repository, U.T.: {UMass Trace Repository}. \url{http://traces.cs.umass.edu/index.php/Network/Network} (2021), [Online; accessed on 21-November-2022]

\bibitem{lime}
Ribeiro, M.T., Singh, S., Guestrin, C.: "why should {I} trust you?": Explaining the predictions of any classifier. CoRR  \textbf{abs/1602.04938} (2016), \url{http://arxiv.org/abs/1602.04938}

\bibitem{colonial_pipeline_ransomwareattack}
Robertson, J., Turton, W.: {}. \url{https://shorturl.at/I7iJA} (May 2021), [Online; accessed 30-October-2021]

\bibitem{roponena2022towards}
Roponena, E., Kampars, J., Grabis, J., Gail{\=\i}tis, A.: Towards a human-in-the-loop intelligent intrusion detection system. In: CEUR Workshop Proceedings. pp. 71--81 (2022)

\bibitem{sabev2020integrated}
Sabev, S.I.: Integrated approach to cyber defence: Human in the loop. technical evaluation report. Information \& Security: An International Journal  \textbf{44},  76--92 (2020)

\bibitem{sharafaldin2018towards}
Sharafaldin, I., Gharib, A., Lashkari, A.H., Ghorbani, A.A.: Towards a reliable intrusion detection benchmark dataset. Software Networking  \textbf{2018}(1),  177--200 (2018)

\bibitem{feat_corr_citation}
Sheikh, R.: {Feature Selection Techniques in Machine Learning with Pytho}. \url{https://shorturl.at/Ow4sT} (October 2018), [Online; accessed 30-April-2023]

\bibitem{GradientInput}
Shrikumar, A., Greenside, P., Kundaje, A.: Learning important features through propagating activation differences. CoRR  \textbf{abs/1704.02685} (2017), \url{http://arxiv.org/abs/1704.02685}

\bibitem{saliency}
Simonyan, K., Vedaldi, A., Zisserman, A.: Deep inside convolutional networks: Visualising image classification models and saliency maps (2014)

\bibitem{adversary}
Slack, D., Hilgard, S., Jia, E., Singh, S., Lakkaraju, H.: Fooling lime and shap: Adversarial attacks on post hoc explanation methods. In: Proceedings of the AAAI/ACM Conference on AI, Ethics, and Society. pp. 180--186 (2020)

\bibitem{smoothgrad}
Smilkov, D., Thorat, N., Kim, B., Viégas, F., Wattenberg, M.: Smoothgrad: removing noise by adding noise (2017)

\bibitem{snapp1992dids}
Snapp, S.R., Smaha, S.E., Teal, D.M., Grance, T.: The $\{$DIDS$\}$(distributed intrusion detection system) prototype. In: USENIX Summer 1992 Technical Conference (USENIX Summer 1992 Technical Conference) (1992)

\bibitem{stone2009your}
Stone-Gross, B., Cova, M., Cavallaro, L., Gilbert, B., Szydlowski, M., Kemmerer, R., Kruegel, C., Vigna, G.: Your botnet is my botnet: analysis of a botnet takeover. In: Proceedings of the 16th ACM conference on Computer and communications security. pp. 635--647 (2009)

\bibitem{strom2018mitre}
Strom, B.E., Applebaum, A., Miller, D.P., Nickels, K.C., Pennington, A.G., Thomas, C.B.: Mitre att\&ck: Design and philosophy. In: Technical report. The MITRE Corporation (2018)

\bibitem{IG}
Sundararajan, M., Taly, A., Yan, Q.: Axiomatic attribution for deep networks (2017)

\bibitem{tabassum2019survey}
Tabassum, A., Erbad, A., Guizani, M.: A survey on recent approaches in intrusion detection system in iots. In: 2019 15th International Wireless Communications \& Mobile Computing Conference (IWCMC). pp. 1190--1197. IEEE (2019)

\bibitem{tang2020saae}
Tang, C., Luktarhan, N., Zhao, Y.: Saae-dnn: Deep learning method on intrusion detection. Symmetry  \textbf{12}(10), ~1695 (2020)

\bibitem{tao2018improved}
Tao, P., Sun, Z., Sun, Z.: An improved intrusion detection algorithm based on ga and svm. Ieee Access  \textbf{6},  13624--13631 (2018)

\bibitem{vasiliadis2008gnort}
Vasiliadis, G., Antonatos, S., Polychronakis, M., Markatos, E.P., Ioannidis, S.: Gnort: High performance network intrusion detection using graphics processors. In: Recent Advances in Intrusion Detection: 11th International Symposium, RAID 2008, Cambridge, MA, USA, September 15-17, 2008. Proceedings 11. pp. 116--134. Springer (2008)

\bibitem{Explainable_Machine_Learning_Framework}
Wang, M., Zheng, K., Yang, Y., Wang, X.: An explainable machine learning framework for intrusion detection systems. IEEE Access  \textbf{8},  73127--73141 (2020). \doi{10.1109/ACCESS.2020.2988359}

\bibitem{evaluating6metrics}
Warnecke, A., Arp, D., Wressnegger, C., Rieck, K.: Evaluating explanation methods for deep learning in security. In: 2020 IEEE european symposium on security and privacy (EuroS\&P). pp. 158--174. IEEE (2020)

\bibitem{waskle2020intrusion}
Waskle, S., Parashar, L., Singh, U.: Intrusion detection system using pca with random forest approach. In: 2020 International Conference on Electronics and Sustainable Communication Systems (ICESC). pp. 803--808. IEEE (2020)

\bibitem{attacks2020}
Week, I.: {The 10 biggest cyber security attacks of 2020}. \url{https://searchsecurity.techtarget.com/news/252494362/10-of-the-biggest-cyber-attacks} (Jan 2021), [Online; accessed 1-March-2021]

\bibitem{Feature_oriented_Design}
Wu, C., Qian, A., Dong, X., Zhang, Y.: Feature-oriented design of visual analytics system for interpretable deep learning based intrusion detection. In: 2020 International Symposium on Theoretical Aspects of Software Engineering (TASE). pp. 73--80 (2020). \doi{10.1109/TASE49443.2020.00019}

\bibitem{yulianto2019improving}
Yulianto, A., Sukarno, P., Suwastika, N.A.: Improving adaboost-based intrusion detection system (ids) performance on cic ids 2017 dataset. In: Journal of Physics: Conference Series. vol.~1192, p. 012018. IOP Publishing (2019)

\bibitem{DeconvNet-Occlusion-}
Zeiler, M.D., Fergus, R.: Visualizing and understanding convolutional networks (2013)

\end{thebibliography}
\appendix
%\vspace{-6mm}

\section{Hyperparameters of AI Models}\label{app:ai_models_hyperparams}
% We now present the full details of the hyperparameter configurations for our different AI models.
The complete specifications of the hyperparameter setups for each of our various AI models are now available.

\textbf{Deep Neural Network (DNN):} 
A deep neural network is the first classifier (DNN). The architecture of this classifier comprises an input layer with the count of neurons according to the employed number of features, and the rectified linear unit (ReLU) activation function. The ReLU activation function, a hidden layer with a size of 16 neurons, and a dropout layer with a dropout set at 0.01 come next. ``softmax'' ends the setup. The ``categorical\_crossentropy'' approach was selected as the optimization algorithm, and adaptive momentum (ADAM) was used as the loss function. To train the model with a batch size of 1024, eleven epochs were required. For the other settings, we utilized their default values.

\textbf{Random Forest (RF):} 
The RandomForest (RF) classifier was utilized to find harmful samples in the network data. The following are the hyperparameters we utilized for this classifier: The maximum tree depth was set to 10, the minimum number of samples needed to separate the internal node was set to 2, and the remaining settings were used as supplied by default. The parameter n\_estimators, which indicates the number of trees used, was set to the value of 100.

\textbf{AdaBoost (ADA)}: 
AdaBoost was employed in this investigation as well. The classifier's parameter configuration is as follows: Decision\_Tree\_Classifier was selected as the base estimator, the maximum number of estimators at which boosting will be finished was set to 50, and the weight assigned to each classifier in each boosting iteration was set to 1.

\textbf{K-nearest Neighbour (KNN)}: 
The KNN classifier was also utilized. In this investigation, we employed KNN using the following default hyperparameters: all weights were uniform, the search method was set to 'auto', and the value of the `$n$' neighbors was set to five.

\textbf{Support Vector Machine (SVM)}: 
The SVM classifier was also utilized. This classifier's parameter setup is as follows: The regularization is set to 0.5, the probability is set to `True,' the kernel to `linear,' and the gamma to 0.5.

\textbf{Multi-layer Perceptron (MLP)}:
MLP was the next classifier to be employed. We used the same DNN configuration for this classifier.

\textbf{LightGBM}: 
LightGBM was a classifier utilized in this investigation. We utilized it with the following default hyper-parameters: n\_splits = 10, n\_repeats = 3, error\_score = `raise', n\_jobs = 1, and scoring = `accuracy' respectively.

\section{Feature Selection Xplique Table}\label{app:feat_selec_details}

Table~\ref{tbl:Xplique_methods} presents the comprehensive outcomes of Xplique for every feature selection technique used to the two datasets under investigation.

\begin{table*}[hbt!]
\caption{Top incursion features under various Xplique techniques for the CICIDS-2017 and RoEduNet-SIMARGL2021 datasets are the results of feature selection. For each dataset, we extract the top-5 (k = 5) and top-10 (k = 10) features.}
\resizebox{\linewidth}{!}{
\begin{tabular}{|cccccc|}
\hline
\textbf{CICIDS-2017}  & \textbf{Saliency}           & \textbf{Integrated Gradients}                               & \textbf{Occlusion}                                 & \textbf{SmoothGrad}    & \textbf{DeconvNet}           \\ \hline
\textbf{1}              & Packet Length Mean               & Destination Port          & Destination Port             & Bwd Packets/s    & Destination Port          \\
\textbf{2}               & Bwd Packets/s                   &  Avg Bwd Segment Size     & Init\_Win\_bytes\_forward    & Total Length of Bwd Packets   & Total Length of Fwd Packets         \\       
\textbf{3}               & Subflow Bwd Bytes               & Bwd Packet Length Mean    & Avg Bwd Segment Size         & Subflow Fwd Bytes     &  Total Length of Bwd Packets      \\
\textbf{4}                & Total Length of Bwd Packets    & Bwd Packet Length Max     & Bwd Packet Length Mean       & Subflow Bwd Bytes    & Bwd Packet Length Max       \\
\textbf{5}                & Init\_Win\_bytes\_forward      & FIN Flag Count            & Down/Up Ratio                & Max Packet Length    & Bwd Packet Length Mean          \\
\hline
\textbf{6}               & Bwd Packet Length Max           &  Max Packet Length        & Bwd Packet Length Max        & Flow IAT Std     & Max Packet Length        \\
\textbf{7}               & Flow IAT Std                    & URG Flag Count            & FIN Flag Count               & Init\_Win\_bytes\_forward     & Packet Length Mean        \\
\textbf{8}                & Average Packet Size            & Packet Length Std         & Max Packet Length            & Avg Bwd Segment Size   & Packet Length Std           \\
\textbf{9}               & Max Packet Length               & Flow IAT Std              & Packet Length Std            & Down/Up Ratio    & Packet Length Variance   \\
\textbf{10}              & Subflow Fwd Bytes               & Down/Up Ratio             & Packet Length Mean           & Bwd Packet Length Max     & Average Packet Size      \\ \hline
\textbf{CICIDS-2017}  & \textbf{VarGrad}            & \textbf{SquareGrad} & \textbf{LIME} & \textbf{GradientInput}    & \textbf{Voting}                                    \\ \hline
\textbf{1}              & Packet Length Mean           & Subflow Bwd Bytes           & Destination Port              & Packet Length Std       & Packet Length Mean     \\
\textbf{2}               & Subflow Bwd Bytes           & Max Packet Length           & Avg Bwd Segment Size          & Packet Length Variance  & Destination Port          \\       
\textbf{3}               & Bwd Packet Length Max       & Bwd Packet Length Max       & Bwd Packet Length Mean        & Bwd Packet Length Max   & Total Length of Fwd Packets         \\
\textbf{4}                & Max Packet Length          & Packet Length Mean          & Bwd Packet Length Max         & Avg Bwd Segment Size     & Total Length of Bwd Packets      \\
\textbf{5}                & Avg Bwd Segment Size       & Avg Bwd Segment Size        & Max Packet Length             &  Bwd Packet Length Mean   & Bwd Packet Length Max       \\
\hline
\textbf{6}               & Bwd Packet Length Mean      & Flow IAT Std                & Packet Length Std             & Flow IAT Std      & Bwd Packet Length Mean      \\
\textbf{7}               & Flow IAT Std                & Bwd Packet Length Mean      & Down/Up Ratio                 & Init\_Win\_bytes\_forward    & Max Packet Length     \\
\textbf{8}                &  Bwd Packets/s             & Bwd Packets/s               & FIN Flag Count                &  Max Packet Length     & Packet Length Std      \\
\textbf{9}               &  Subflow Fwd Packets        & Total Length of Bwd Packets & Init\_Win\_bytes\_forward     & Packet Length Mean     & Packet Length Variance     \\
\textbf{10}              & Total Length of Bwd Packets & Init\_Win\_bytes\_forward   & Flow IAT Std                  &  Down/Up Ratio           & Average Packet Size   \\ \hline \hline

\textbf{SIMARGL2021} & \textbf{Saliency}           & \textbf{Integrated Gradients}  & \textbf{Occlusion}    & \textbf{SmoothGrad}      & \textbf{DeconvNet}         \\ \hline
\textbf{1}              & TCP\_WIN\_SCALE\_IN          & TCP\_WIN\_SCALE\_IN       & TCP\_WIN\_SCALE\_IN          &  TCP\_WIN\_SCALE\_IN & TCP\_WIN\_SCALE\_IN     \\
\textbf{2}               & L4\_SRC\_PORT               & TCP\_WIN\_MSS\_IN         & TCP\_WIN\_MSS\_IN            & L4\_SRC\_PORT        & L4\_SRC\_PORT         \\       
\textbf{3}               & FLOW\_ID                    & TCP\_WIN\_MIN\_IN         & TCP\_WIN\_MIN\_IN            & FLOW\_ID             &  FLOW\_ID       \\
\textbf{4}                & TCP\_WIN\_MIN\_OUT         &  L4\_SRC\_PORT            & L4\_SRC\_PORT                & TCP\_WIN\_MIN\_OUT   & TCP\_WIN\_MIN\_OUT          \\
\textbf{5}                & TCP\_WIN\_MIN\_IN          & FLOW\_ID                  & FLOW\_ID                     & TCP\_WIN\_MIN\_IN    &  TCP\_WIN\_MIN\_IN          \\
\hline
\textbf{6}               & TCP\_WIN\_SCALE\_OUT        & TCP\_WIN\_MAX\_IN         & TCP\_WIN\_MAX\_IN            & TCP\_WIN\_SCALE\_OUT & TCP\_WIN\_SCALE\_OUT         \\
\textbf{7}               & TCP\_WIN\_MAX\_OUT          & PROTOCOL                  & PROTOCOL                     & TCP\_WIN\_MAX\_OUT   & TCP\_WIN\_MAX\_OUT       \\
\textbf{8}                & TOTAL\_FLOWS\_EXP          & FIRST\_SWITCHED           & FIRST\_SWITCHED              & TOTAL\_FLOWS\_EXP    & TOTAL\_FLOWS\_EXP      \\
\textbf{9}               & PROTOCOL                    &  TCP\_FLAGS               & TCP\_FLAGS                   & PROTOCOL             & PROTOCOL     \\
\textbf{10}              & TCP\_FLAGS                  & TCP\_WIN\_SCALE\_OUT      & TCP\_WIN\_SCALE\_OUT         & -TCP\_FLAGS          & TCP\_FLAGS    \\ \hline
\textbf{SIMARGL2021} & \textbf{VarGrad}             & \textbf{SquareGrad}   & \textbf{LIME}                    & \textbf{GradientInput}                & \textbf{Voting}                          \\ \hline
\textbf{1}           & TCP\_WIN\_SCALE\_IN          & TCP\_WIN\_SCALE\_IN   & TCP\_WIN\_SCALE\_IN              & TCP\_WIN\_SCALE\_IN     & FLOW\_DURATION\_MILLISECONDS        \\
\textbf{2}           & FLOW\_ID                     & L4\_SRC\_PORT         & TCP\_WIN\_MSS\_IN                &  TCP\_WIN\_MSS\_IN      & FIRST\_SWITCHED       \\       
\textbf{3}           & L4\_SRC\_PORT                & FLOW\_ID              & TCP\_WIN\_MIN\_IN                & TCP\_WIN\_MIN\_IN       & TOTAL\_FLOWS\_EXP       \\
\textbf{4}           & TCP\_WIN\_MIN\_IN            & TCP\_WIN\_MIN\_OUT    & L4\_SRC\_PORT                    & L4\_SRC\_PORT           & TCP\_WIN\_MSS\_IN       \\
\textbf{5}           & TCP\_WIN\_SCALE\_OUT         & TCP\_WIN\_MIN\_IN     & FLOW\_ID                         & FLOW\_ID                & LAST\_SWITCHED       \\
\hline
\textbf{6}           &  TCP\_WIN\_MIN\_OUT          & TCP\_WIN\_SCALE\_OUT  & TCP\_WIN\_MAX\_IN                & TCP\_WIN\_MAX\_IN       & TCP\_WIN\_MAX\_IN      \\
\textbf{7}           & TCP\_WIN\_MAX\_OUT           & TCP\_WIN\_MAX\_OUT    & PROTOCOL                         & PROTOCOL                & TCP\_WIN\_MIN\_IN      \\
\textbf{8}           & PROTOCOL                     & TOTAL\_FLOWS\_EXP     & FIRST\_SWITCHED                  & FIRST\_SWITCHED         & TCP\_WIN\_MIN\_OUT     \\
\textbf{9}           &  TCP\_WIN\_MSS\_IN           & PROTOCOL              & TCP\_FLAGS                       & TCP\_FLAGS              & PROTOCOL       \\
\textbf{10}          &  TCP\_FLAGS                  & TCP\_FLAGS            & TCP\_WIN\_SCALE\_OUT             & TCP\_WIN\_SCALE\_OUT    & TCP\_WIN\_MAX\_OUT        \\ \hline %\hline                             
\end{tabular}
}
\label{tbl:Xplique_methods}
\end{table*}

\end{document}